\RequirePackage[hyphens]{url}
\documentclass[12pt]{article}
\usepackage[margin=1in]{geometry}
\usepackage{amsmath}
\usepackage{amsthm}
\usepackage{graphicx}
\usepackage[official]{eurosym}
\usepackage{url}
\usepackage{tablefootnote}
\usepackage{multirow}
\usepackage{subfig}
\usepackage{float}
\usepackage[hidelinks]{hyperref}
\usepackage{url}
\usepackage{breakurl}
\hypersetup{
    colorlinks=true,
    linkcolor=blue,
    filecolor=magenta,      
    urlcolor=cyan,
}
\usepackage{caption}
\usepackage{booktabs}
\usepackage{float}
\providecommand{\keywords}[1]{\textbf{keywords ---} #1}
\usepackage{tikz}
\usetikzlibrary{automata} 
\usepackage{graphics}
\title{Forecasting the levels of disability in the older population of England: Application of neural nets}

\begin{document}

\author{Marjan Qazvini \footnote{Correspondence to: Marjan Qazvini, marjan.qazvini@gmail.com}}
\date{}

\maketitle

\begin{abstract}
Deep neural networks are powerful tools for modelling non-linear patterns and are very effective when the input data is homogeneous such as images and texts. In recent years, there have been attempts to apply neural nets to heterogeneous data, such as tabular and multimodal data with mixed categories. Transformation methods, specialised architectures such as hybrid models, and regularisation models are three approaches to applying neural nets to this type of data. 
In this study, first, we apply $K$-modes clustering algorithm to define different levels of disability based on responses related to mobility impairments, difficulty in performing Activities of Daily Livings (ADLs), and Instrumental Activities of Daily Livings (IADLs). We consider three cases, namely binary, 3-level, and 4-level disability. We then try Wide \& Deep,   TabTransformer, and TabNet models to predict these levels using socio-demographic, health, and lifestyle factors. We show that all models predict different levels of disability reasonably well with TabNet outperforming other models in the case of binary disability and in terms of 4 metrics. We also find that factors such as urinary incontinence, ever smoking, exercise, and education are important features selected by TabNet that affect disability.
\end{abstract}

\keywords{Neural Nets, NLP, ELSA, Transformers, $K$-modes algorithm}

\section{Introduction}
According to World Health Organisation (WHO), about $1.3$ billion people experience significant disability, which is equivalent to $16\%$ of the world's population. Disability reduces life expectancy, and disabled people suffer from physical and mental health conditions, such as heart diseases, dementia, etc. The number of people with disability is growing, and this is affected by socio-economic factors such as age, gender, employment status, etc.\footnote{https://www.who.int/news-room/fact-sheets/detail/disability-and-health -- accessed March 2023} The International Classification of Functioning, Disability and Health (ICF) describes disability as impairments of body functions and structures, activity limitations and participation restrictions.\footnote{https://www.cdc.gov/nchs/data/icd/icfoverview\_finalforwho10sept.pdf -- accessed March 2023} On the other hand, Washington Group Short Set on Functioning (WG-SS) classifies disability based on 6 questions on functioning.\footnote{https://www.washingtongroup-disability.com/question-sets/wg-short-set-on-functioning-wg-ss accessed March 2023} One problem with surveys, is that they use different definition and classification of disability (Freedman et al., 2004). Palmer and Harley (2012) compare impairment and functioning disabilities in ICF and WG-SS and the shortcomings of such questions. In this study, we investigate disability among participants in the English Longitudinal Study of Ageing (ELSA) survey. In this survey, which is carried out every two years, participants are asked to self-rate their health conditions and their ability in performing activities such as walking 100 yards, stooping, lifting (physical activity), dressing, bathing, eating (Activities of Daily Livings (ADL)), using a map, preparing a hot meal, and shopping (Instrumental Activities of Daily Livings (IADL)). However, one issue with these disability factors is that they do not indicate the severity of the disability. There are different methods to measure the severity of disability. For example, Manton and Gu (2005) use Grade of Membership (GoM) models to categorise the levels of disability in the US elderly population. They classify disability into different levels  of non-disabled (active, modest impairment, moderate impairment), disabled (IADL, ADL, frail), and institutionalised. They consider the impact of disability on medical costs and find an improvement in disability trends over the years 1982-1999. Gobbens et al. (2014) use frailty factors such as gait speed, physical activity, hand grip strength, and fatigue to predict ADL and IADL disability among Dutch elderly. Pongiglione et al. (2015) investigate the association between disability and mortality using ELSA. They apply latent variable models to identify disability categories and discrete-time survival analysis to study the relationship between disability and mortality. Their results show that disabled women survive longer than disabled men. In another study, Pongiglione et al. (2017) compare binary and multi-level disability and suggest a 4-category disability for disability classification.
\par Some studies consider disability forecasting. Understanding factors that affect disability and the prevalent categories of disability in the future can help with prevention and decisions regarding health care and facility provision in society. The factors used for the prediction of disability fall into three categories of social, physical, and psychological factors. Vermeulen et al. (2011) provide a review of studies that use physical frailty factors to predict ADL disability. Using regression analysis, Gobbens et al. (2012) also find that physical frailty has the largest predictive power among other factors. They use the Tilburg Frailty Indicator (TFI) to predict disability, quality of life, and health care utilisation among the older population of a municipality in the Netherlands. Chen et al. (2016) use a state-transition simulation model to predict disability among the older population of Japan by 2040. They measure disability based on the number of ADLs and IADLs that the population may have difficulty performing. Fong et al. (2015) use a stochastic frailty model to forecast disability among older populations of Australia. They find frailty among older females happen faster than among older men and that people with disability who need care services will be more than the government's estimation by 2050. 
\par In this study, we use neural networks (NNs) to forecast disability among the older population of England. ELSA is a tabular dataset. A tabular dataset, can be represented in a spreadsheet and includes both nominal and categorical features (variables) as opposed to texts, images, and audio, which are homogeneous. NNs have proved to be very successful with homogeneous data. In recent years, attempts have been made to apply shallow NNs and deep NNs (DNNs) to tabular data. One approach is regularisation and the assignment of a penalty to all weights. The justification for such methods is that in a tabular dataset, all features are important, and regularisation can reduce the number of features and help with interpretation (Shavitt and Segal, 2018). Early stopping (Yao et al., 2007), dropout (Srivastava et al., 2014), the dynamic learning rate of an optimiser, batch normalisation (BN) (Ioffe and Szegedy, 2015), etc. are other methods of regularisation. Kadra et al. (2021) apply a combination of these methods to 40 tabular datasets and demonstrate that a combination of these methods not only improves the performance of both Multilayer Perceptrons (MLPs) and networks structures tailored to tabular data but also, in some cases, outperform Gradient Boosting Decision Trees (GBDTs). Some studies introduce network architectures specific to tabular data. These architectures are mainly inspired by the networks applied to Natural Language Processing (NLP). Some of these architectures are Transformers (Vaswani et al., 2017) with attention-based blocks, such as TabTransformer (Huang et al., 2020), TabNet (Arik and Pfister, 2021), etc. TabNet is a sequence network with an attention mask to select features. Huang et al. (2020) show that TabTransformer performs better than MLP and is as good as GBDTs. Somepalli et al. (2021) introduced SAINT, which is a network with inter-sample attention over both rows and columns. Adaptive relation modelling network (ARM-Net) is another architecture for tabular data. It is introduced by Cai et al. (2021) and is composed of multi-head gated attention computation and exponential transformation. Their model considers the interaction between features by looking at interaction order and interaction weights for each cross feature. Some studies introduce hybrid models, which are a combination of different models and algorithms with DNNs. For example, Cheng et al. (2016) propose Wide \& Deep model, which is a combination of a generalised linear model and a DNN for recommender systems. By combining these two models, they introduce feature interaction into their models. (See also Lian et al., 2018 and Huang et al., 2019). Some models combine DT algorithms and DNNs. The idea is that as DTs are good at interpretability, combining DTs with DNNs can help with the interpretation of the results. For example, Frosst and Hinton (2017) propose a soft DT where the results of NNs can be interpreted by a DT. They find that their model performs better than a DT but worse than a NN. On the other hand, TabNN, introduced by Ke et al. (2018), uses the results from GBDT to train a NN. Fiedler (2021) considers some modifications, such as ghost batch normalisation, leaky gates, etc., to existing networks and demonstrates that the modified networks can perform as well as the tailor-made networks and even GBDTs. See Borisov et al. (2022) for a comprehensive review of the application of DNNs to tabular data.    

\par In this study, we first use $K$-modes algorithm (Huang, 1998, and He et al., 2006) to categorise disability. Similar to Pongiglione et al. (2017), we consider disability with 2, 3, and 4 levels. We then apply Wide \& Deep, TabTransformer, and TabNet models to our dataset to predict these levels of disability. The rest of this paper is organised as follows: Section 2 analyses the ELSA dataset and explains how we change an unsupervised problem into a supervised problem. Section 3 describes the models and algorithms. Section 4 discusses the results, and Section 5 concludes.  

\begin{table}[h]
\caption{Demographic factors and disability levels}
\label{tab:demog}
\footnotesize
\centering
\scalebox{0.7}{
\begin{tabular}{lccccccc}\toprule
				& Binary disability &  \multicolumn{2}{c}{ 2-level disability}&   \multicolumn{3}{c}{3-level disability}       \\\midrule
				& Disability $y=1$ &  Mild $y=1$ & Severe $y=2$ & Mild  $y=1$ & Moderate  $y=2$ & Severe  $y=3$  & Total\\\midrule
Female			& 1,074		    &  3,889	& 950		& 3,889	    & 950		        & 1,058	        & 6,113\\
Male				& 673		    &  3,766	& 603		& 3,766	    & 603			& 598		& 5,106\\\midrule
Married			& 935		    &  5,370	& 829		& 5,370	    & 829			& 1,052		& 7,463\\
Single			& 99			    &  420		& 91			& 420	    & 91			& 88			& 617\\
Divorced			& 203		    & 809		& 179		& 809	    & 179			& 163		& 1,189\\
Widowed			& 510		    & 1,056		& 454		& 1,056	    & 454			& 353		& 1,950\\\midrule
Retired			& 1,057		    & 3,507		& 936		& 3,507	    & 936			& 951		& 5,623\\
Employed			& 68			    & 2,549		& 52			& 2,549	    & 52			& 317		& 2,951\\
Care for family	 	& 135		    & 717		& 111		& 717	    & 111			& 199		& 1,080\\
Permanently sick	& 458		    & 135		& 432		& 135	    & 432			& 106		& 701\\
Self-employed		& 14			    & 551		& 10			& 551	    & 10			& 56			& 625\\
Unemployed		& 3			    & 99		& 3			& 99		    & 3			& 13			& 115\\
Other			& 12			    & 97		& 9			& 97		    & 9			& 14			& 124\\\midrule
Education (high)	& 170		    & 1,983		& 154		& 1,983	    & 154			& 302		& 2,472\\
Middle			& 332		    & 2,203		& 301		& 2,203	    & 301			& 428		& 2,987\\
Low				& 1,245		    & 3,469		& 1,098		& 3,469	    & 1,098			& 926		& 5,760\\\midrule
Disabled (\%)		& 15.57		    & 68.23	 	& 13.84		& 68.23	    & 13.84		 	& 14.76		& 	    \\\bottomrule
\end{tabular}}
\end{table}

\section{Description of the data}
\label{sec:data}

\begin{figure}[h]
\centering
\includegraphics[scale = 0.57]{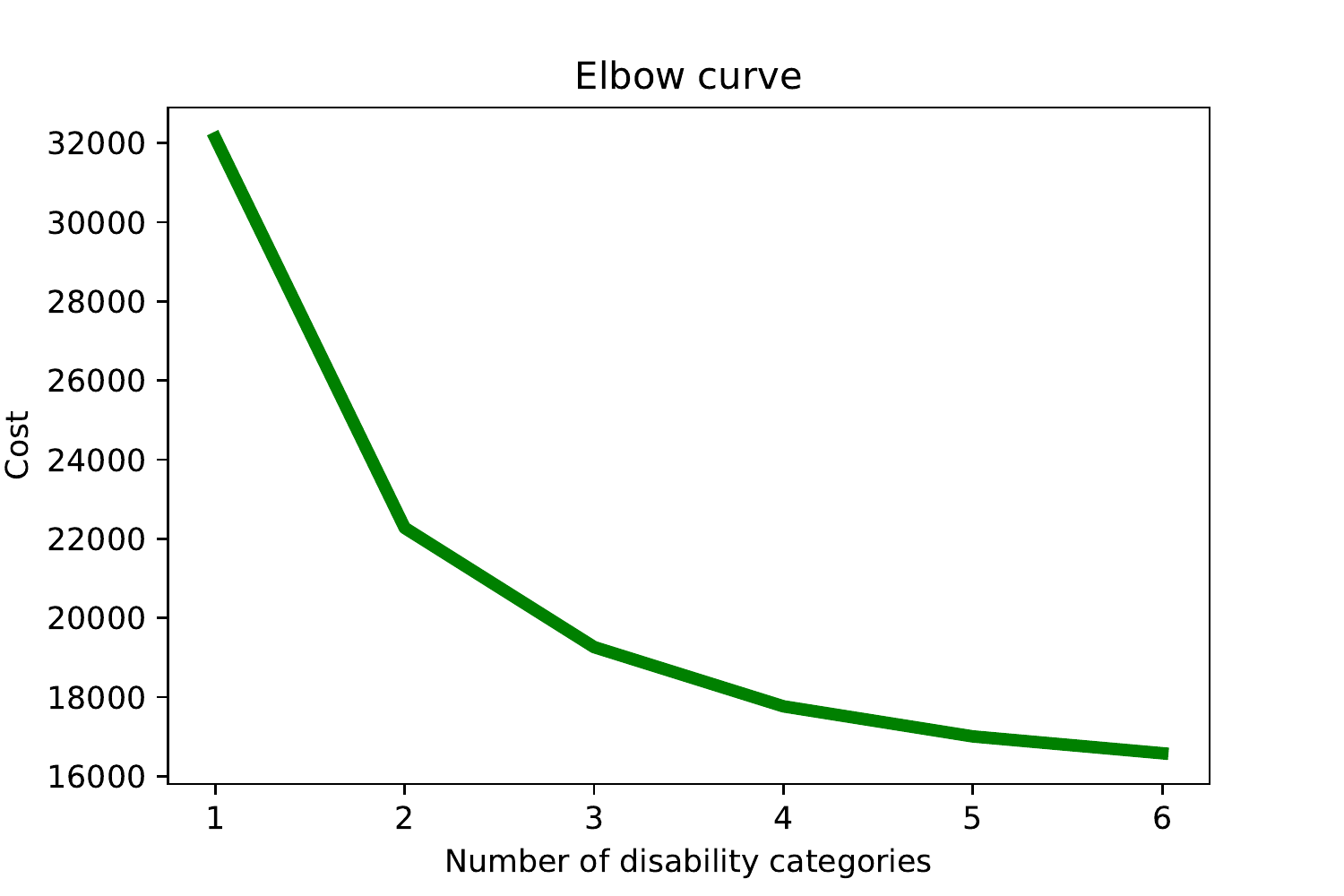}
\caption{Elbow curve to determine the optimal levels of disability}
\label{fig:elbow}
\end{figure}

The English Longitudinal Study of Ageing (ELSA) is a collection of economic, social, psychological, cognitive, health, biologica,l and genetic data. The study commenced in 2002, and the sample has been followed up every 2 years. The first cohort was selected from respondents to the Health Survey for England (HSE) in 1998, 1999, and 2001 and included people born on or before February 29, 1952, i.e., aged 50 and older. The first ELSA wave was in 2002-2003. (For more information on ELSA, sampling, and interview process, see, for example, Steptoe et al., 2013 and Blake et al., 2015). In this study, we only consider the \textit{core members} from the first wave, i.e., age-eligible members who participated in the HSE and are interviewed in the first wave of ELSA when invited to join (Bridges et al. 2015). There are $11,391$ core members in wave 1. We consider questions related to disability, health, socio-economic, and lifestyle factors as our features and questions related to mobility impairments and difficulty performing ADLs and IADLs (see, Table \ref{tab:ADL}, for details) as target features. Our purpose is to predict the levels of disability among participants. In ELSA, however, the participants are only asked whether they have difficulty performing these activities or not, and there is no indication as to how much it is difficult for them to carry out such activities. To classify participants according to different levels of disability, we use $K$-modes clustering algorithm, which is an unsupervised learning method and suitable for categorical features (Huang, 1998, and He et al., 2006 ). We use the elbow curve, illustrated by Figure \ref{fig:elbow}, to find the optimal number of clusters, i.e., disability levels. We use these levels and group participants accordingly. The resulting categories are our class labels (target feature). We consider 3 cases: 2 levels of disability with categories disability (1) and no disability (0), 3 levels of disability with categories severe (2), mild (1), and no disability (0), 4 levels of disability with categories severe (3), moderate (2), mild (1), and no disability (0). Further, we only consider individuals whose responses to questions related to mobility impairment, ADL, and IADL difficulties, i.e., questions that formed our target feature are complete. Therefore, we are left with 11,219 participants (data points). We have 1 numerical feature (age) that we need to standardise and 24 categorical features. Responses such as \textit{refusal}, \textit{don't know}, and \textit{not applicable} are taken as missing values. We use the mean to 
fill in missing values of numerical features and the most frequent category to fill in missing values of categorical features. Table \ref{tab:demog} presents the frequency of participants with different levels of disability according to sex, marital status, employment status, and education level. We observe that in the case of binary disability, $15.57\%$ of participants are categorised as disabled, in the case of 3-level disability, $68.23\%$ of participants have mild disability, and $13.84\%$ suffer from severe disability.

\begin{figure}[h]
\centering
\includegraphics[scale = 0.25]{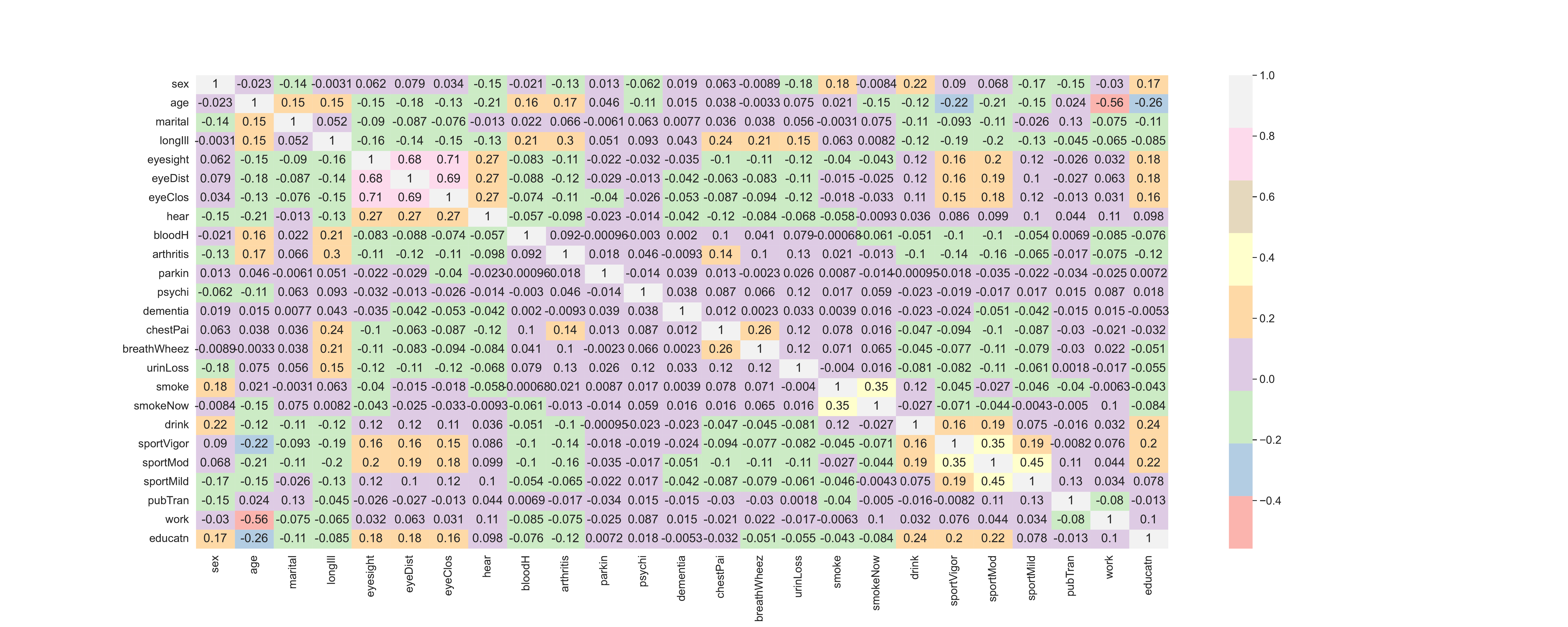}
\caption{Spearman's correlation}
\label{fig:heat}
\end{figure}

\begin{figure}[H]
\centering
\includegraphics[scale = 0.3]{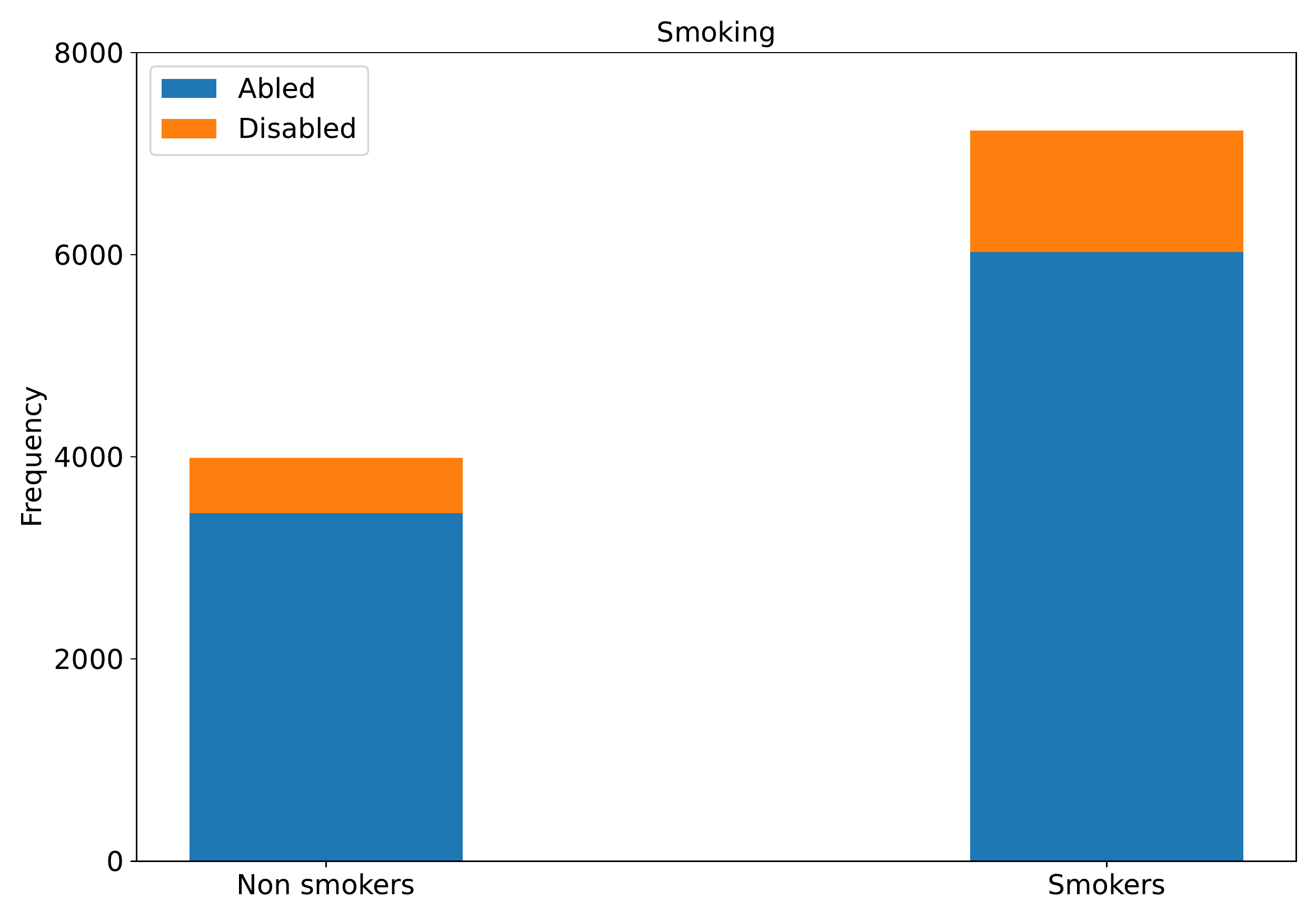}
\includegraphics[scale = 0.3]{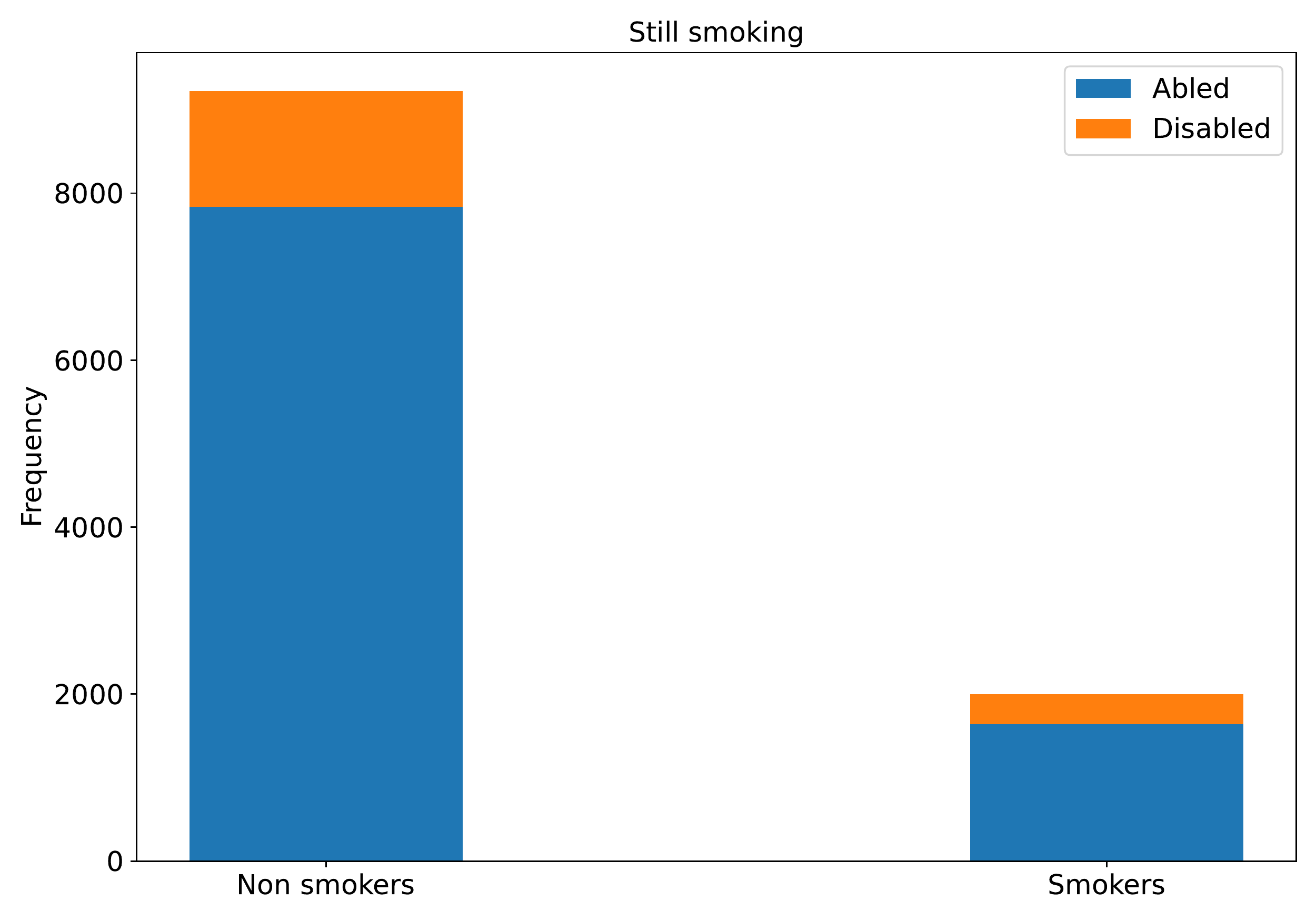}
\includegraphics[scale = 0.3]{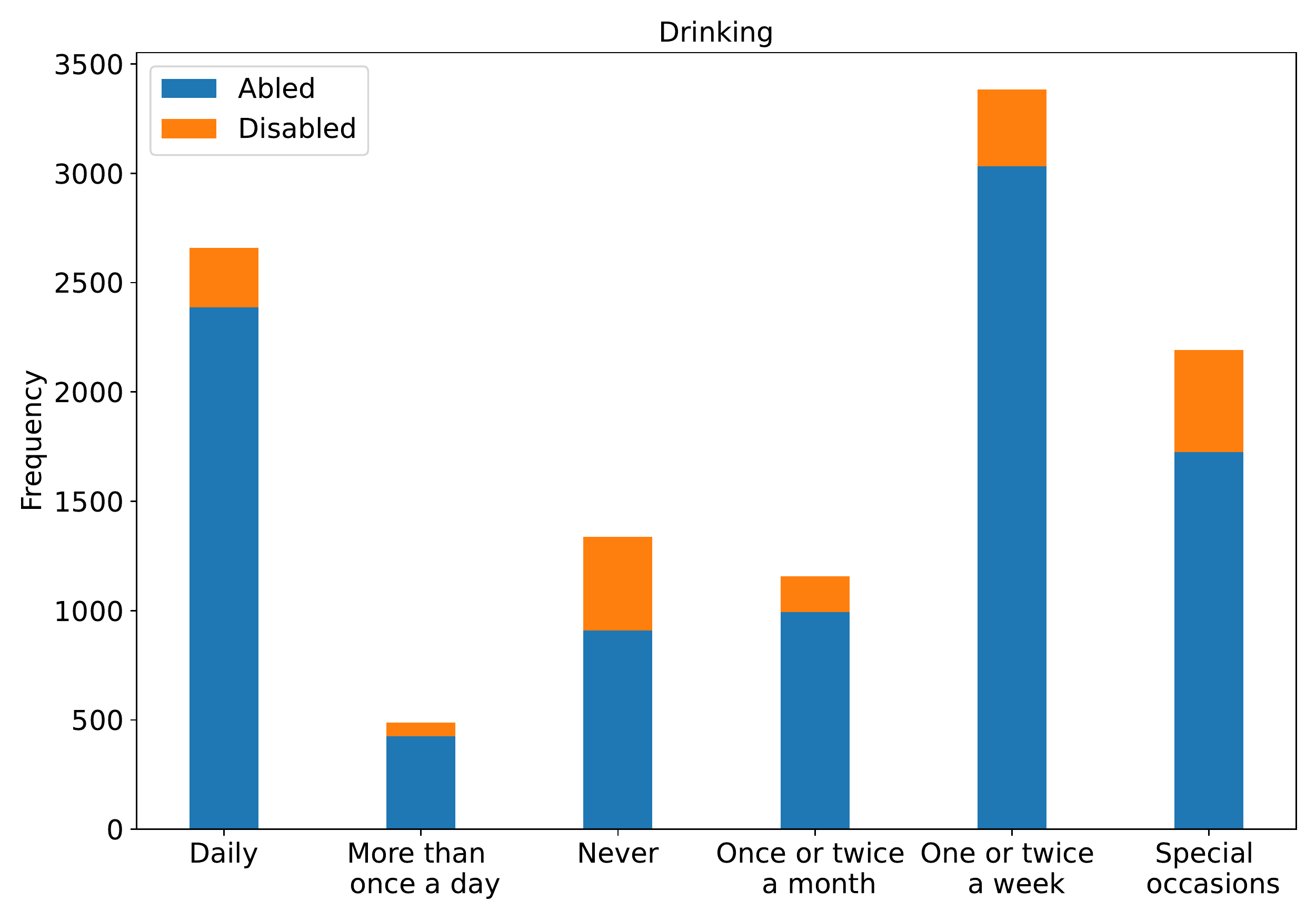}
\includegraphics[scale = 0.3]{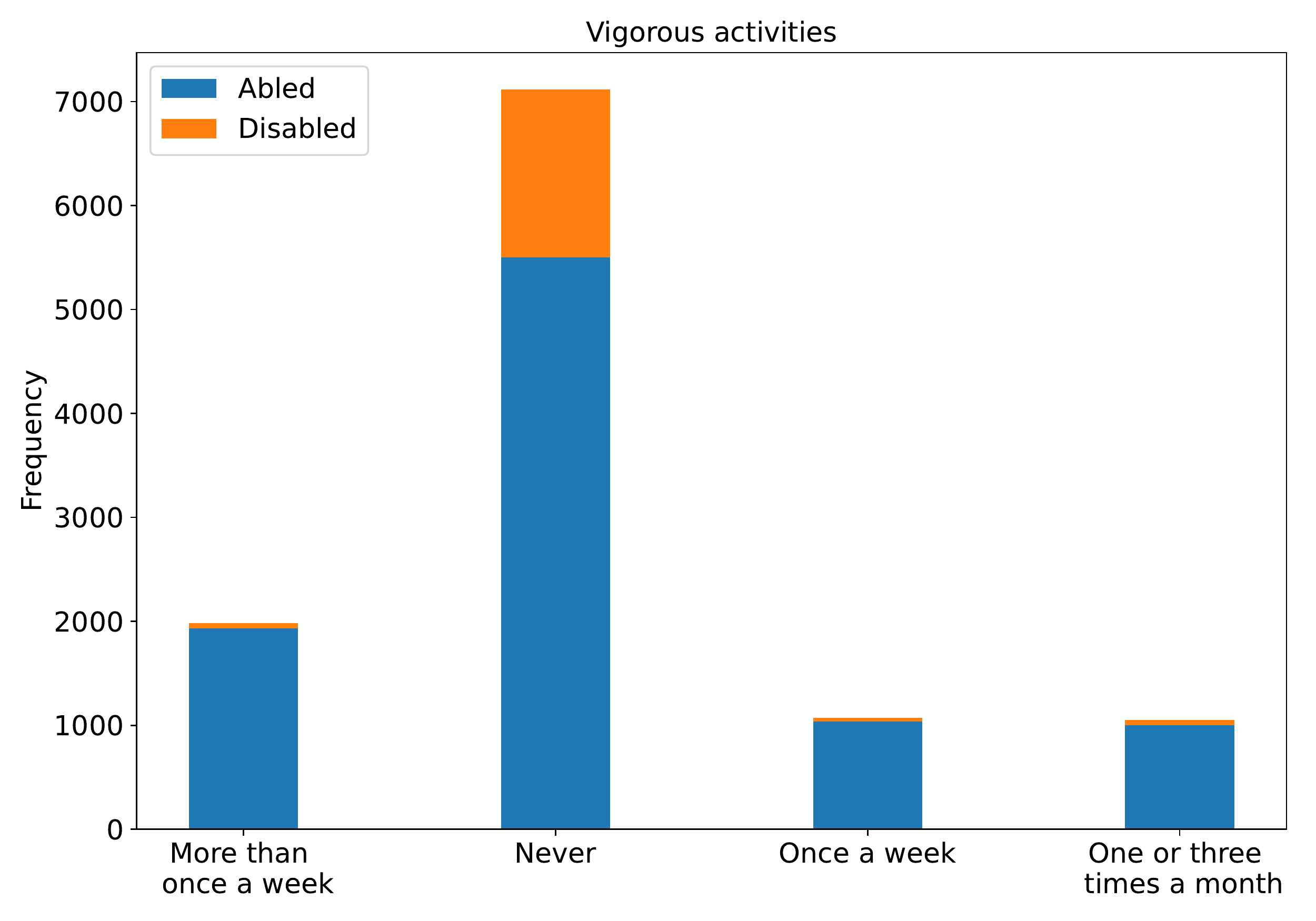}
\includegraphics[scale = 0.3]{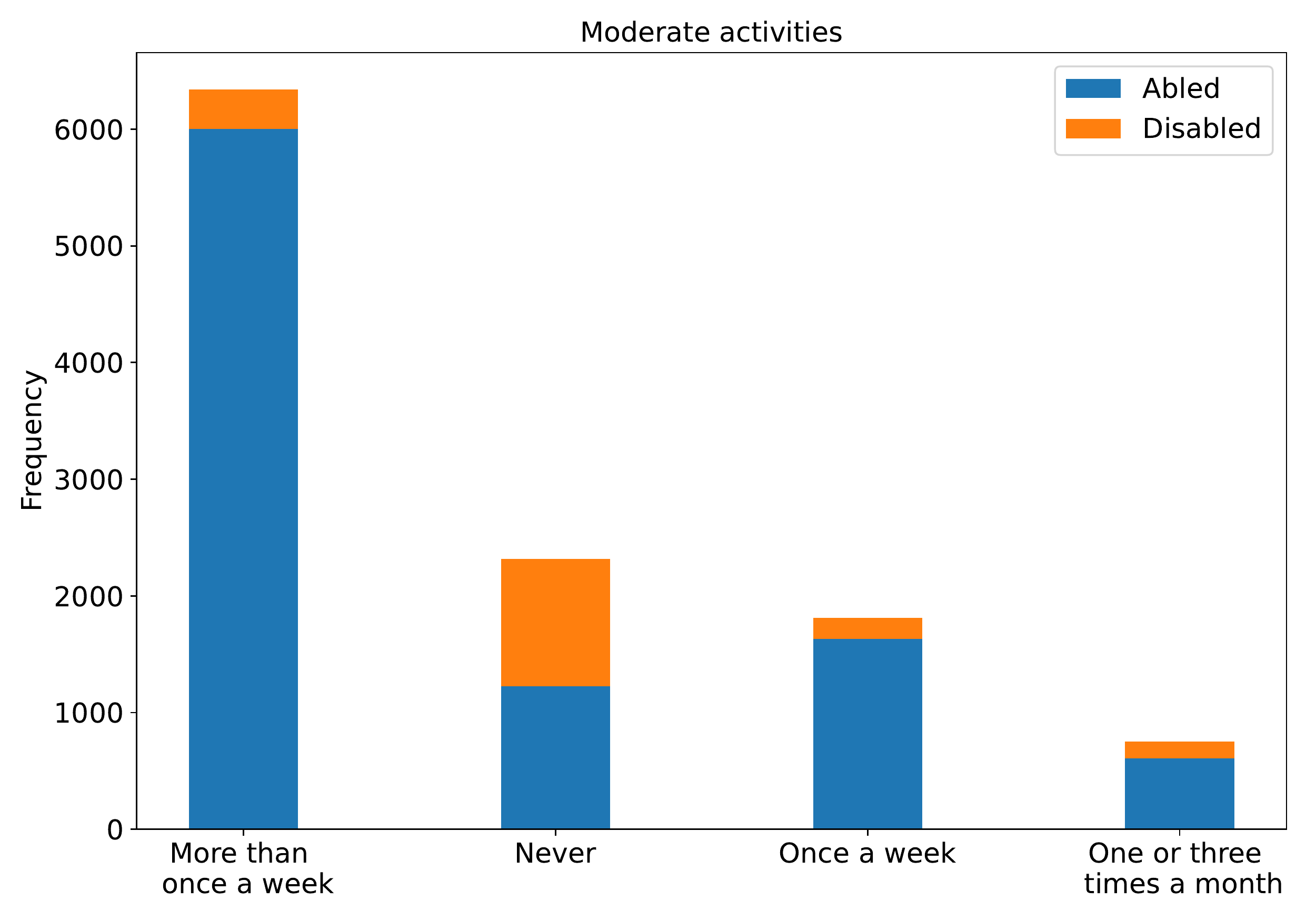}
\includegraphics[scale = 0.3]{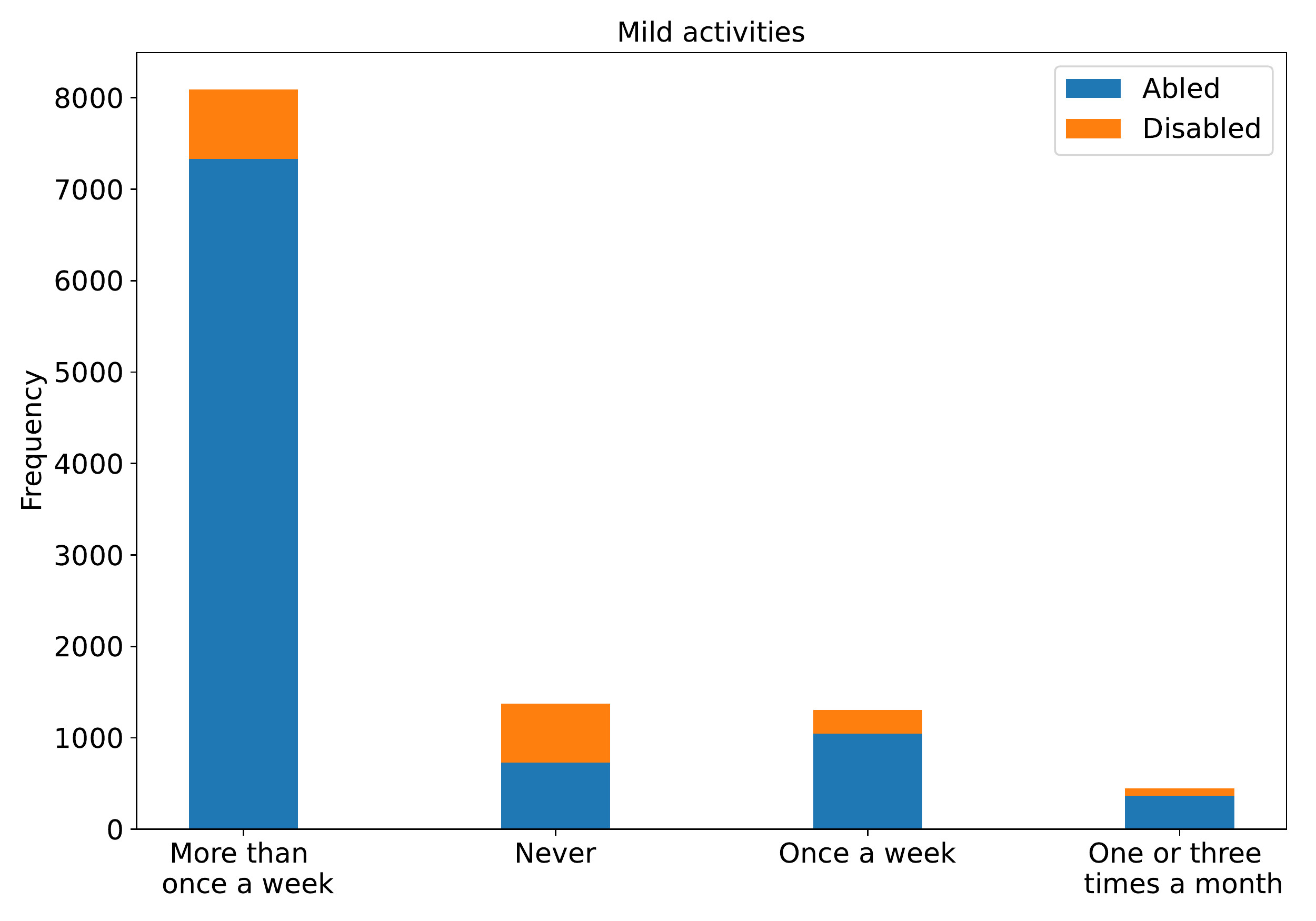}
\includegraphics[scale = 0.3]{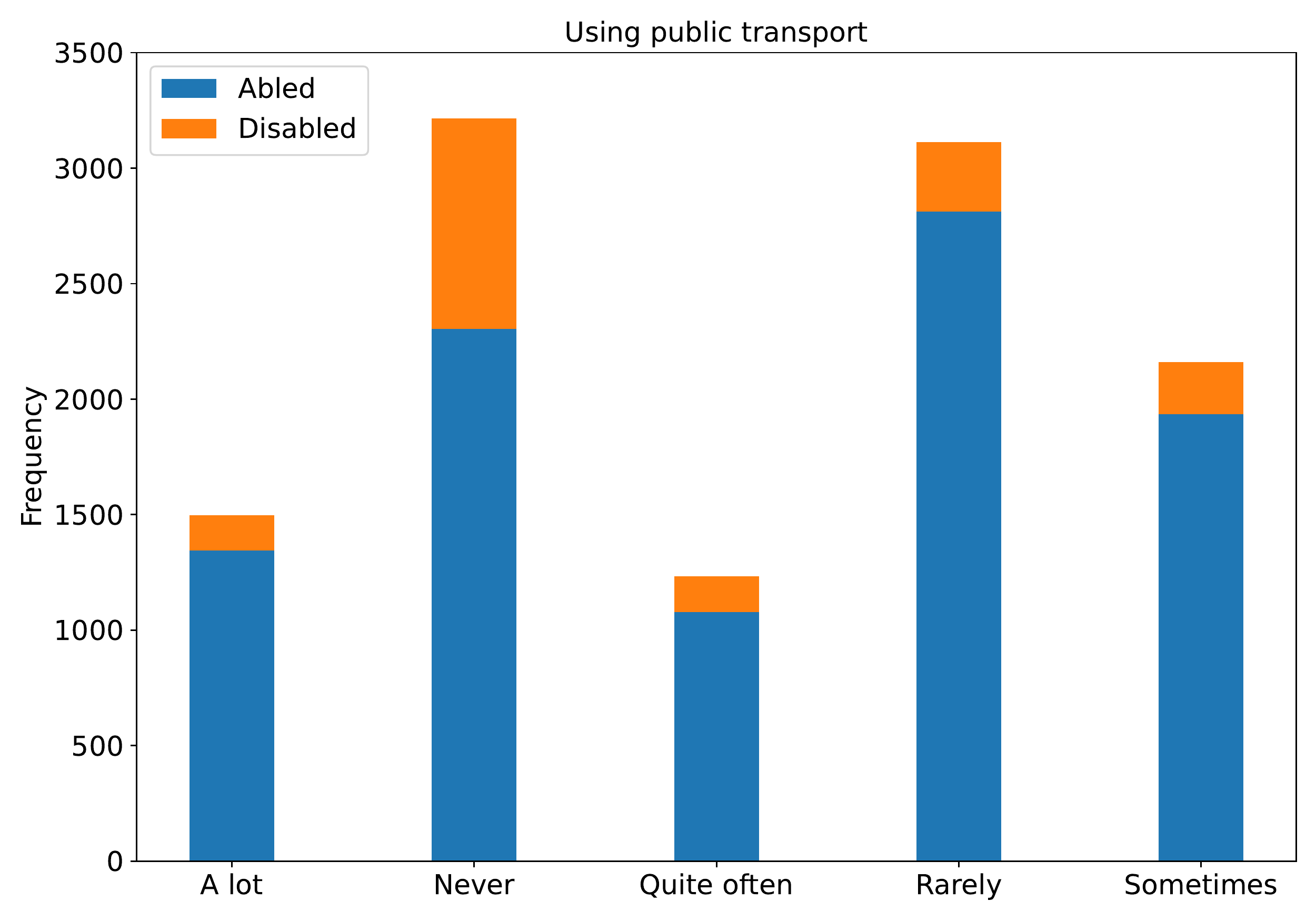}
\caption{Lifestyle factors and the frequency of participants with and without disability}
\label{fig:lifestyle}
\end{figure}

\noindent With a 4-level disability, $68.23\%$ of participants have mild disability, $13.84\%$ have moderate, and $14.76\%$ have severe disability. Figure \ref{fig:heat} shows Spearman's rank correlation (Denuit et al. 2006) heatmap of all features. As we can see, there are relationships between some features that need to be addressed in the models that we use in Section \ref{sec:res}. Figure \ref{fig:lifestyle} shows the number of participants with different lifestyle factors such as ``smoking", ``drinking", and ``exercising" and the portion of disabled and abled people with each of these factors. This is also presented in Table \ref{tab:lifstyl} for different levels of disability. The last pre-processing step before feeding our data into models, is to encode categorical features. We use the label encoding method to convert categorical features into numerical values.

\section{Models and algorithms}
In this section, we explain the models and algorithms that we use in this study. Let $D = \{(\mathbf{x}_i, \mathbf{y}_i)\}_{i=1}^{N}$ denote our dataset with $N$ samples, $\mathbf{x}_i = (x_{i1}, x_{i2}, \dots, x_{im})$ denote a vector of $m$ features, and $\mathbf{y}_i$ denote the labels (target features) in a supervised learning model. In a binary classification problem, $\mathbf{y}_i \in \{0,1\}$ and in a multi-class classification $\mathbf{y}_i \in \{1, 2, \dots, c\}$. Suppose our training set is drawn from the joint probability distribution $P_{\theta}(\mathbf{x}, \mathbf{y})$. Then, in the case of binary classification, we have a logistic function, given by
\begin{eqnarray}
\label{eq:logit}
 P_{\theta}(y_i = 1 | \mathbf{x} ) = \frac{1}{1 + \exp(- \theta^{T} \mathbf{x}_i)},
\end{eqnarray}
and in the case of multi-class classification, a softmax function, given by
\begin{eqnarray}
\label{eq:softmax}
P_{\theta}(y_i = c | \mathbf{x}) = \frac{\exp(\theta^T \mathbf{x}_i)}{ \sum_{j=1}^N \exp(\theta^T \mathbf{x}_j)}.
\end{eqnarray}
We use logistic as the output activation function in the case of binary disability and softmax in the case of 3 and 4-level disability.

\subsection{$K$-modes clustering algorithm}
\label{sec:Kmode}
As explained in Section \ref{sec:data}, in order to determine different levels of disability, we need to group participants based on their responses to questions related to mobility (10 questions), ADLs (6 questions), and IADLs (7 questions). We can apply $K$-modes clustering algorithm, which similar to $K$-means algorithm, is an unsupervised method. The only difference is that the latter applies to numerical data, whereas the former applies to categorical data. Suppose we have $K$ clusters, say, disability levels. Denote $Q = (q_{k1}, q_{k2}, \dots, q_{k m})$ be the vector of the $k$-th cluster with $m$ attributes (23 questions in our case). The vector $Q$ can be any observation. The optimal number of clusters is then obtained by minimising
\begin{eqnarray}
D(\mathbf{X}, \mathbf{Q}) = \sum_{i=1}^N \sum_{j=1}^m \delta(x_{ij}, q_{kj}), \quad 		k = 1, 2, \dots, K \nonumber,
\end{eqnarray}  
where 
\begin{eqnarray}
\delta(x, y) =
\begin{cases}
0	&  (x = y) \nonumber\\
1 	& (x \neq y), 
\end{cases}
\end{eqnarray}
and $D$ is a measure of dissimilarity between the vector $Q$ and each data point. Then, each participant (datapoint) is assigned to a cluster with the least dissimilarity. After allocating our participants into $K$ clusters, we need to update $Q$. In each group, we select the most frequent category of the features to build new vectors. We repeat this process until the cluster becomes stable, i.e. there is no change in the allocated data points (Huang, 1998, and He et al., 2006).

\subsection{DNN architectures}
Deep learning is a mathematical framework for learning representations from data. It includes successive layers (data transformations) that are learned through neural networks. Vanilla neural networks (NNs) are multilayer perceptrons (MLPs), also called feed-forward neural networks (FNN), Dense layer, and Fully-connect (FC). It is a mapping $\mathbf{y} = f_{\theta}(\mathbf{x})$. A FNN consists of different layers, denoted by $f^{(l)}$ for $l \in \{1, 2, \dots, L\}$ and the information flows through the input layer, hidden layers, and the output layer, i.e.
\begin{eqnarray}
\label{eq:fFNN}
f_{\theta}(\mathbf{x}) = f_{\theta}^{(l)} \left(f_{\theta}^{(l-1)}\left(\dots \left(f_{\theta}^{(1)} (\mathbf{x}) \right) \right)  \right),
\end{eqnarray}
where $l$ is the number of layers that determines the depth of the network. Each layer consists of different units, which in a dense network, all units of two consecutive layers are connected. Let $\mathbf{z}^{(l)}$ be the input vector to layer $l$, and $\mathbf{a}^{(l)}$ be the output vector from layer $l$ with $\mathbf{a}^{(0)} = \mathbf{x}$ being the input layer and $\mathbf{a}^{L} = \mathbf{y}$, the output layer. Then, the feed-forward propagation for layer $l$ and unit $i$ is given by
\begin{eqnarray}
\label{eq:ff}
z_i^{(l)} &=& \mathbf{w}_i^{(l)} \mathbf{a}^{(l-1)} + b_i^{(l)}, \nonumber\\
a_i^{(l)} &=&  f_{\theta}(z_i^{(l)})\nonumber,
\end{eqnarray}
where $\theta = \{\mathbf{w}, b\}$ are the weights and bias, respectively, and $f$ is the activation function. The model's parameters can be learned through a process, called back propagation algorithm (Rumelhart et al., 1986). Let $J(\theta)$ be the cost function that compares the output $\hat{y}$ with the actual $y$, and is given by
\begin{eqnarray}
\label{eq:cost}
J(\theta) = \frac{1}{N} \sum_{i=1}^N L_{\theta}(\hat{y}_i, y_i),
\end{eqnarray}
where $L$ is the loss function. Our aim is to minimise the cost function by making the necessary adjustments to the parameters. The updated parameter using the gradient descent algorithm is given by
\begin{eqnarray}
\label{eq:sgd}
\theta := \theta - \alpha \nabla_{\theta} J(\theta),
\end{eqnarray}  
where $\alpha$ is the learning rate and the hyper-parameter that can be fixed or adjusted during training. (See Goodfellow et al., 2016, for more details)

\subsection{Regularisation}
DNNs are highly prone to overfitting, meaning that the network can fully learn the training set while not generalising well to unseen datasets. In this section, we explain some methods to regularise DNNs. 

\subsubsection{Penalty}
One way to regularise a function is by adding a penalty term to \ref{eq:cost}. This shrinks the weights and helps with the prediction and interpretation of the model (Tibshirani, 1996). The regularised cost function is then given by
\begin{eqnarray}
\tilde{J}(\mathbf{w}) = J(\mathbf{w}) + \lambda R(\mathbf{w}),
\end{eqnarray}
where $\lambda \in [0, \infty)$ and $R(\mathbf{w})$ can take the following forms:
\begin{itemize}
\item $L_1$ penalty: $\|\mathbf{w}\|_1$ (Lasso regression)
\item $L_2$ penalty: $\frac{1}{2}\|\mathbf{w}\|_2^2=\frac{1}{2} \mathbf{w}^T\mathbf{w}$ (ridge regression).
\end{itemize}
The hyper-parameter $\lambda$ controls the strength of the penalty term. Larger $\lambda$ penalises the weights (not biases) more heavily, and therefore the function becomes less complex and generalises well.

\subsubsection{Early stopping}
In Machine Learning algorithms, we can use validation curves to tune hyper-parameters. This curve compares the training and validation errors to find an optimal value of hyper-parameters. In early stopping, the algorithm monitors the training and validation errors at each epoch (iteration step). If after a predetermined number of steps, there is no improvement based on a predetermined value, the algorithm terminates or the optimisation learning rate is modified. Yao et al. (2007) study early stopping in gradient descent algorithms and demonstrate its relationship to Tikhonov regularisation ($L_2$ regularisation), boosting (Freunde and Shapire, 1997), etc. Early stopping can be implemented in TensorFlow using ``Callbacks".

\subsubsection{Dropout}
When designing a DNN, we can add dropout layers. A dropout layer randomly eliminates some units according to a pre-determined probability. The idea is that since DNNs require large training datasets, constructing an ensemble of DNNs similar to DTs is not feasible. A dropout layer can add an element of randomness, and therefore the algorithm can better generalise to unseen datasets. With dropout, the feed-forward propagation in \ref{eq:ff} is modified as
\begin{eqnarray}
r^{(l-1)} & \sim & \textnormal{Bernoulli}(p) \nonumber\\
\tilde{\mathbf{a}}^{(l-1)} &=& \mathbf{r}^{(l-1)} \odot \mathbf{a}^{(l-1)}\nonumber\\
z_i^{(l)} & =& \mathbf{w}_i^{(l)} \tilde{\mathbf{a}}^{(l-1)} + b_i^{(l)}\nonumber\\
a_i^{(l)} &=& f_{\theta}(z_i^{(l)})\nonumber,
\end{eqnarray}
where $\odot$ is the element-wise product of $\mathbf{r}$ and $\mathbf{a}^{(l-1)}$. As we can see in a network with a dropout layer, the output from layer $l-1$ is a thinned vector $\tilde{\mathbf{a}}^{(l-1)}$. 
This method works very well with different types of datasets and NNs, including convolutional NNs (CNNs) (Srivastava et al., 2014).

\subsubsection{Batch and layer normalisation}
When a dataset is large, in order to speed up the training algorithm, we can divide the set into mini batches. That means instead of updating the network's parameters by going through the whole training set, we can update the parameters for each batch. In this case, the gradient of the loss over a mini batch is an estimate of the gradient over the training set. Therefore, \ref{eq:sgd} can be modified as
\begin{eqnarray}
\theta := \theta - \alpha \frac{1}{m} \sum_{i=1}^m \nabla_{\theta} J(x_i, \theta),
\end{eqnarray}
where $m$ is the batch size and a hyper-parameter. Batch normalisation (BN), similar to dropout, is another method to reduce the internal covariate, which arises when each layer is affected by the previous layer. Therefore, when applying BN, we can even remove the dropout layer. In this method, for a batch size of $m$ with mean $\mu$, and variance $\sigma^2$, the inputs to a layer in \ref{eq:ff} is modified as
\begin{eqnarray}
 z_i^{(l)} & =& \mathbf{w}_i^{(l)} \mathbf{a}^{(l-1)} + b_i^{(l)}\nonumber\\
 z_{i , \textnormal{norm}}^{(l)} &=& \frac{z_i^{(l)} - \mu}{\sqrt{\sigma^2 + \epsilon}}\\
\textnormal{BN}(z_i^{(l)}) &=& \tilde{z}_i^{(l)} = \gamma z_{i, \textnormal{norm}}^{(l)} + \beta\\
a_i^{(l)} &=& f_{\theta}(\tilde{z}_i^{(l)})\nonumber,
\end{eqnarray}
where $\gamma$ and $\beta$ are new scale and location parameters and will be updated in the back-propagation process, $z_{i , \textnormal{norm}}^{(l)}$ is the normalised vector with zero mean and unit variance, and $\epsilon$ is a very small non-negative number. BN reduces the effect of initialisation and vanishing gradients, which happens when the gradients get very small as we go through deeper layers. It also helps the algorithm to converge faster as it is possible to use a higher learning rate (Ioffe and Szegedy, 2015). BN is effective in FNNs, but not in other NNs. For example, in Recurrent NNs (RNNs), (Rumelhart et al., 1986) layer normalisation is applied. In this method, the inputs to all units of a layer are normalised. Unlike BN, layer normalisation considers normalisation per iteration step rather than one batch, and all units in a layer share the same mean and variance. Similar to BN, new scale and location parameters are added to the units after the normalisation but before the non-linearity (activation function) (Bat et al. 2016).

\subsection{Transformers}
In Natural Language Processing (NLP), words are considered as categorical features. Some models, like RNNs, process words (categorical features) as an input sequence, while other models, like Bag-of-Words, process categorical features without considering their orders. The third group is Transformers (Vaswani et al. 2017), which is a combination of both, i.e., the order of categorical features is not important, but the model adds positional vectors. Transformers are based on neural Attention. Attention layers determine which features should be paid more attention to and which features should be paid less attention to. By assigning scores to different features, Attention takes into account the interaction between categorical features and identifies significant features. Before passing input sequences through a Transformer, we need to convert them into numerical representations. In language processing, this can be done through vectorisation, and in our case, through, for example, one-hot encoding or label encoding. Then, the numerical vectors are passed through an embedding layer. This layer reduces a very long one-hot vector into a more dense vector, called embedding in NLP. In our study, the embedding layer looks at unique categories in each feature vector and converts them into embeddings. After the embedding layer, our inputs are ready to go through the Transformer block, which as illustrated in Figure \ref{fig:TT}, begins with the Multi-head Attention layer and ends with the second normalisation layer:

\begin{itemize}
\item Multi-head attention layer: This is a set of parallel self-attention layers. A self-attention layer compares each input vector with other vectors. It works like a query-key-value in a database. Let $\mathbf{x}_1$ be an embedded vector for feature 1. The length of this vector is equal to the embedding size. We want to compare $\mathbf{x}_1$ (query) with other embedded feature vectors (keys). Queries, keys, and values are defined as:
\begin{itemize}
\item Query: $\mathbf{Q} = \mathbf{W}_q \mathbf{x}_i$
\item Key: $\mathbf{K} = \mathbf{W}_k \mathbf{x}_i$
\item Value: $\mathbf{V} = \mathbf{W}_v \mathbf{x}_i$,
\end{itemize} 
where $\mathbf{W}_q, \mathbf{W}_k$, and $\mathbf{W}_v$ are the trainable weights and will be adjusted in the back-propagation process. In order to measure the similarity between our feature vector $\mathbf{x}_1$ and other vectors, we can use the dot-product. We then scale the result and pass it through a softmax function. This gives us the associated weights of our query with each key. The attention score is then given by 
\begin{eqnarray}
\textnormal{attention}(\mathbf{Q}, \mathbf{K}, \mathbf{V}) = \textnormal{softmax}\left(\frac{\mathbf{K}^T \mathbf{Q}}{\sqrt{d}}\right) \mathbf{V} ,
\end{eqnarray}
where $d$ is the length of the input vector. The higher the score, the more relevant are the features. The result is a new vector representation for our feature $\mathbf{x}_1$. In multi-head self-attention, this process is done in each head and in parallel, where the number of heads is a hyper-parameter. We define the weight matrix for queries, keys, and values, multiple times, i.e., $\mathbf{W}_{q1}, \mathbf{W}_{q2}, \dots$. Transformers usually consist of an encoder and a decoder. The input sequence is first encoded to a set of feature vectors,  and a decoder uses these vectors to produce a set of new sequences, say, words in another language. The output from the last encoder goes to the decoder. Decoder is comprised of masked multi-head attention, where subsequent sequence elements are masked by letting softmax values approach $-\infty$. The purpose of this mask is to stop the model from using future input sequences that are to be predicted by the model. 
\item Normalisation layer: This normalises our new vectors by subtracting the mean and dividing by standard deviation, where mean and standard deviation are calculated across one layer. The input to this layer consists of the outputs from the previous layers, i.e., LayerNomalization$(x+ \textnormal{Sublayer}(x))$, where the Sublayer is the previous layer and $x$ is the initial input. This is called skip connection or residual connection (He et al., 2016) and prevents losing information in a deep structure due to vanishing gradients, which may happen in the back-propagation process as we go deep into the model. This is particularly true in sequence models as in the back-propagation process, we need to differentiate the cost function with respect to all parameters at different points in time. There are 2 inputs in this layer: one is the output from the multi-head attention layer, and the other is input vectors to the Transformer block. In Figure \ref{fig:TT}, the first skip connection is illustrated by a shortcut from the concatenation layer to the addition layer and the second one from the normalisation layer to the addition layer.
\item FNN: There are two dense layers (MLPs) with ``ReLU" (Nair and Hinton, 2010) activation function, i.e., $\max(0, \mathbf{w}\mathbf{x}+b)$. The number of hidden units in the first FNN is a hyper-parameter, and in the second one is equal to the embedding dimension to make sure that the dimension of the sequence throughout the Transformer is fixed and is the same as the embedding dimension. The next layer is another normalisation layer with a skip connection. Therefore, the input to this layer is the sum of the input to the first dense layer and the output of the second dense layer. 
\end{itemize}

\subsection{Classification metrics}
\label{app:class}
A confusion matrix is a tool for binary and multi-class classification analysis. Table \ref{tab:modes} shows a confusion matrix for binary classification problems. In a multi-class classification this can be computed for one class versus the others. We can obtain the following measures from this table:

\begin{table}{}
\caption{Confusion matrix}
\centering
\footnotesize
\begin{tabular}{ccc}\toprule
				& Predicted positive 							& Predicted negative	  \\\midrule
Actual positive     	& TP	(correctly predicted as positive)				& FN	 (incorrectly predicted as negative)	 \\
Actual negative		& FP (incorrectly predicted as positive)			& TN	 (correctly predicted as negative)	 \\\bottomrule
\end{tabular}
\label{tab:modes}
\end{table}

\begin{itemize}
\item Accuracy: $\frac{TP+TN}{TP+TN+FP+FN}$. This measures the overall performance of an algorithm. In a multi-class problem, this refers to total accurate predictions to the overall predictions.
\item Sensitivity (recall): $\frac{TP}{TP+FN}$. This measures the performance of an algorithm in predicting positive classes. In a multi-class problem, we can define sensitivity for each class separately. In this case, the positive class is the accurate predictions of that class and the negative class is the incorrect predictions of other classes.
\item Precision: $\frac{TP}{TP+FP}$. This measures the quality of prediction and similar to recall, in a multi-class problem can be defined for each class.
\item Specificity: $\frac{TN}{TN+FP}$. This measures the performance of an algorithm in predicting negative classes and can be defined for each class separately. 
\item Receiver operating characteristics (ROC): It plots sensitivity against specificity across different thresholds. 
\item AUC: it measures the area under ROC curve. 
\item $F_1$ score (harmonic mean of precision and recall): $2 \times \frac{\textnormal{Precision} \times \textnormal{Recall}}{\textnormal{Precision} + \textnormal{Precision}}$. 
\end{itemize}
These classification measures are usually applied to binary cases and in multi-class classification, we need to calculate these measures for each class separately. They can be expressed as micro and macro. Macro-averaged measures are the average scores across all classes, and micro-averaged measures are obtained after aggregating scores across all classes. The macro-averaged score is more suitable when the classes are imbalanced, i.e., the observation of one class is more than the other classes. For more details, see Murphy (2012).

\section{Results and discussions}
\label{sec:res}

\par In this section, we try Wide \& Deep (Cheng et al. 2016), TabTransformer (Huang et al. 2020), and TabNet (Arik and Pfister (2021) models on ELSA to predict the levels of disability.
We choose Wide \& Deep because it is a combination of wide (linear models) and deep (NNs) and is used in recommender systems. TabTransformer and TabNet are two Transformer-based models designed for tabular data. TabTransformer has a simple architecture. TabNet's architecture is similar to DTs and selects significant features at each step and at the aggregate level. We implement these models in TensorFlow 2.12 using Keras Functional API and in a CPU. (See, for example, Chollet, 2021, and Ya, and Wang, 2023). We divide our dataset into 2 subsets: $80\%$ training $(8975)$, and $20\%$ test sets $(2244)$, and apply $K$-fold cross-validation to build validation sets by dividing our training set into $K$ subsets. We choose $K=5$ and each time use $K$th set as the validation set. We use the training subset to train our models, the validation subset to tune hyper-parameters, including the number of epochs, and the test subset for prediction. 
\\ Let $\mathbf{x}_i = (x_{i1}, x_{i2}, \dots, x_{im})$ denote a vector of $m$ features. We then divide our features into two vectors 
$\{\mathbf{x}_{\textnormal{cat}}, \mathbf{x}_{\textnormal{cont}}\}$, where $\mathbf{x}_{\textnormal{cat}}$ is a vector of categorical features and $\mathbf{x}_{\textnormal{cont}}$ is a vector of continuous features. All categorical features are passed through an Embedding layer separately, where they are given a new vector representation.
Let $\mathbf{E}_{\mathbf{W}}(\mathbf{x}_{\textnormal{cat}}) = \{\mathbf{e}_1(x_1), \mathbf{e}_2(x_2), \dots, \mathbf{e}_m(x_m)\}$ denote all embedded categorical features. This is the first step in all models, including the Deep part of Wide \& Deep model.  As explained in Section \ref{sec:data} we label encoded categorical features but did not normalise numerical features.   

\subsection*{Wide \& Deep}
We begin with normalising numerical features.

\begin{itemize}
\item Architecture: The wide part is a logistic regression, which is an FNN with 1 layer and 1 unit in the case of binary disability, and 3 and 4 units, in the case of 3-level and 4-level disability, respectively. The deep part is an embedding layer with dimension 32  for all 24 categorical features, and a 3-layer FNN with 32 units.
\item Input layer: In the wide part, categorical and numerical features are combined and passed through as one input vector (we do not consider any interactions between features). In the deep part categorical and numerical features are separated and passed through the model as 2 separate inputs.
\item Output layer: It is 1 dimension in the case of binary disability and 3 and 4 dimensions in the case of 3-level and 4-level disability. 
\end{itemize}
Finally, we combine Wide and Deep models and pass them through WideDeepModel in TensorFlow. Hence, the estimated $\mathbf{y}$ is given by

\begin{eqnarray}
\hat{\mathbf{y}} =  \sigma\left( \mathbf{w}_{\textnormal{wide}} \mathbf{x} + f_{\theta}\left( \mathbf{E}_{\mathbf{W}}(\mathbf{x}_{\textnormal{cat}})   , \mathbf{x}_{\textnormal{cont}} \right)\right)\nonumber,
\end{eqnarray}
where $\sigma$ is the sigmoid function, $\mathbf{w}_{\textnormal{wide}}$ is the weights of the logistic regression. The second part is the Deep model, where $\mathbf{E}$ is the embedding layer, $\mathbf{W}$, the trainable parameters, and $f$ is FNN with parameter $\theta$.

\subsection*{TabTransformer}
\begin{figure}[H]
\centering
\includegraphics[scale = 0.25]{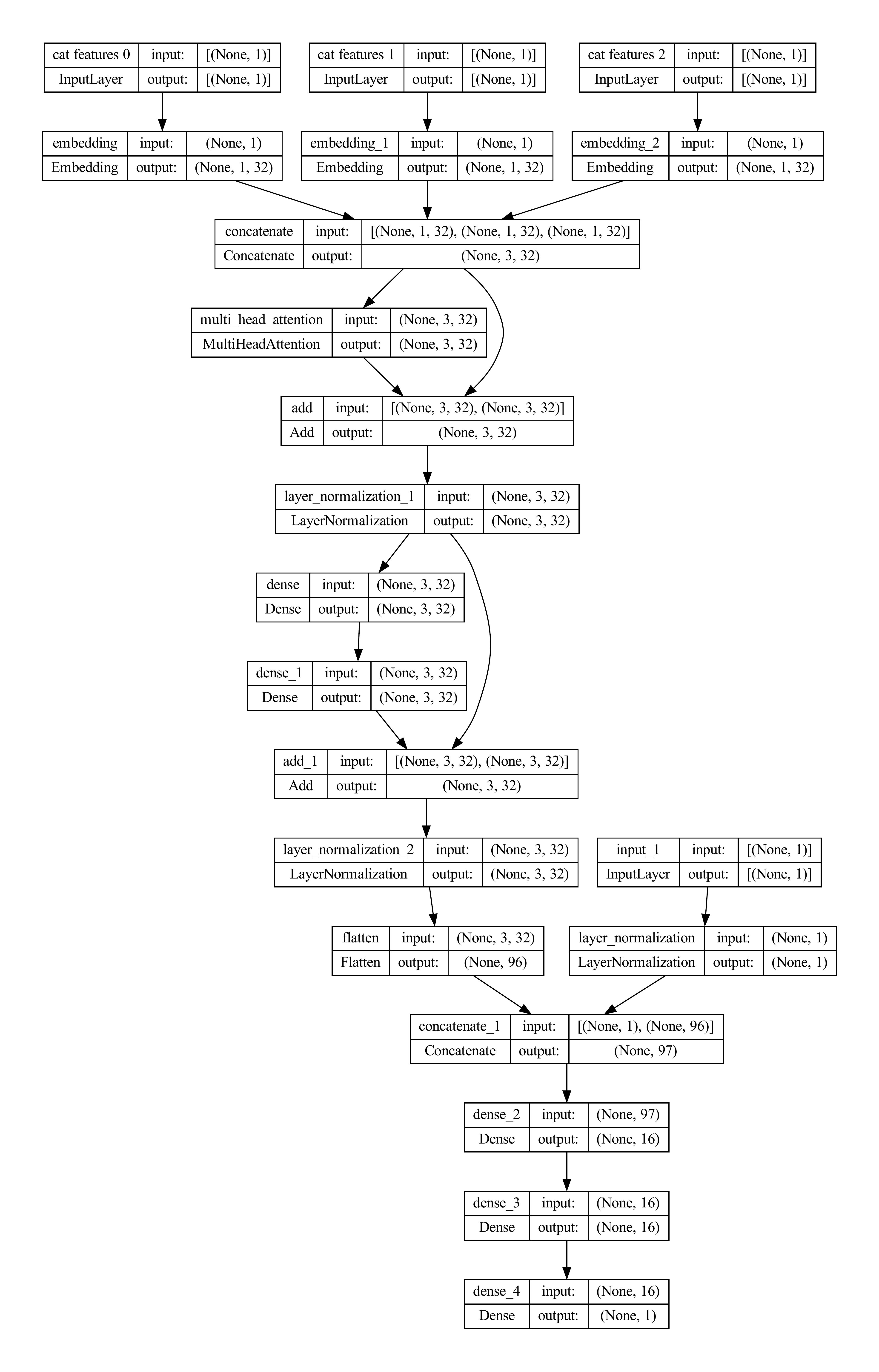}
\caption{TabTransformer architecture for 3 categorical, and 1 numerical features, 1 Transformer with 1 multi-head Attention layer, and 2 FNN with 16 units}
\label{fig:TT}
\end{figure}

In this model, we do not need to normalise numerical features. Figure \ref{fig:TT} shows the architecture of this model with the input shapes of tensors. We have 24 categorical features and 1 numerical feature. But in this figure, we only show 3 categorical features and 1 numerical feature. Also, for brevity, this figure only shows 1 Transformer with 1 Multi-head Attention and 2 FNN layers with 16 units. The parameters of TabTransformer include the trainable weights in the Embedding layer, parameters of Transformers, and parameters of FNN outside the Transformer. Therefore, the estimated $\mathbf{y}$ is given by
\begin{eqnarray}
\hat{\mathbf{y}} = f_{\theta}\left(g_{\psi} \left( \mathbf{E}_{\mathbf{W}}(\mathbf{x}_{\textnormal{cat}})\right)   , \mathbf{x}_{\textnormal{cont}} \right),
\end{eqnarray}
where $\mathbf{E}$ is the Embedding layer, $\mathbf{W}$ is the trainable weights, $g$ is the Transformer layer, $\psi$ is the Transformer's parameters, $f$ is the FNN outside the Transformer and $\theta$ is the parameters of FNN. 

\begin{itemize}
\item Architecture: The first layer for numerical features is the LayerNormalization, and for categorical features is the Embedding layer, where the input dimension is equal to the number of unique categories and the output dimension (embedding dimension) is 32, which can be adjusted later. Then, we concatenate all the outputs from the Embedding layer, which is 24 embedded vectors with a length of 32. The output from the concatenation layer is passed through the Transformer block. In this block, we have a MultiHeadAttention layer with 4 heads and the key dimension is equal to the embedding dimension. Then, there is a skip connection, represented in Figure \ref{fig:TT} by Add layer. The output then goes through the LayerNormalization and then 2 Dense layers (MLPs) with the dense dimension (number of units) of 32 in both layers. After that, there is another skip connection between the previous LayerNormalization and the next one. At this stage, we finish the first Transformer block. We have 4 Transformers with 4 heads in the case of binary and 4-level disability and 6 Transformers with 8 heads in the case of 3-level disability. After going through all Transformers, we then use a Flatten layer, which flattens the inputs and concatenates its output with the normalised numerical features. The output is then passed through a 4-layer FNN with 16 units each.
\item Input layer: We separate numerical and categorical features. Numerical features are tensors with lengths equal to the number of numerical features, which in our case is 1. Each categorical feature is passed through an Input layer. Therefore, we have 24 Input layers. 
\item Output layer: For binary disability, the output has 1 dimension with sigmoid activation function and for 3, and 4-level disability, the output has 3 and 4 dimensions with softmax activation function. 
\end{itemize}

\subsection*{TabNet}
We use package tf-TabNet, which is a TensorFlow implementation of TabNet. TabNet encoder consists of two types of Transformers, namely, Feature Transformer and Attentive Transformer. It also includes Attention Masks. One important property of TabNet is its capacity for interpretability. Similar to DTs, it is based on sequential decision steps and at each step, the model decides which feature to use and aggregates the processed feature representation. 

\begin{itemize}
\item Architecture:
\begin{itemize}
\item Feature Transformer: It consists of 4 blocks of Fully-connected FC layer (FNN, Dense layer), a BN, and a Gated Linear Unit (GLU) (Dauphin et al. 2017), where GLU$(x) = x \sigma(x)$ with $\sigma$ being the logistic function. GLU is an activation function that acts like a gate and decides what percentage of features can be passed through. The first two blocks are shared, in the sense that the weights are shared across the decision steps. These two blocks are connected through a normalised skip connection with $\sqrt{0.5}$ to maintain numerical stability. The next two blocks are independent, in the sense that they have different weights at each decision step. These two are connected with each other and the previous block through normalised skip connection. The output of this Transformer is then passed through the Attentive Transformer. 
\item Attentive Transformer: In this block, the model decides which feature should be selected and which feature should not be selected. The first layer is an FC layer, followed by a BN layer. The output is then multiplied by prior scales. Let $M_i \in R^{B \times D}$ be a trainable mask at decision step $i$, where $B$ is the batch size, and $D$ is the feature dimension. Also, let $P_i$ be the prior scale at decision step $i$ with $P_0 = 1$, i.e., all features are feasible at $i=0$. The prior scale represents how much a feature has been used previously. It is then updated by
\begin{eqnarray}
P_i := P_i (\gamma - M_j), \quad \textnormal{for} j = 1, \dots i\nonumber,
\end{eqnarray}
where $\gamma$ is the relaxation parameter. For example, if $\gamma=1$, a feature can only be used at one decision step since if it has been selected in the previous step, i.e. $M_j = 1$, then $P_i = 0$. The mask at decision step $i$ is then given by
\begin{eqnarray}
M_i = \textnormal{sparsemax}(P_{i-1} h_i(a_{i-1}))\nonumber,
\end{eqnarray}
where $h_i$ is a function trained by an FC layer and BN, $a_{i-1}$ is the output from the previous decision step. Sparsemax is an activation function similar to softmax, but unlike softmax, which only gets very close to 0, it can produce sparse probabilities (Martins and Astudillo, 2016). 
\item Attention Mask: Attentive Transformer generates a Mask from the previous decision step. If $M^{b,d}_i = 0$, then the $dth$ feature of $bth$ sample does not contribute to decision step $i$. Let $\eta^b_i$ be the contribution of aggregate decisions at decision step $i$ and for sample $b$, and is given by
\begin{eqnarray}
\eta^b_i = \sum_{c=1}^{N_d} \textnormal{ReLU}(d_i^{b,c})\nonumber.
\end{eqnarray}
Then, the aggregate feature importance Mask is given by
\begin{eqnarray}
M^{b,j}_{\textnormal{agg}} = \frac{\sum_{i=1}^{N_{\textnormal{steps}}}  \eta^b_i M^{b,j}_i}{\sum_{j=1}^D \sum_{i=1}^{N_{\textnormal{steps}}} \eta^b_i M^{b,j}_i}\nonumber.
\end{eqnarray}
The next step is another Feature Transformer. The output of this Feature Transformer is split into two parts: $N_a$ amount of $a_i$ goes to the Attentive Transformer, and $N_d$ amount of $d_i$ is passed through a ReLU activation function and then aggregated across all decision outputs, giving
\begin{eqnarray}
d = \sum_{i=1}^{N_{\textnormal{steps}}} \textnormal{ReLU}(d_i)\nonumber.
\end{eqnarray}
\end{itemize}
\item Input layer: Categorical features are passed through an embedding layer, and numerical features are not normalised. Therefore, the first layer is BatchNormalization layer. Then its output is passed through the Feature Transformer.
\item Output layer: The output is the sum of outputs from all decision steps, passed through the final layer, which is an FC (Dense) layer with a softmax activation function. 
\end{itemize}

\subsection*{Compilation and fitting}
To compile these models, we use RMSprop optimiser (Tieleman and Hinton, 2012) with BinaryCrossentropy loss function for binary disability and CategoricalCrossentropy for disabilities with 3 and 4 levels, where we need to one-hot encode our target feature in the case of the latter. To monitor the performance of the model, we use a list of metrics, including Accuracy, AUC, Recall, Precision, and $F_1$ score. We also use ModelCheckpoint callback to monitor validation accuracy and save the weights of the model, EarlyStopping to stop training if there was no improvement in validation loss after 10 epochs, and ReducLROnPlateau to reduce the learning rate if there was no change in validation loss after 10 epochs. We first train our models with 5-fold cross-validation to find the optimal number of epochs (iterations) and then evaluate them using our test dataset. Tables \ref{tab:results1}, \ref{tab:results2}, and \ref{tab:results3} show the performance of our models in the case of binary, 3-level, and 4-level disability.

\begin{table}[h]
\caption{The metrics for binary disability based on our computation}
\label{tab:results1}
\centering
\footnotesize
\scalebox{0.8}{
\begin{tabular}{lccccc}\toprule
Models				&  TabTransformer    & TabNet   & Wide \& Deep        \\\toprule
Training accuracy		 &	0.9094	  	&  0.9051	&	0.9045		\\
Training loss			 & 	0.2086	  	&  0.2231	&	0.2147		\\
Training AUC			&      0.9421		&  0.9696	&	0.9382		\\
Training Recall			&      0.6071		&  0.9051	&	0.5674		\\
Training precision		&      0.7586		&  0.9051	&	0.7538		\\
$F_1$ score			&      0.6744		&  0.9051	&	0.6475		\\
Test accuracy			&	0.8980	 	 & 0.8944	&	0.9024		\\
Test loss				&	0.2416		& 0.2457	&	0.2335		\\
Test AUC				&	0.9262		& 0.9634	&	0.9311		\\
Test recall				&	0.5389		& 0.8944	&	0.5556		\\
Test precision			&	0.7549		& 0.8944	&	0.7722		\\
$F_1$ score			&	0.6289		& 0.8944	&	0.6462		\\\bottomrule
\end{tabular}}
\end{table}

\begin{table}[h]
\caption{The metrics for 3-level disability based on our computation}
\label{tab:results2}
\centering
\footnotesize
\scalebox{0.7}{
\begin{tabular}{lccccc}\toprule
Models				&  TabTransformer 	& TabNet   & Wide \& Deep  \\\toprule
Training accuracy		 &	0.7658		& 0.7671	&      0.7670		\\
Training loss			 & 	0.5560		& 0.5546	&      0.5560		\\
Training AUC			&      	0.9157		& 0.9161	&	0.9158		\\
Training Recall	$(y=0)$	&     	0.0069		& 0.0743	&	0.0762		\\
Training Recall	$(y=1)$	&     	0.9008		& 0.9084	&	0.9033		\\
Training Recall	$(y=2)$	&     	0.5990		& 0.5628	&	0.5950		\\
Training Recall	(macro)    &     	0.5022		& 0.5152	&	0.5248		\\
Training Recall	(micro)	&     	0.6996		& 0.7118	&	0.7131		\\
Training precision $(y=0)$	&      	0.7333		& 0.5749	&	0.5191		\\
Training precision $(y=1)$	&      	0.8414		& 0.8317	&	0.8399		\\
Training precision $(y=2)$	&      	0.7477		& 0.7698	&	0.7231		\\
Training precision (macro)	&      	0.7741		& 0.7255	&	0.6940		\\
Training precision (micro)	&      	0.8289		& 0.8177	&	0.8151		\\
$F_1$ score (macro)		&	0.6010		& 0.6048	&	0.6112		\\
$F_1$ score (micro)		&	0.7658		& 0.7671	&	0.7670		\\
Test accuracy			&	0.7611		& 0.7478	&	0.7678		\\
Test loss				&	0.5648		& 0.5854	&	0.5628		\\
Test AUC				&	0.9139		& 0.9074	&	0.9142		\\
Test Recall $(y=0)$		&	0.0049		& 0.0610	&	0.0805		\\
Test Recall $(y=1)$		&	0.9081		& 0.9081	&	0.9153		\\
Test Recall $(y=2)$		&	0.5820		& 0.4920	&	0.5852		\\
Test Recall (macro)		&	0.4983		& 0.4870	&	0.5270		\\
Test Recall (micro)		&	0.6979		& 0.6956	&	0.7170		\\
Test Precision $(y=0)$	&	0.2222		& 0.4545	&	0.4925		\\
Test Precision $(y=1)$	&	0.8286		& 0.8203	&	0.8367		\\
Test Precision $(y=2)$	&	0.7328		& 0.7116	&	0.7398		\\
Test Precision (macro)	&	0.5945		& 0.6621	&	0.6897		\\
Test Precision (micro)	&	0.8135		&0.7981	&	0.8130		\\
$F_1$ score (macro)		&	0.5916		& 0.5705	&	0.6167		\\
$F_1$ score (micro)		&	0.7611		& 0.7478	&	0.7678		\\\bottomrule
\end{tabular}}
\end{table}

\subsection{Results}
\begin{figure}[H]
\centering
\includegraphics[scale = 0.3]{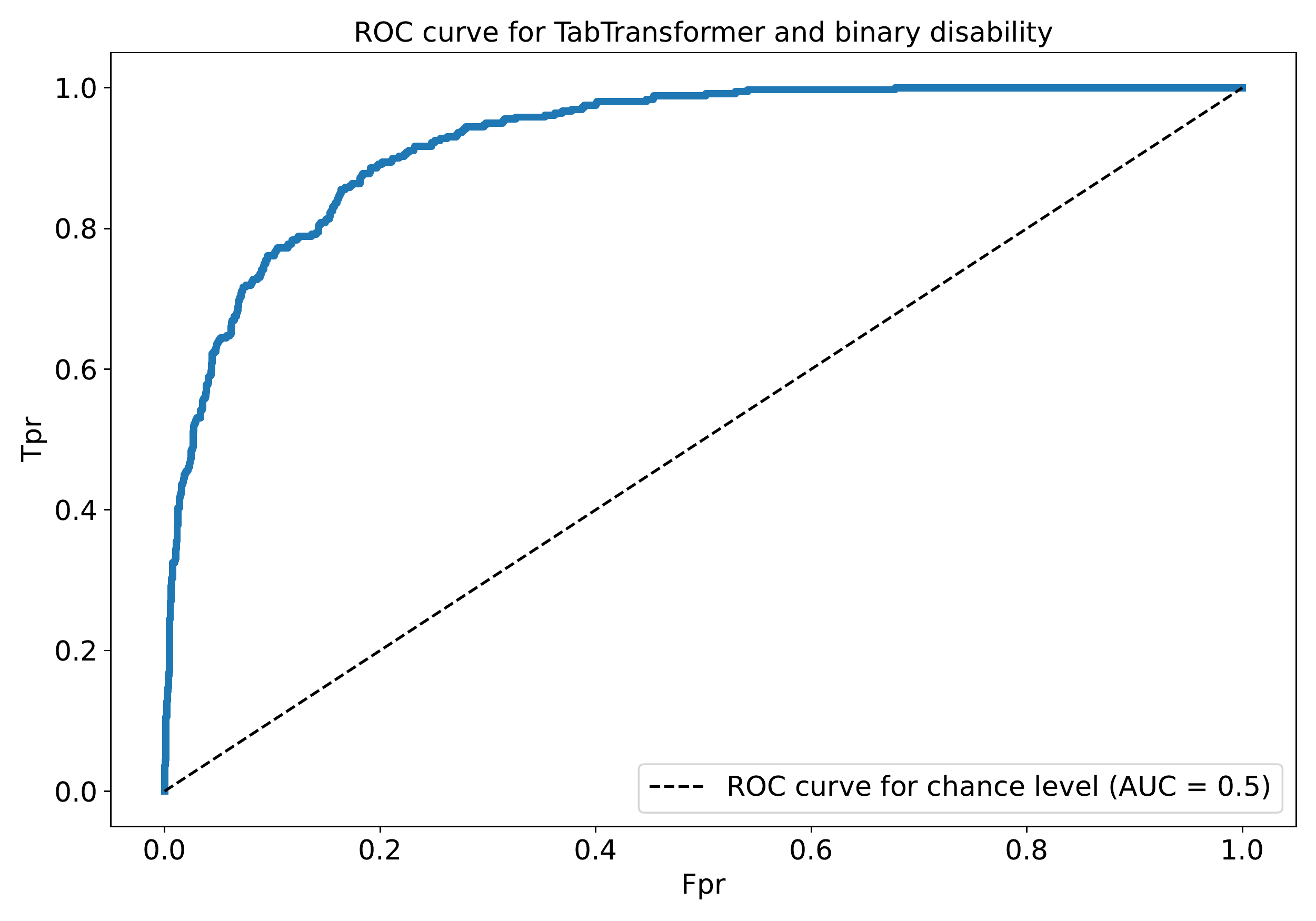}
\includegraphics[scale = 0.3]{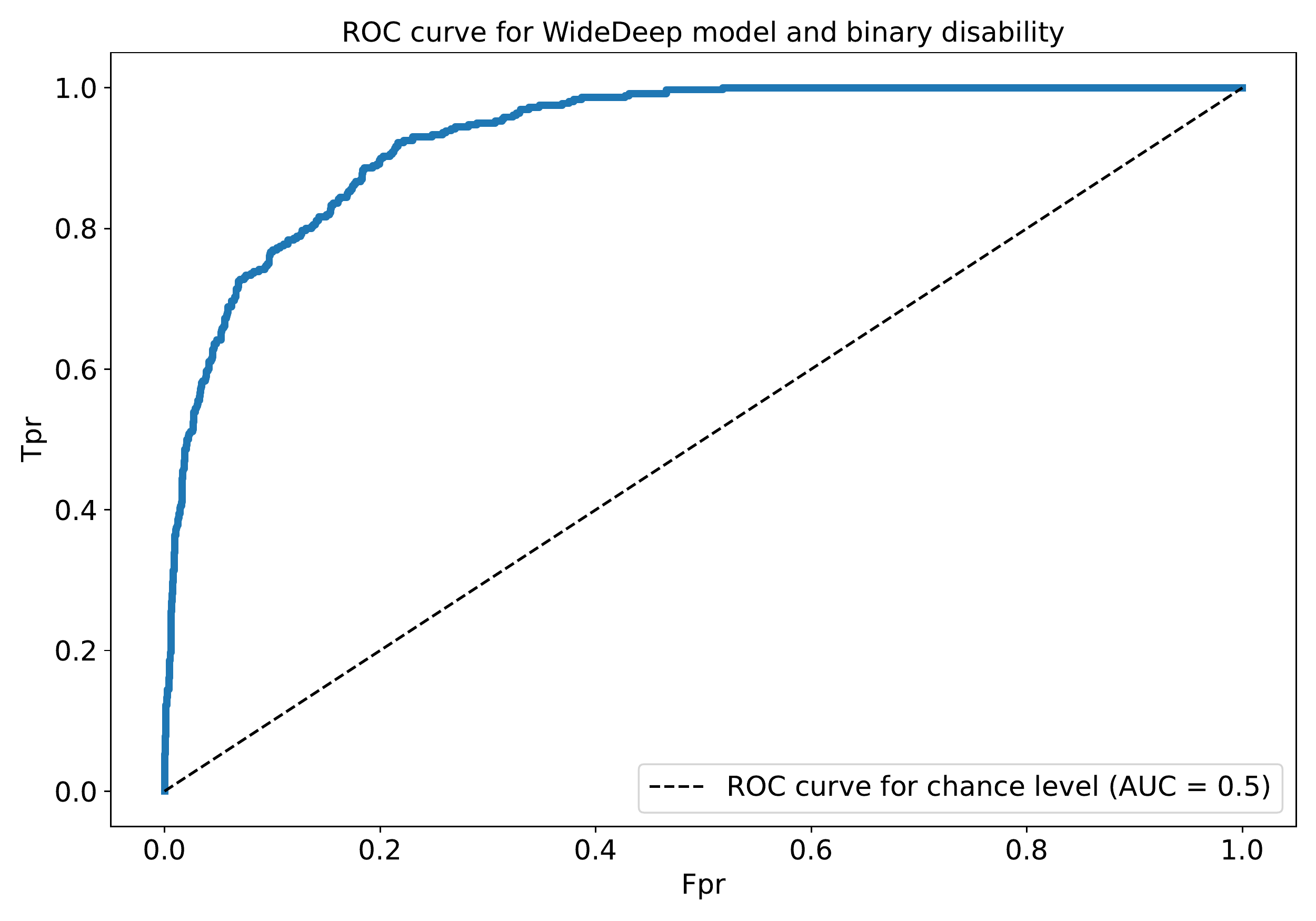}
\includegraphics[scale = 0.3]{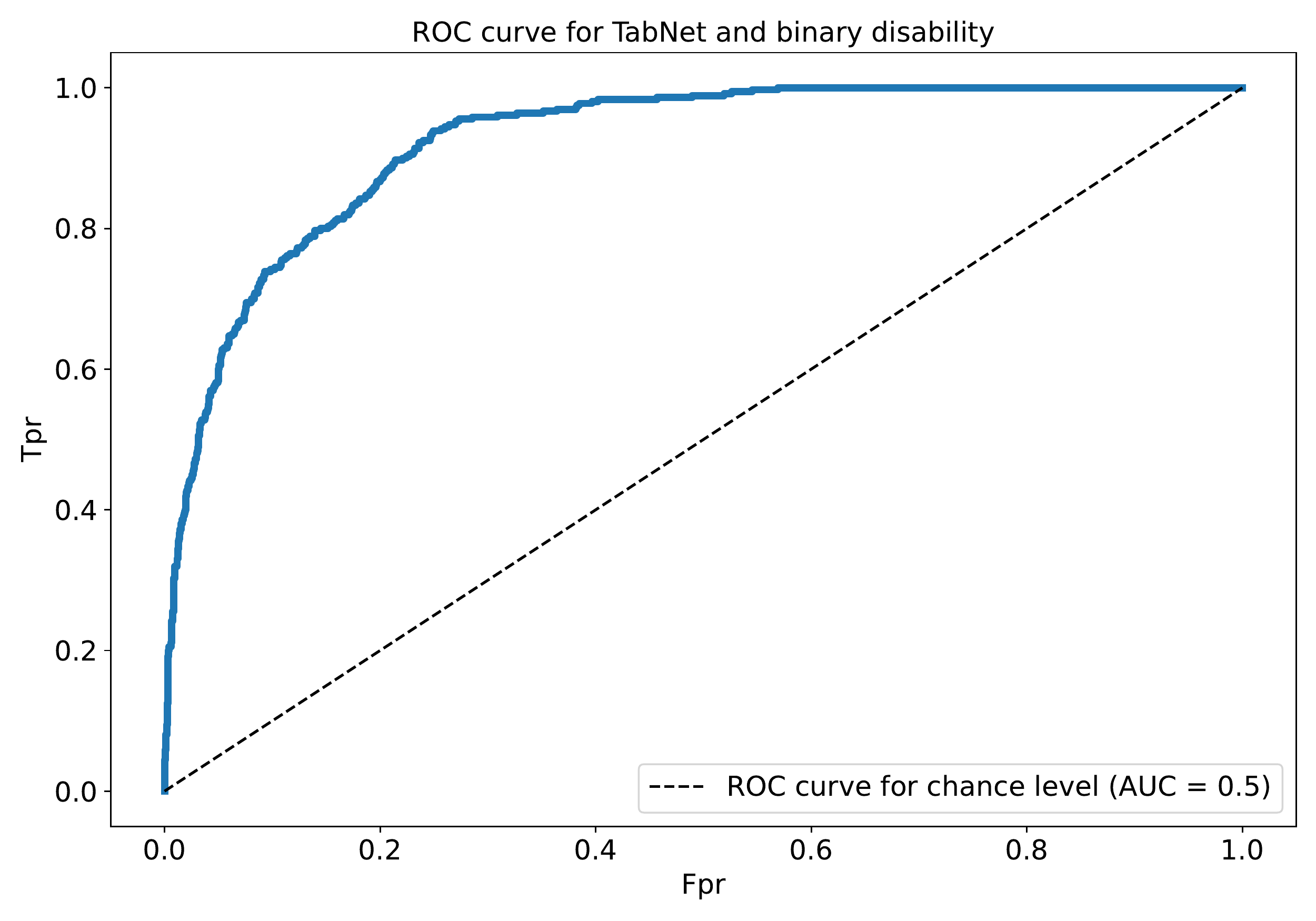}
\includegraphics[scale = 0.33]{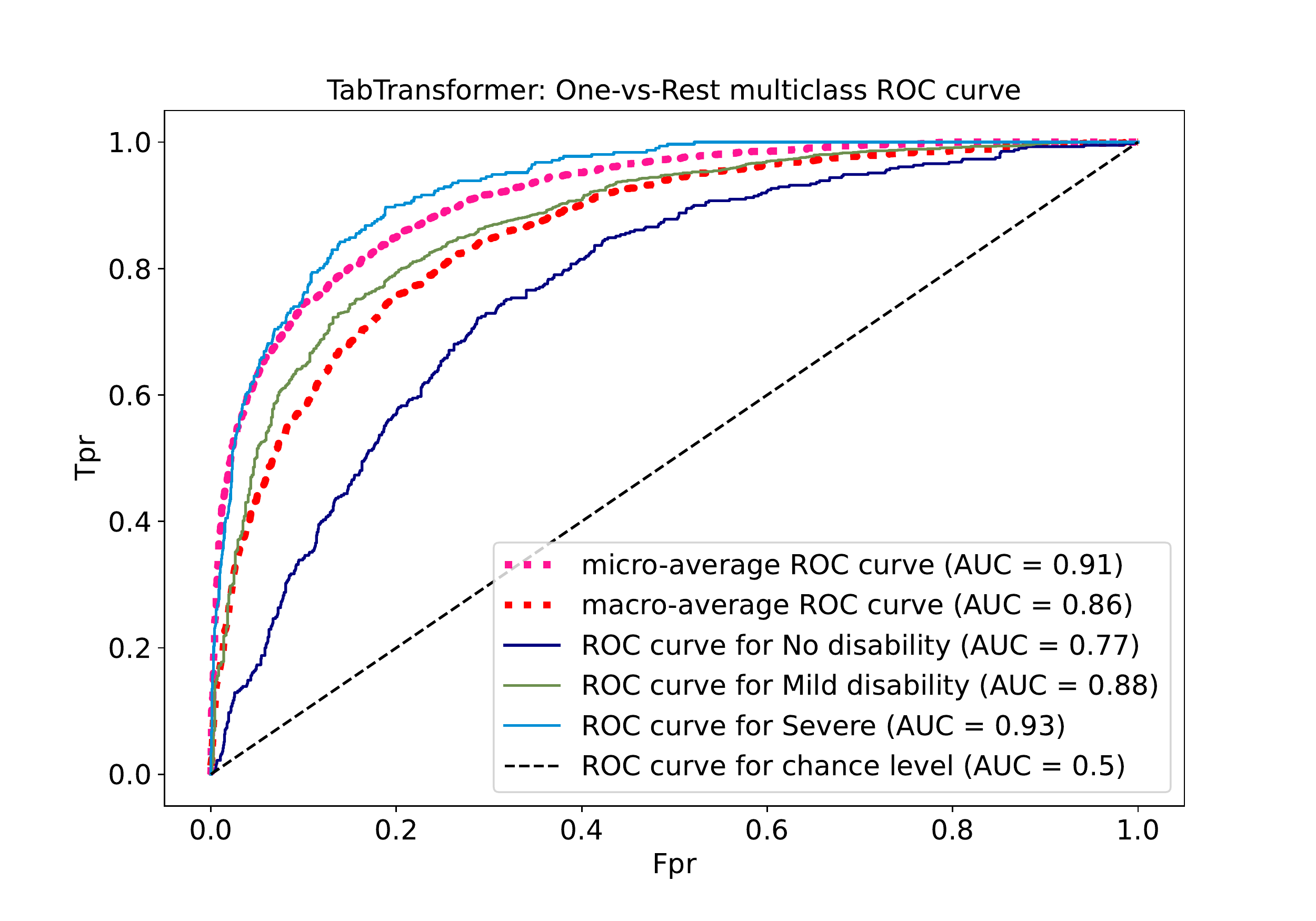}
\includegraphics[scale = 0.32]{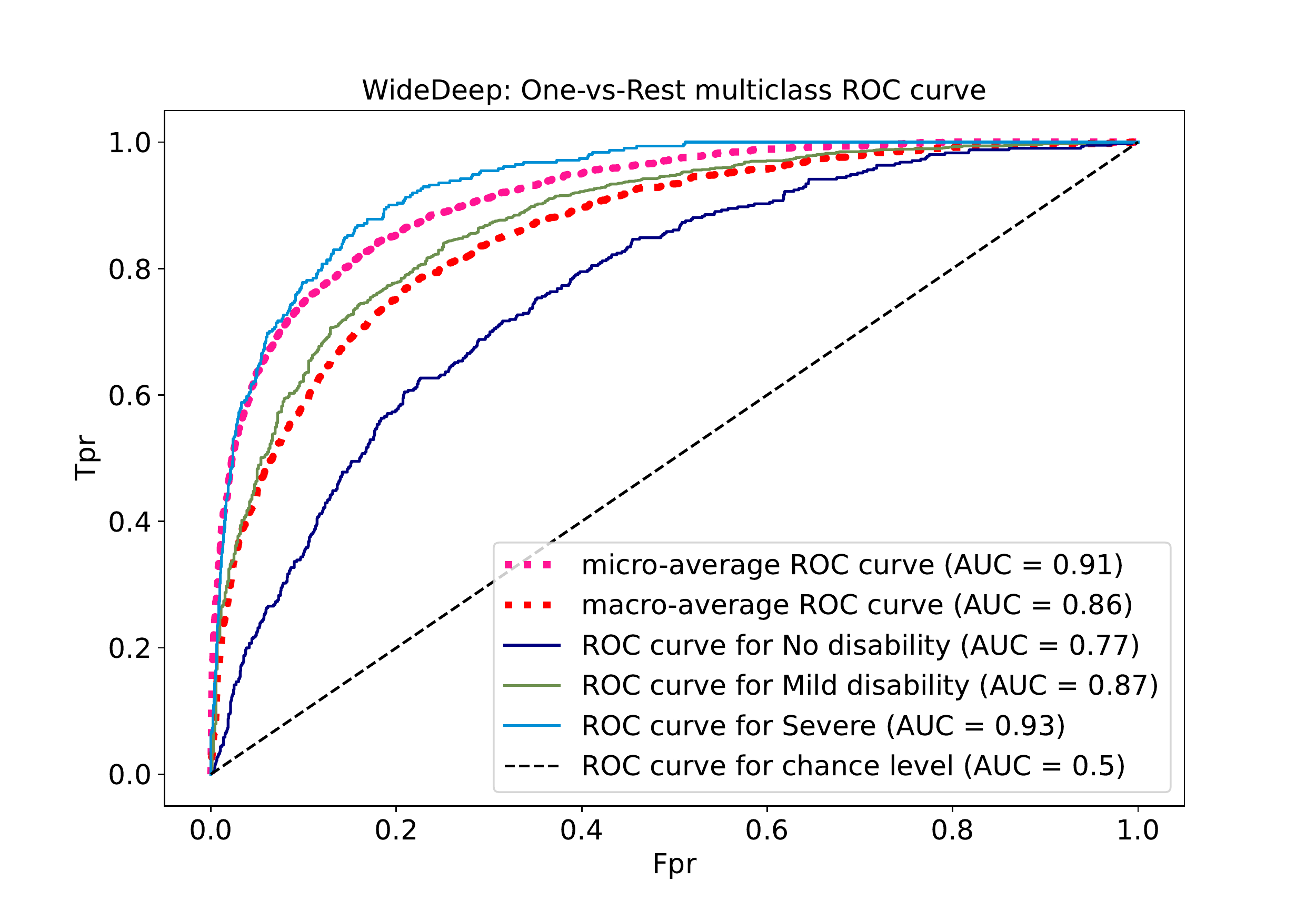}
\includegraphics[scale = 0.32]{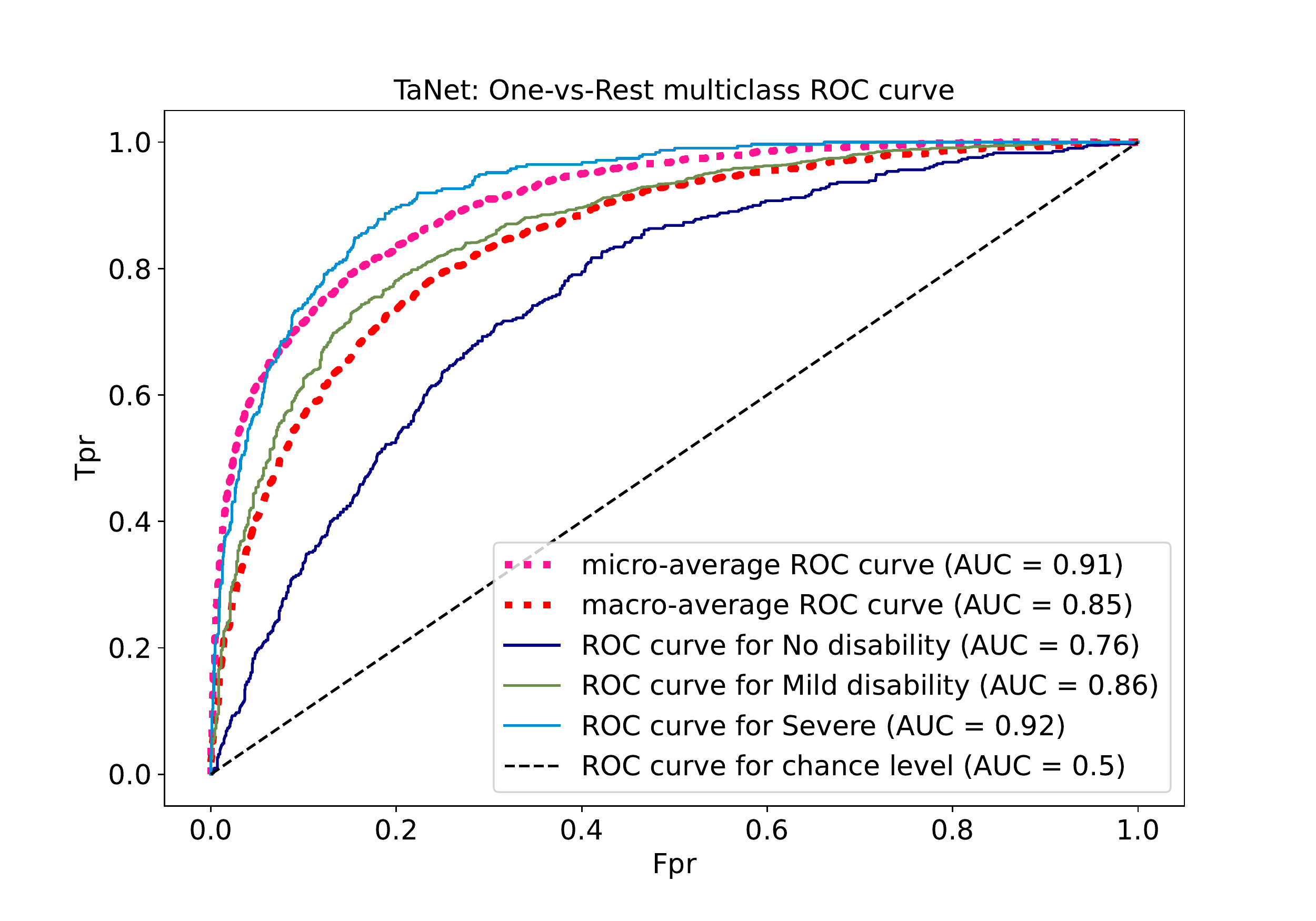}
\caption{ROC curve for 3 models based on test set}
\label{fig:ROC2}
\end{figure}

Table \ref{tab:results1} presents the metrics for the training set and test set in the case of binary disability. After dividing our dataset into training and test set, we have about $84\%$ participants without disability and about $16\%$ participants with disability in both sets. We can see that training and test accuracy for all models are relatively close with TabTransformer performing slightly better than the other two models. In terms of AUC, Recall, Precision, and $F_1$, TabNet outperforms other models. Figure \ref{fig:ROC2} also shows the ROC curve for these three models for binary disability based on the test set. Table \ref{tab:results2} presents the metrics for 3-level disability. We have about $68\%$ participants without disability, $18\%$ with mild disability, and $14\%$ with severe disability. This time we have 3 classes, and therefore, we need to look at some of these metrics in each class. As we can see, in the case of 3-level disability, TabNet and Wide \& Deep perform slightly better than TabTransformer in terms of both training accuracy and training loss, but in terms of test accuracy and test loss, Wide \& Deep and TabTransformer outperform TabNet. In terms of training Precision score for the without disability class, TabTransformer outperforms the other two, while in terms of test Precision, it underperforms. All models have reasonably high levels of AUC, which again indicates that all of them have a high predictability power. Macro and micro $F_1$ scores are reasonably close for all three models. However, Wide \& Deep performs better in terms of macro $F_1$ score in the test test. As we can see, Precision and Recall scores for the mild disability class are higher than other classes across all models, and the worst performance belongs to the without disability class. This can also be observed in Figure \ref{fig:ROC2}, where the closest ROC curve to $45^{\circ}$ line belongs to no disability class. In this Figure, we can see that, unlike Precision and Recall, the ROC curve for the severe disability class is placed above the mild class. Table \ref{tab:results3} provides results for 4-level disability. 

\begin{table}[H]
\caption{The metrics for 4-level disability based on our computation}
\label{tab:results3}
\centering
\footnotesize
\scalebox{0.7}{
\begin{tabular}{lccccc}\toprule
Models				&  TabTransformer & TabNet   & Wide \& Deep 	 \\\toprule
Training accuracy		 &	0.7662		& 0.7562	&	0.7669	\\
Training loss			 & 	0.6236		 &0.6465	&	0.6170	\\
Training AUC			&      	0.9331		& 0.9279	&	0.9346	\\
Training Recall	$(y=0)$	&     	0.0000		& 0.0000	&	0.0000	\\
Training Recall	$(y=1)$	&     	0.9087		& 0.9085	&	0.9026	\\
Training Recall	$(y=2)$	&     	0.5862		& 0.5515	&	0.6119	\\
Training Recall	$(y=3)$	&     	0.0061		& 0.0030	&	0.0137	\\
Training Recall	(macro)	&     	0.3752		& 0.3657	&	0.3821	\\
Training Recall	(micro)	&     	0.7028		& 0.6975	&	0.7034	\\
Training precision $(y=0)$	&      	0.0000		& 0.0000	&	0.0000	\\
Training precision $(y=1)$	&      	0.8369		& 0.8258	&	0.8430	\\
Training precision $(y=2)$	&      	0.7568		& 0.7478	&	0.7495	\\
Training precision $(y=3)$	&      	0.8000		& 0.5714	&	0.6000	\\
Training precision (macro)	&      	0.5984		& 0.5363	&	0.5481	\\
Training precision (micro)	&      	0.8267		& 0.8162	&	0.8296	\\
$F_1$ score (macro)		&	0.4329		& 0.4042 	&	0.4386	\\
$F_1$ score (micro)		&	0.7662		& 0.7562	&	0.7669	\\
Test accuracy			&	0.7571		 &0.7442	&	0.7518	\\
Test loss				&	0.6405		& 0.6665	&	0.6396	\\
Test AUC				&	0.9295		& 0.9238	&	0.9297	\\
Test Recall $(y=0)$		&	0.0000		& 0.0000	&	0.0000	\\
Test Recall $(y=1)$		&	0.9186		& 0.9048	&	0.9153	\\
Test Recall $(y=2)$		&	0.5563		& 0.4920	&	0.5627	\\
Test Recall $(y=3)$		&	0.0088		& 0.0029	&	0.0117	\\
Test Recall (macro)		&	0.3709		& 0.3499	&	0.3724	\\
Test Recall (micro)		&	0.7019		& 0.6827	&	0.7010	\\
Test Precision $(y=0)$	&	0.0000		& 0.0000	&	0.0000	\\
Test Precision $(y=1)$	&	0.8298		& 0.8183	&	0.8382	\\
Test Precision $(y=2)$	&	0.7588		& 0.6986	&	0.7384	\\
Test Precision $(y=3)$	&	0.6000		& 0.5000	&	0.3333	\\
Test Precision (macro)	&	0.5472		& 0.5042	&	0.4775	\\
Test Precision (micro)	&	0.8207		& 0.8042	&	0.8227	\\
Test $F_1$ score (macro)	&	0.4194		& 0.3893	&	0.4174	\\
Test $F_1$ score (micro)	&	0.7571		& 0.7442	&	0.7518	\\\bottomrule
\end{tabular}}
\end{table}

\begin{figure}[h]
\centering
\includegraphics[scale = 0.32]{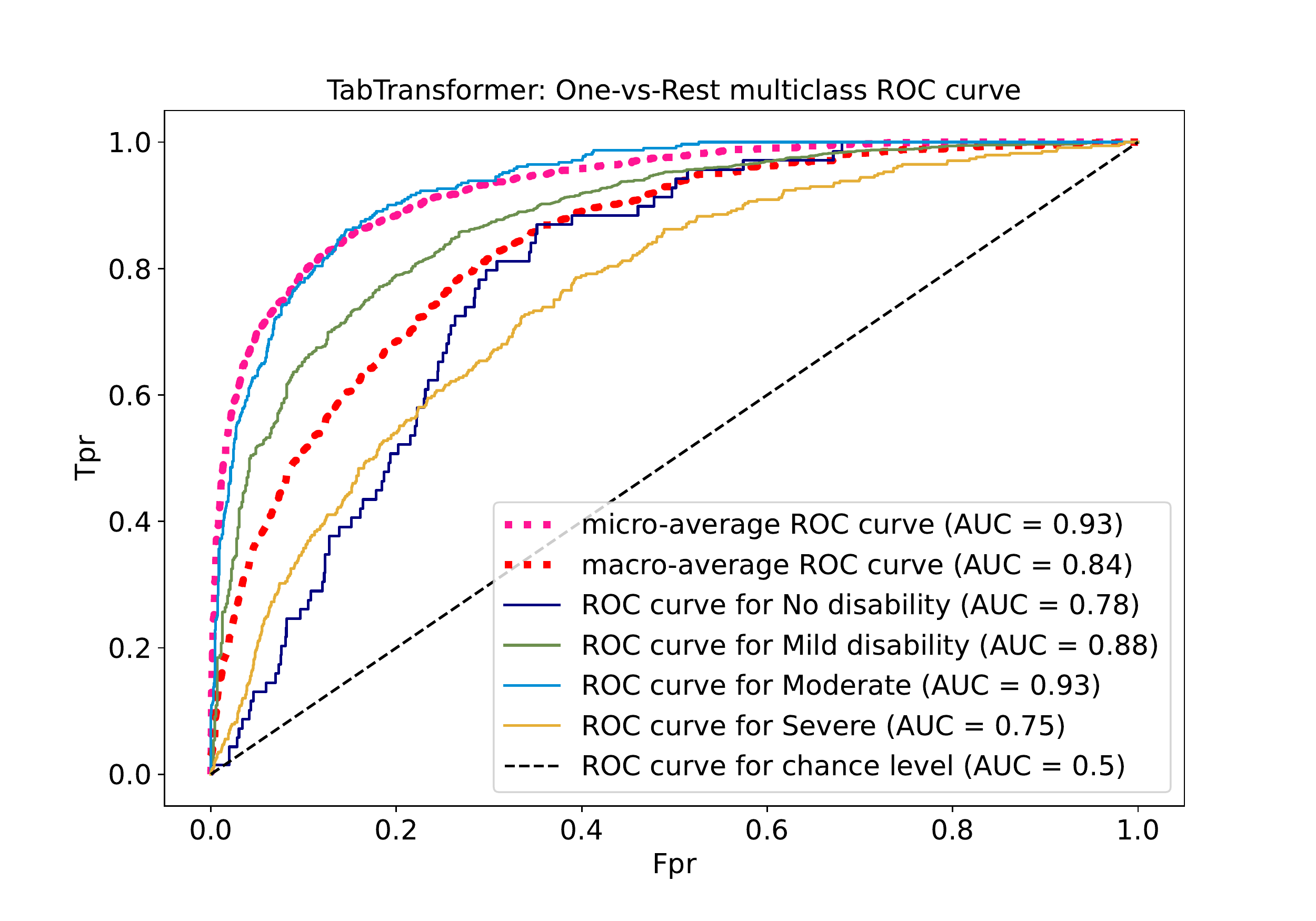}
\includegraphics[scale = 0.32]{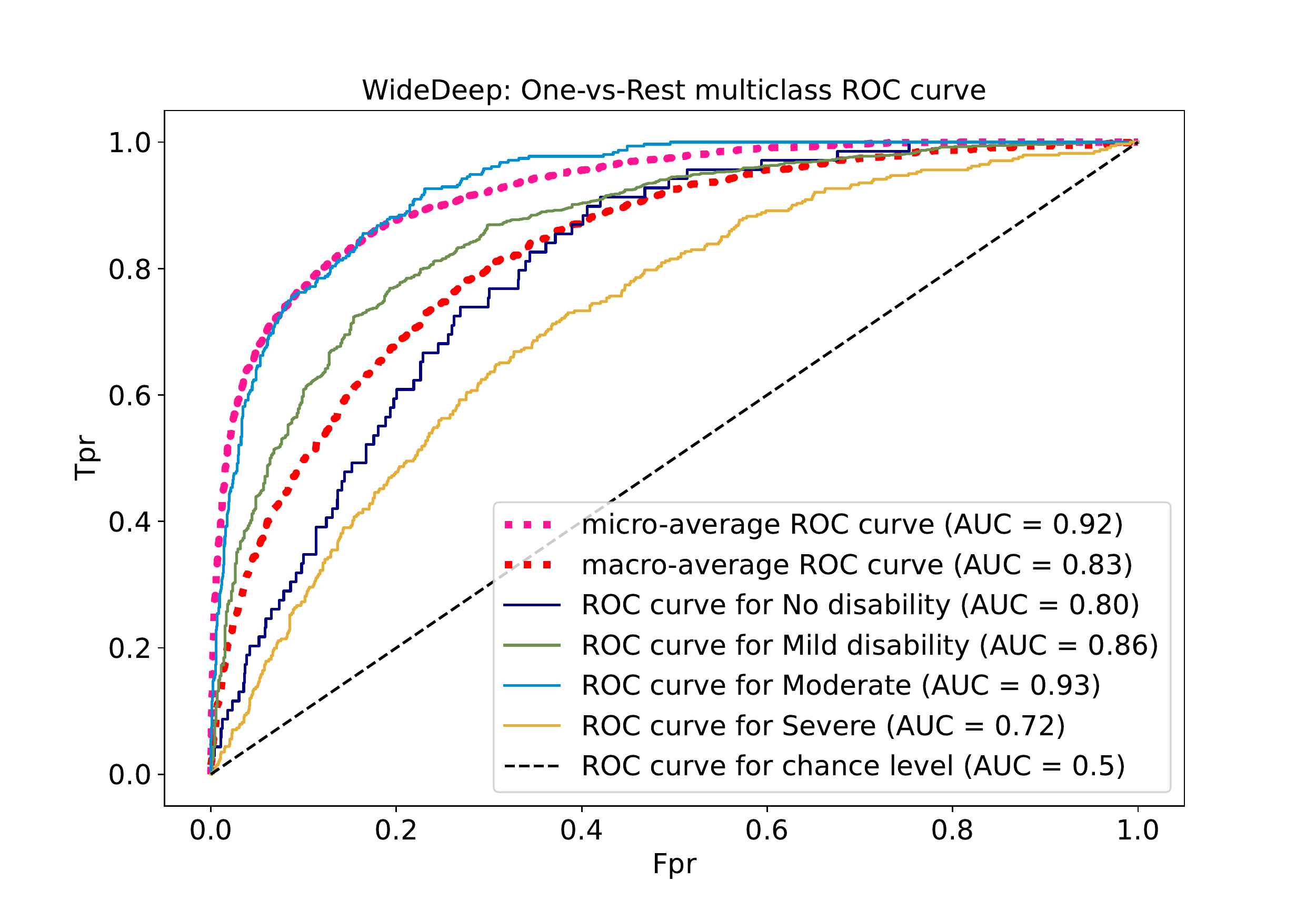}
\includegraphics[scale = 0.32]{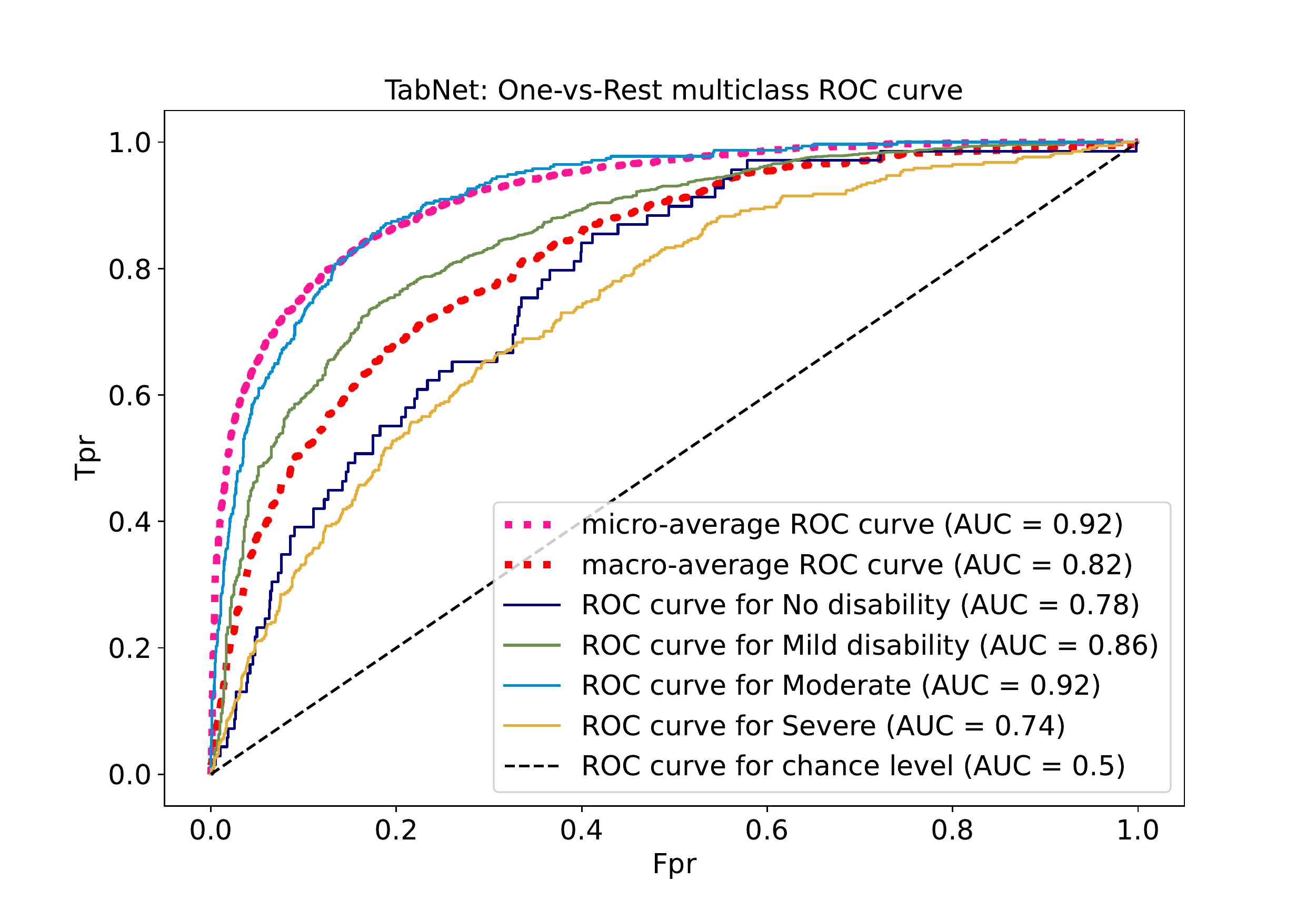}
\caption{}
\label{fig:ROC4}
\end{figure}

\noindent About $68\%$ of participants do not experience any kind of disability, $15\%$ experience a mild level of disability, $14\%$ experience moderate disability, and $3\%$ severe disability. Similar to the previous table, in terms of training and test accuracy, and training and test loss all models have roughly similar performance. As we can see, all models fail to produce any results for 0 level of disability, i.e., no disability, and the Recall for $y=3$, i.e., severe level of disability is close to 0 for all models. Like the 3-level case, the best performance is obtained for the mild class of disability, followed by the moderate class. However, in Figure \ref{fig:ROC4} we can see that the ROC curve for the moderate class is located above the mild class for all thresholds.      
\par
\begin{figure}[H]
\centering
\includegraphics[scale = 0.30]{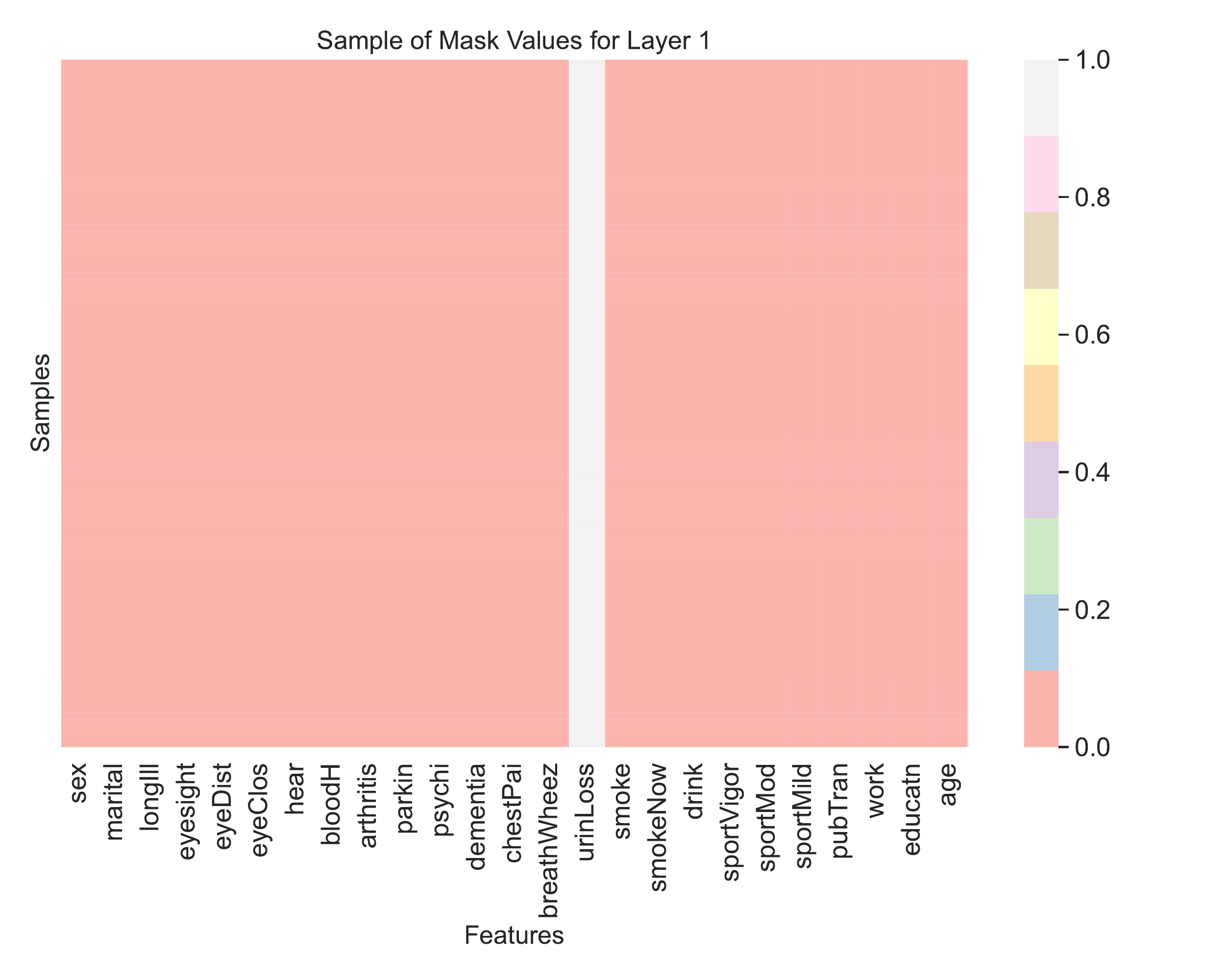}
\includegraphics[scale = 0.30]{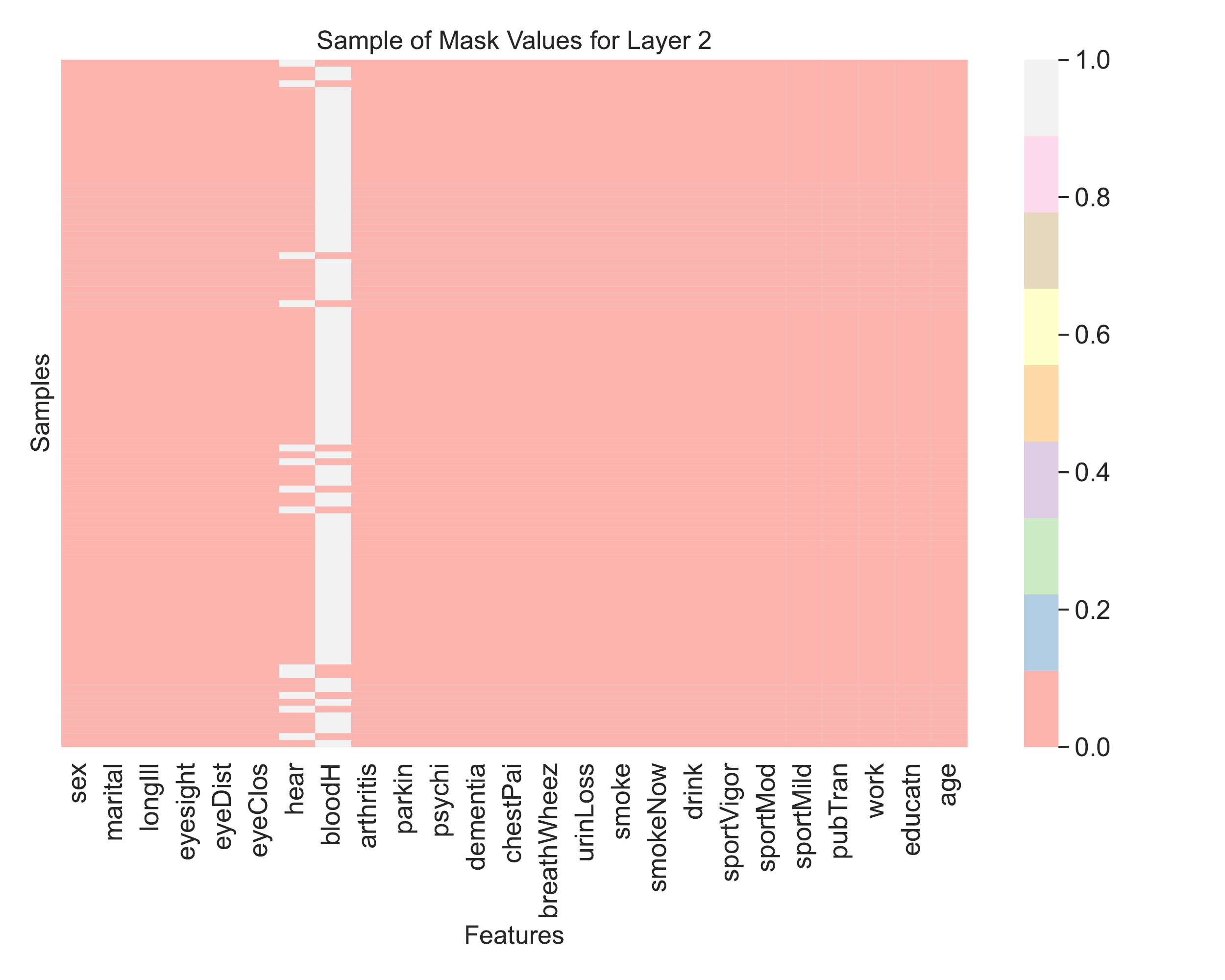}
\includegraphics[scale = 0.30]{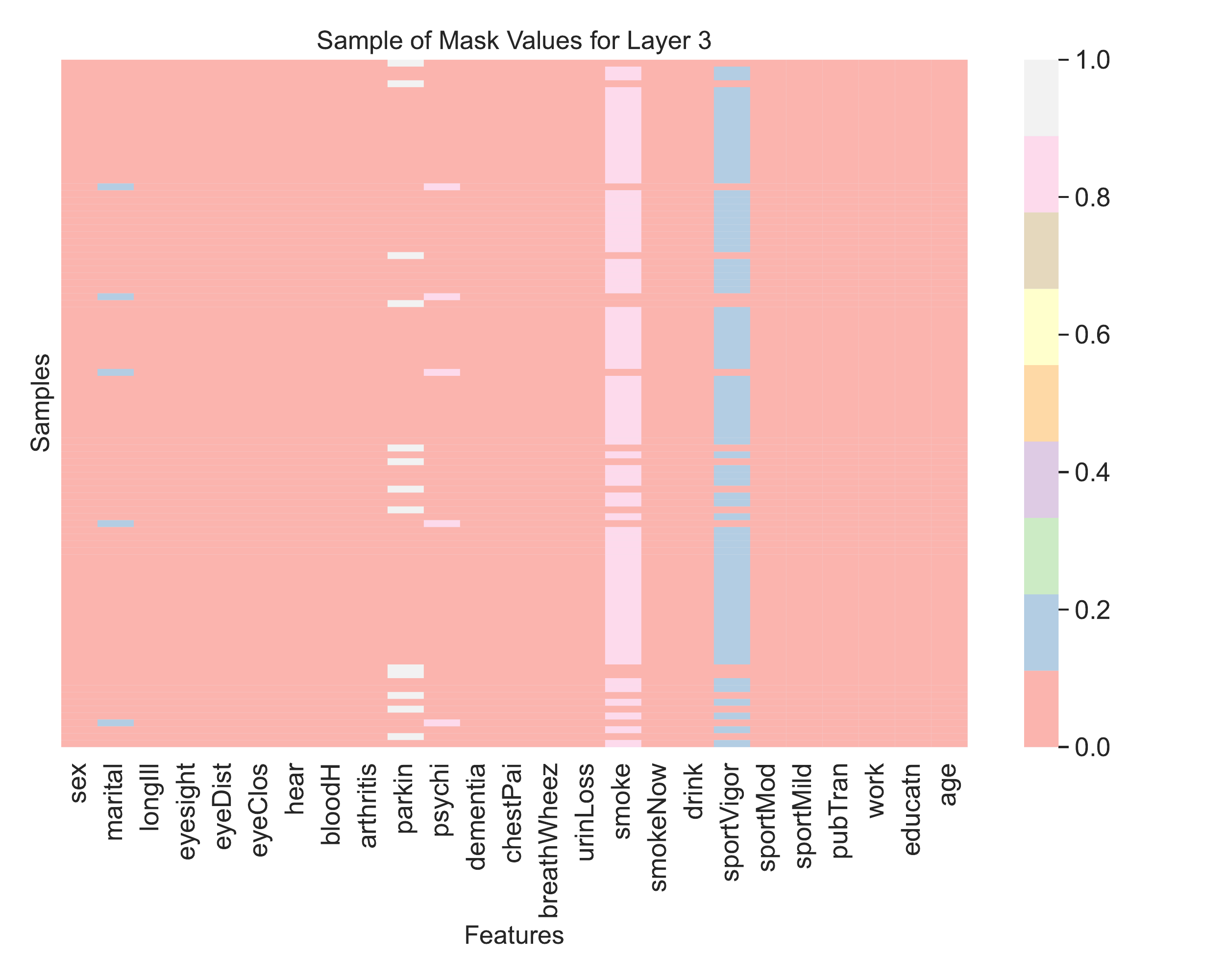}
\includegraphics[scale = 0.30]{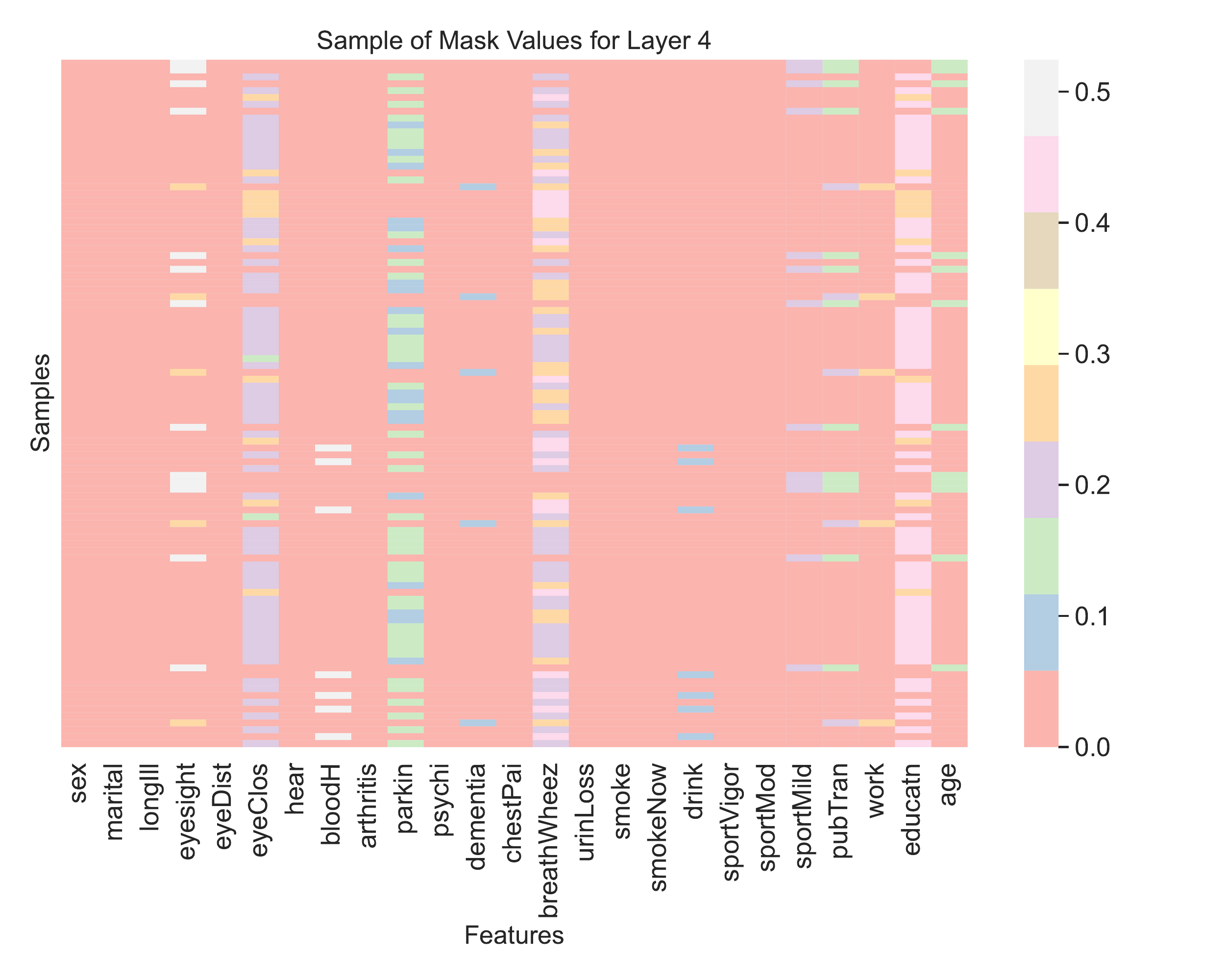}
\includegraphics[scale = 0.30]{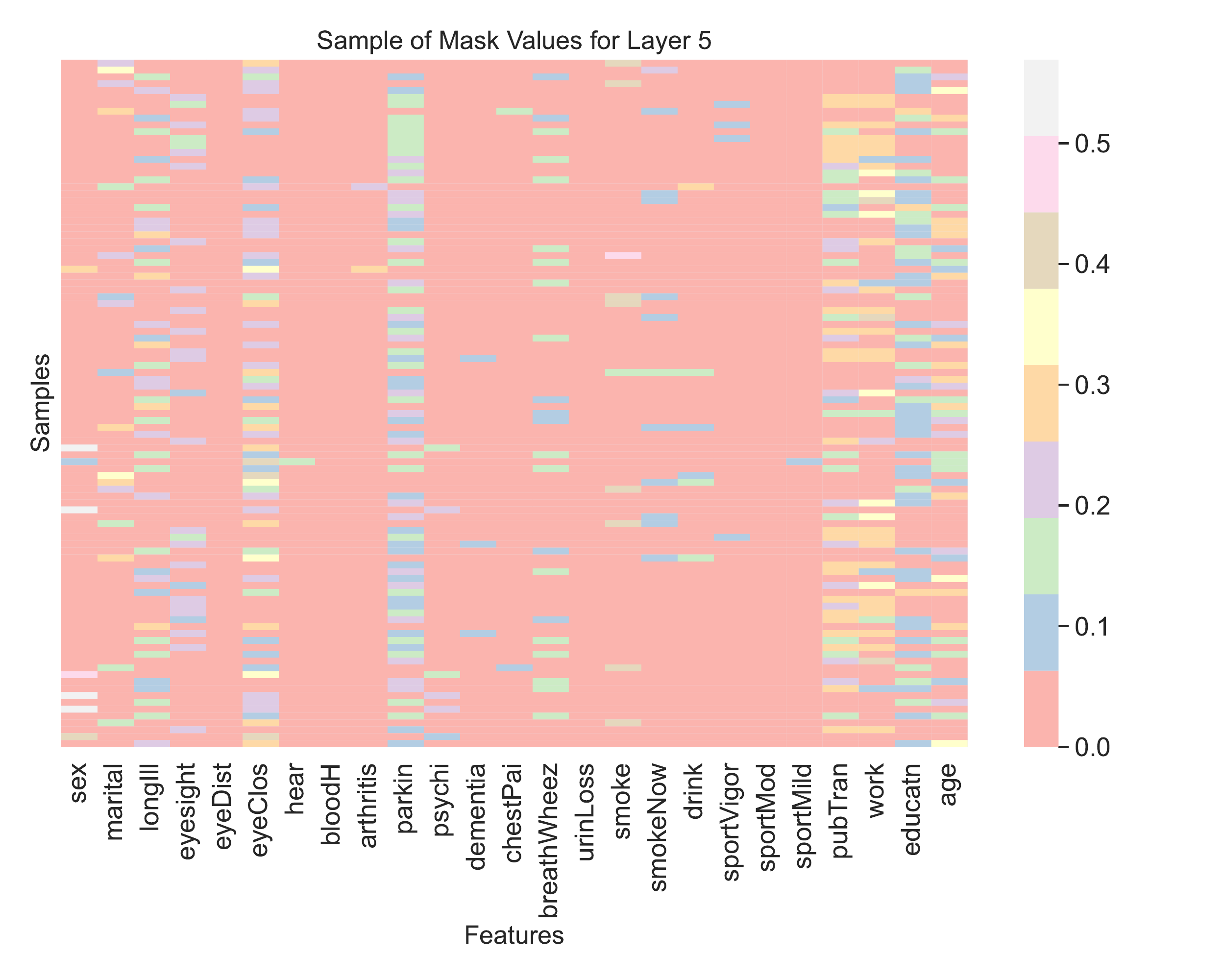}
\includegraphics[scale = 0.30]{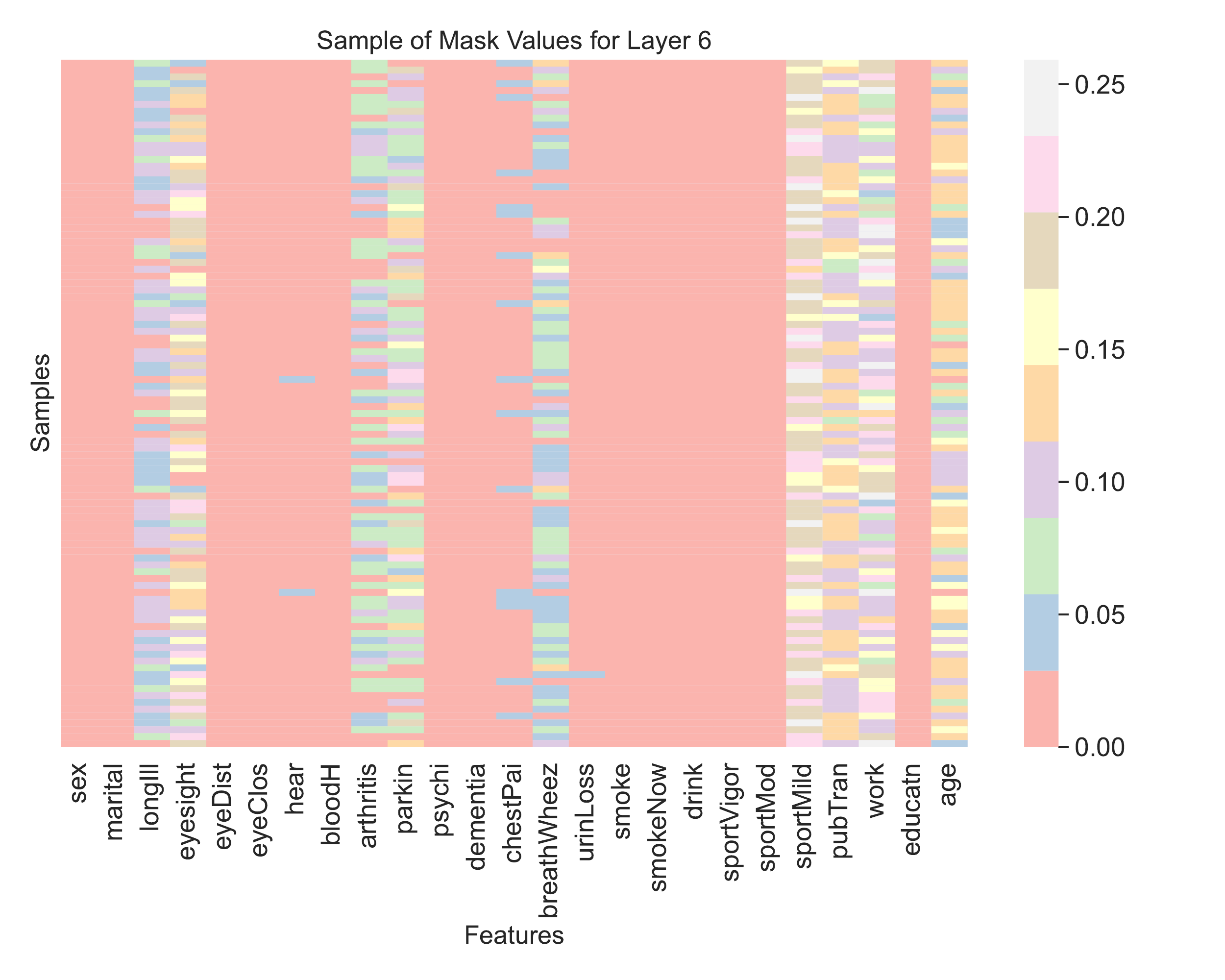}
\caption{TabNet: binary disability. features selection at layers 1, 2, 3, 4, 5, 6}
\label{fig:TN2}
\end{figure}

\begin{figure}[H]
\centering
\includegraphics[scale = 0.30]{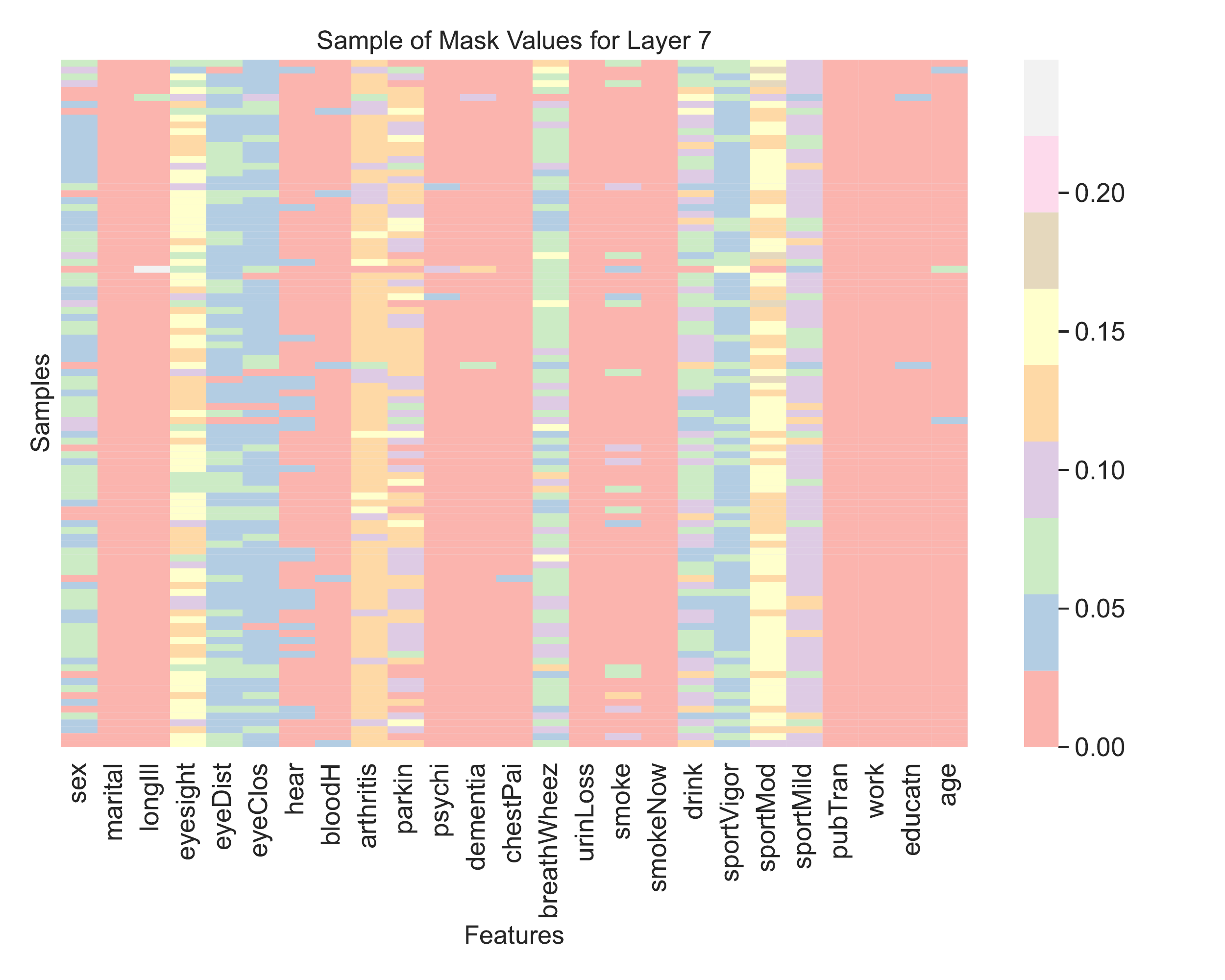}
\includegraphics[scale = 0.30]{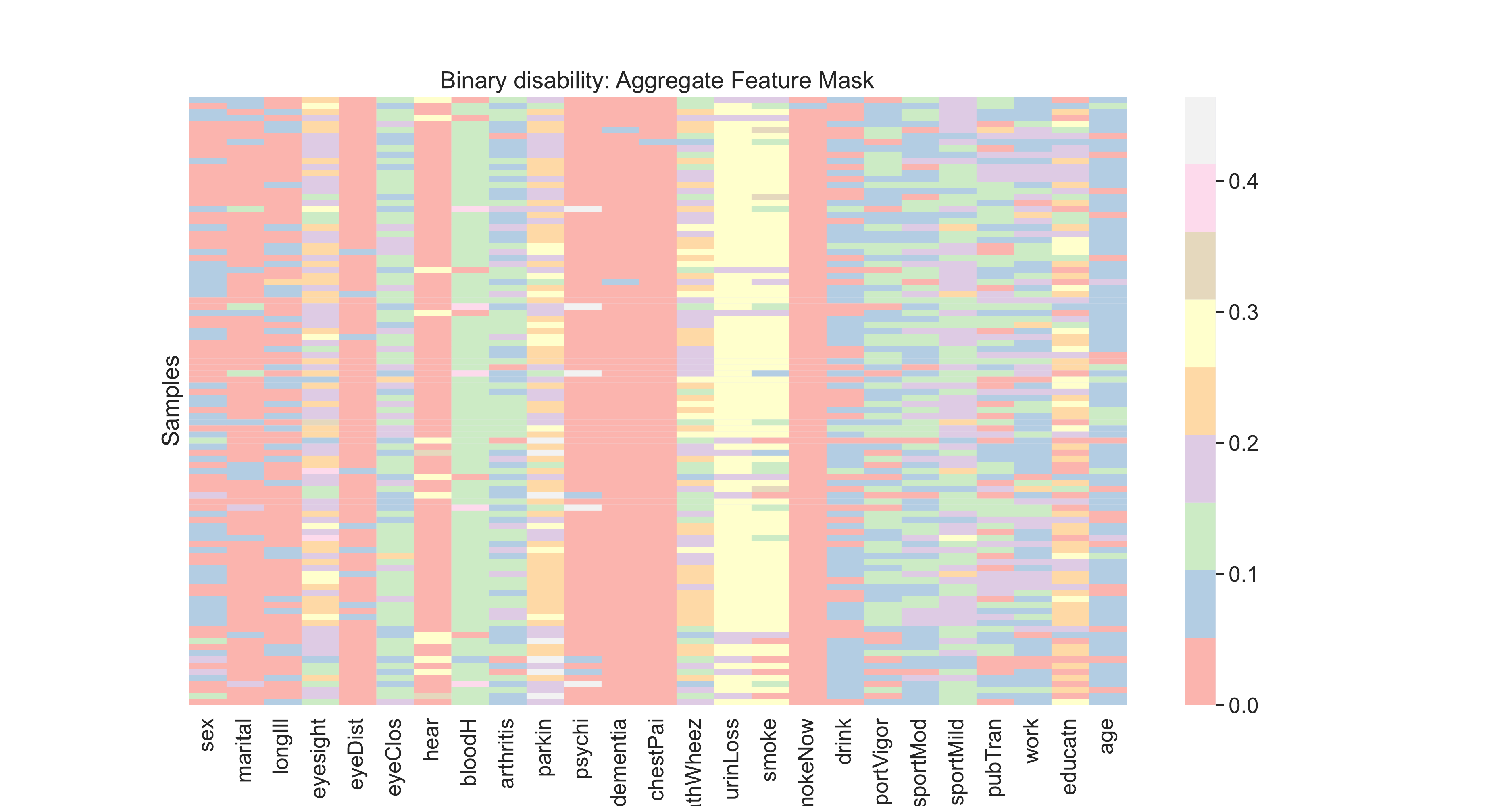}
\caption{TabNet: binary disability, features selection at layer 7 and aggregate layers}
\label{fig:TN2level}
\end{figure}

\begin{figure}[H]
\centering
\includegraphics[scale = 0.30]{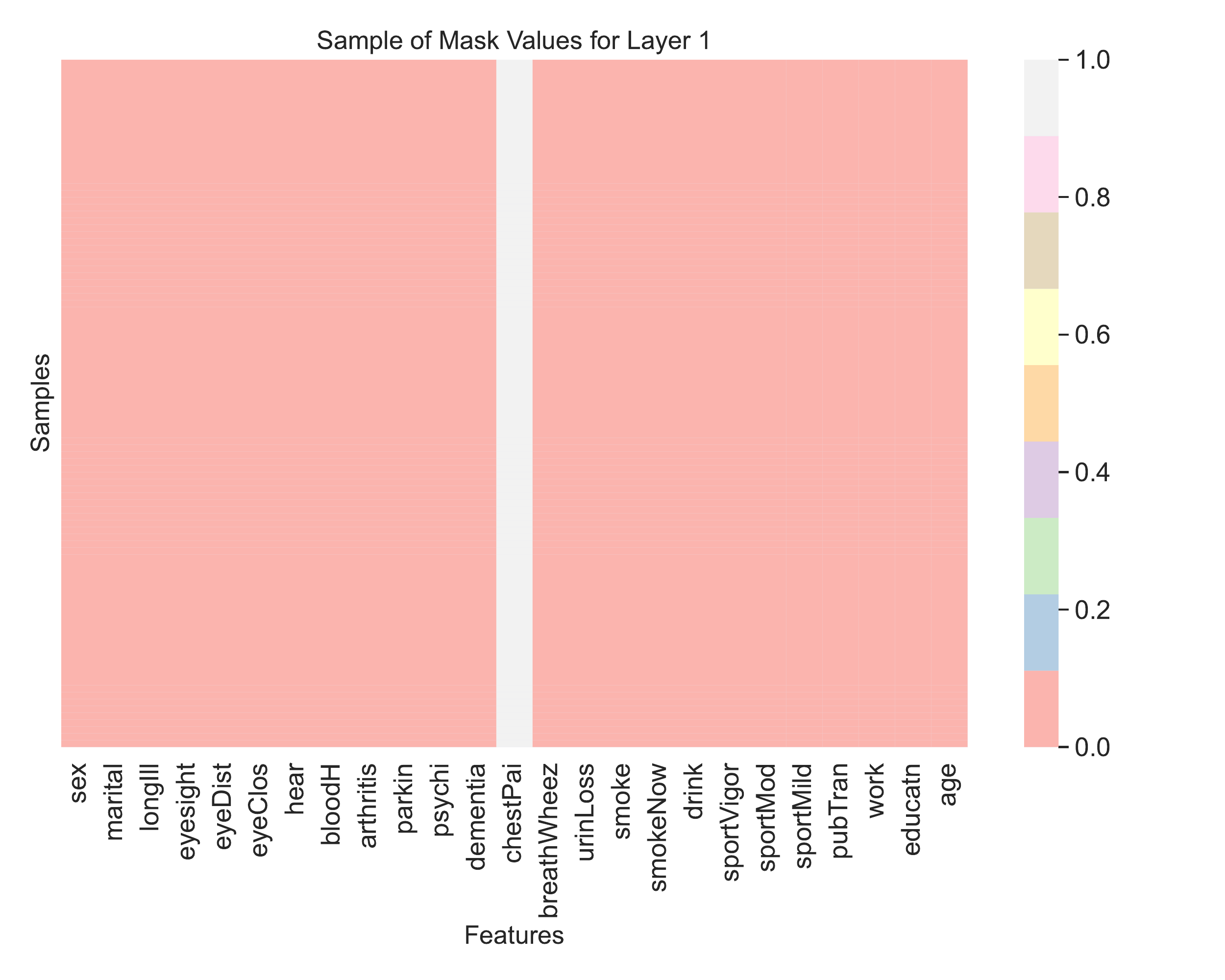}
\includegraphics[scale = 0.30]{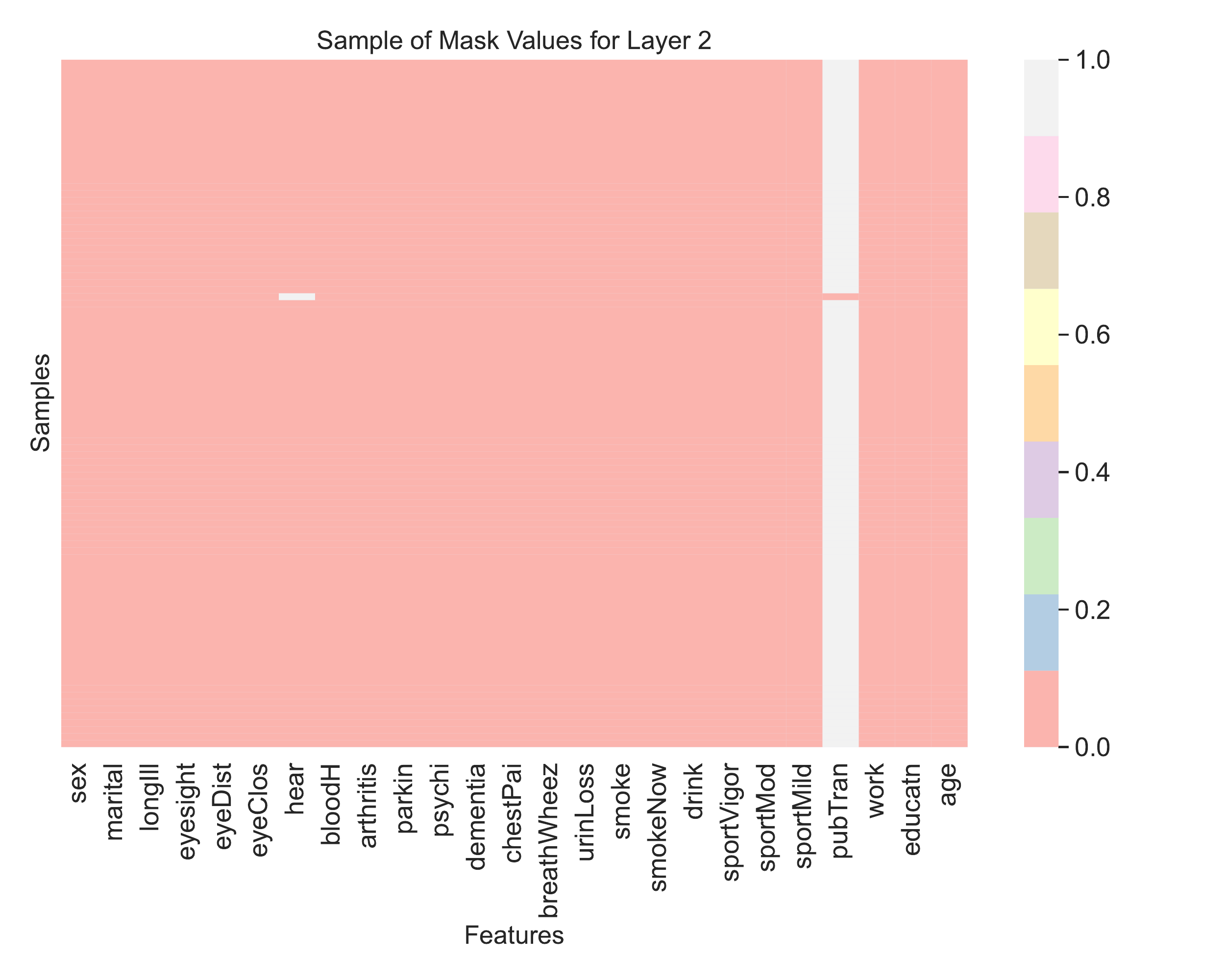}
\includegraphics[scale = 0.30]{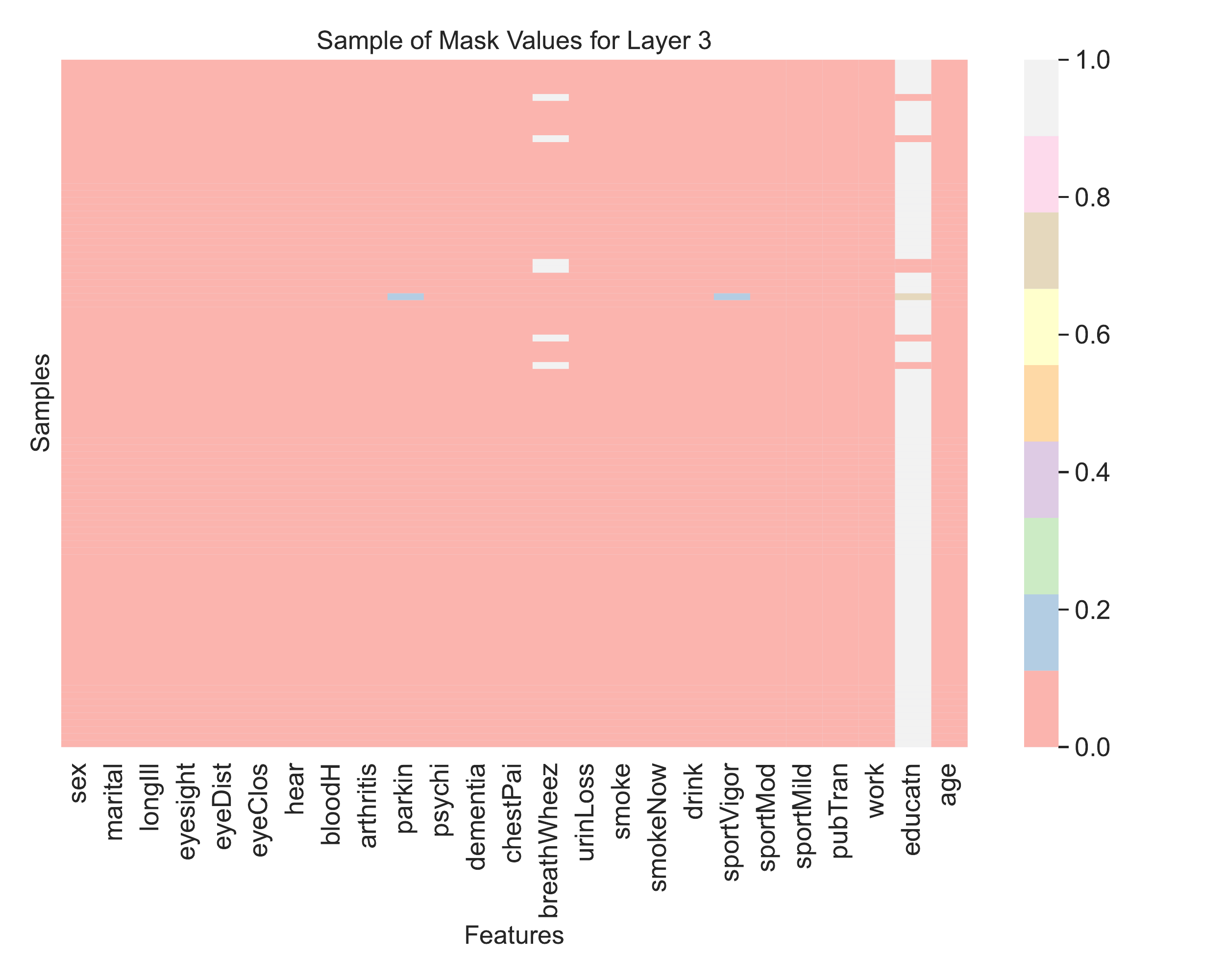}
\includegraphics[scale = 0.30]{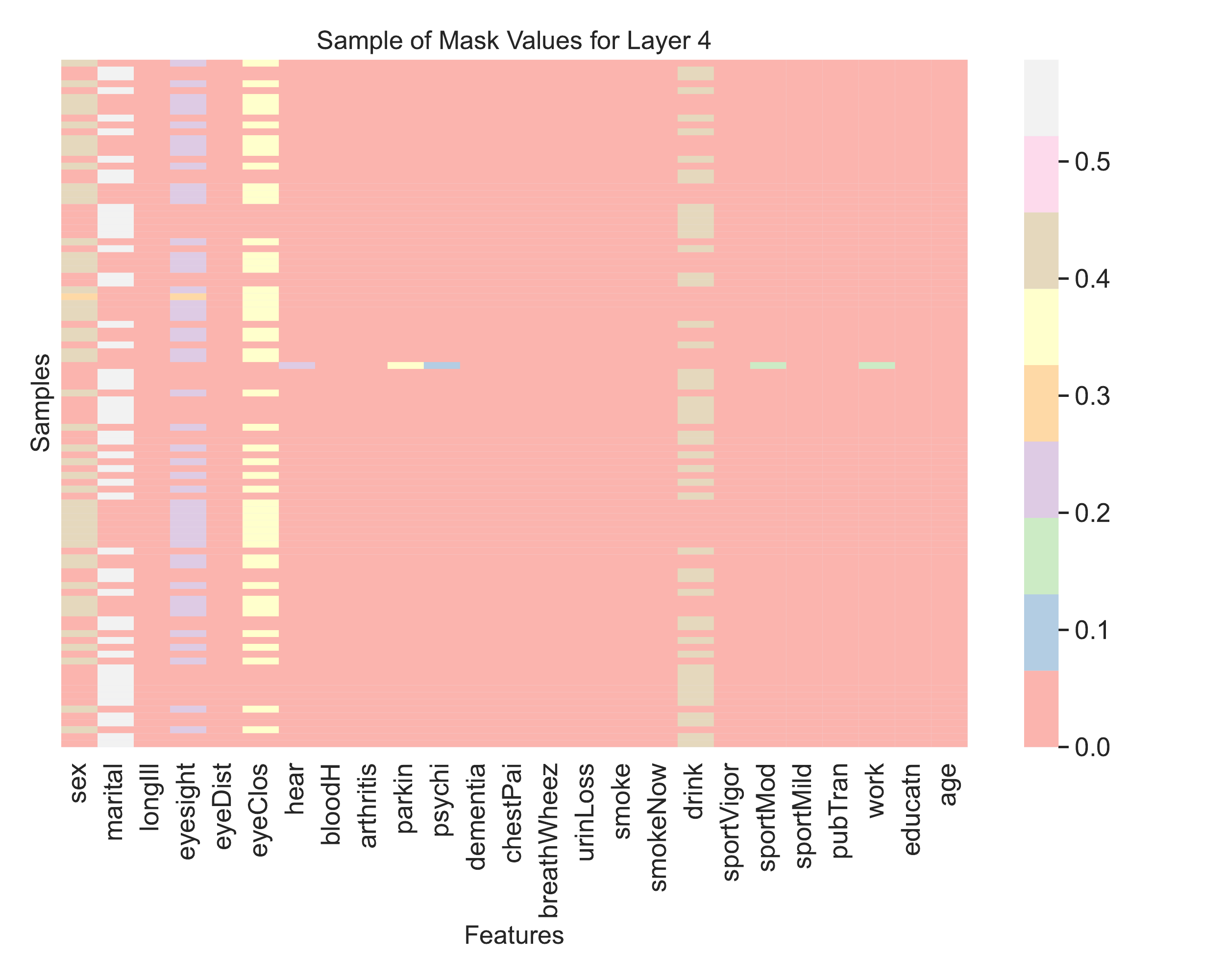}
\includegraphics[scale = 0.30]{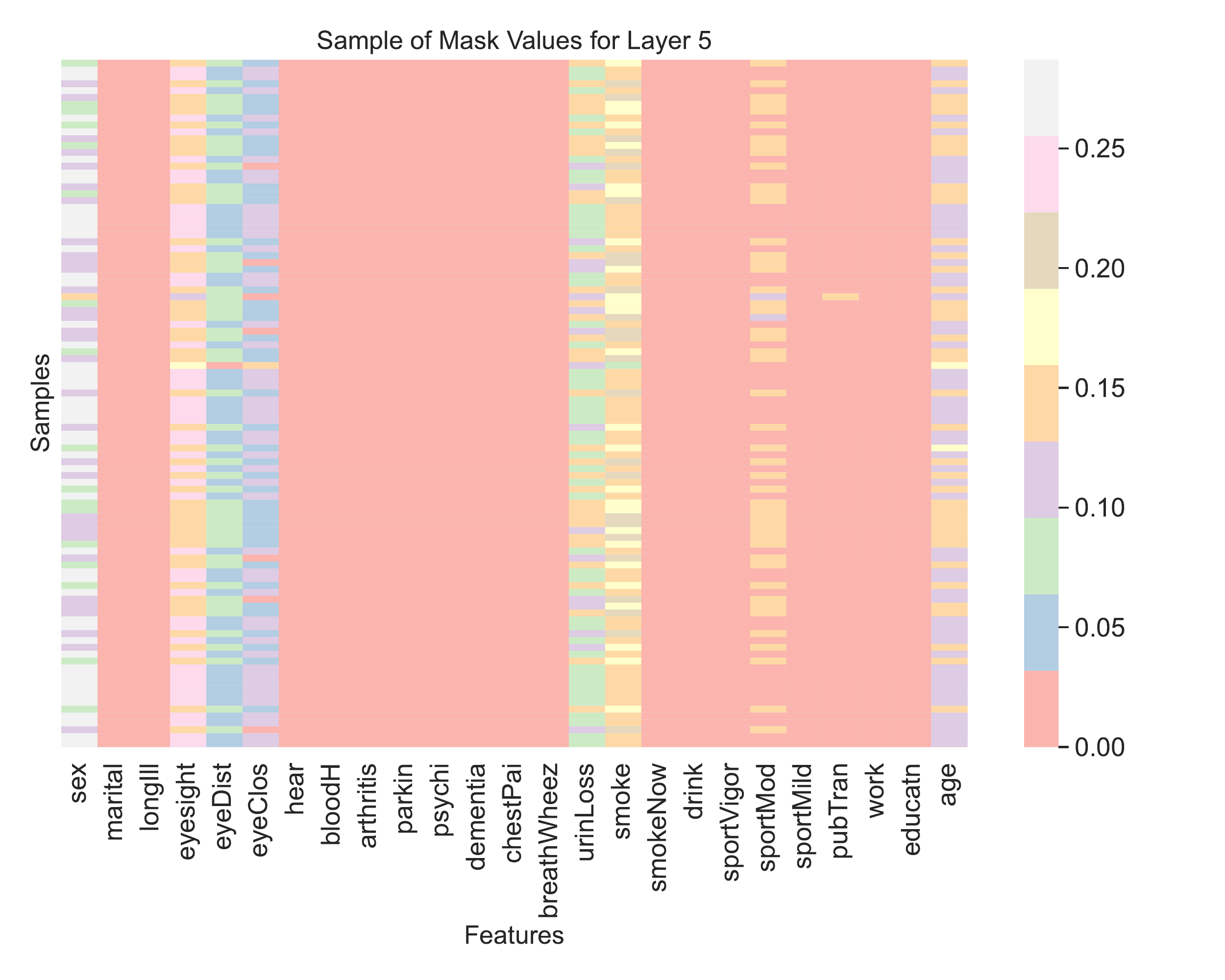}
\includegraphics[scale = 0.30]{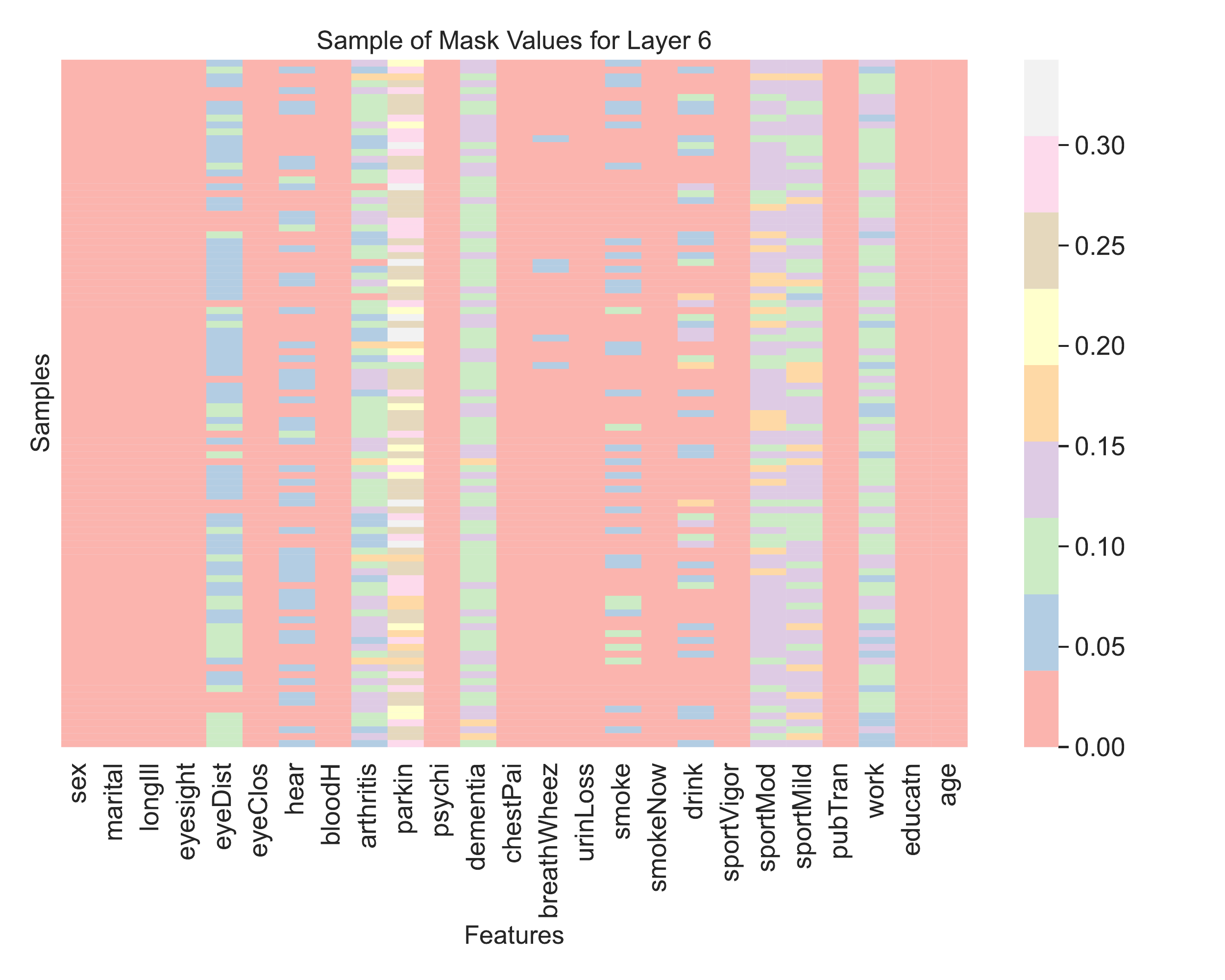}
\caption{TabNet: 3-level disability. features selection at layers 1, 2, 3, 4, 5, 6}
\label{fig:TN3}
\end{figure}

\begin{figure}[H]
\centering
\includegraphics[scale = 0.30]{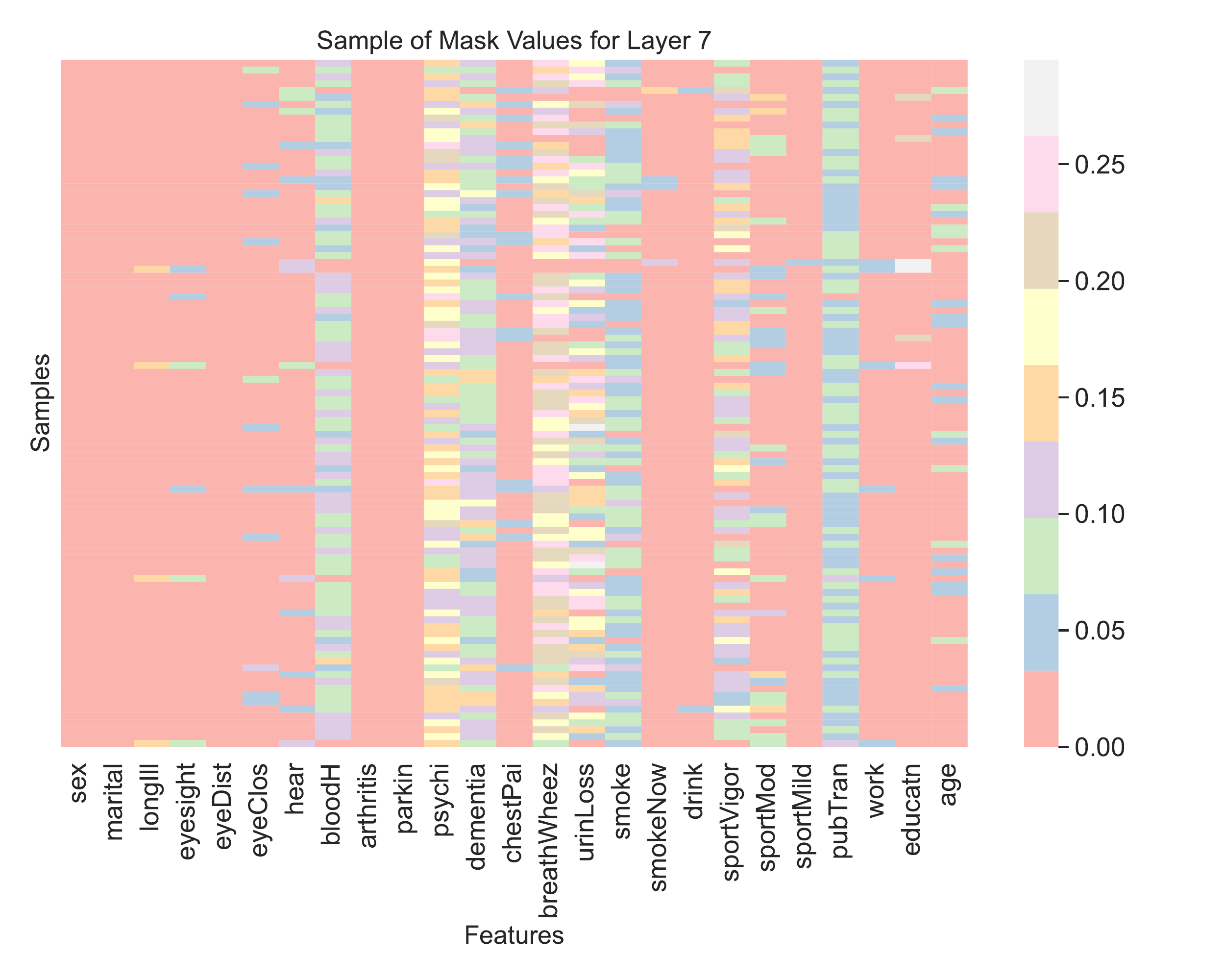}
\includegraphics[scale = 0.30]{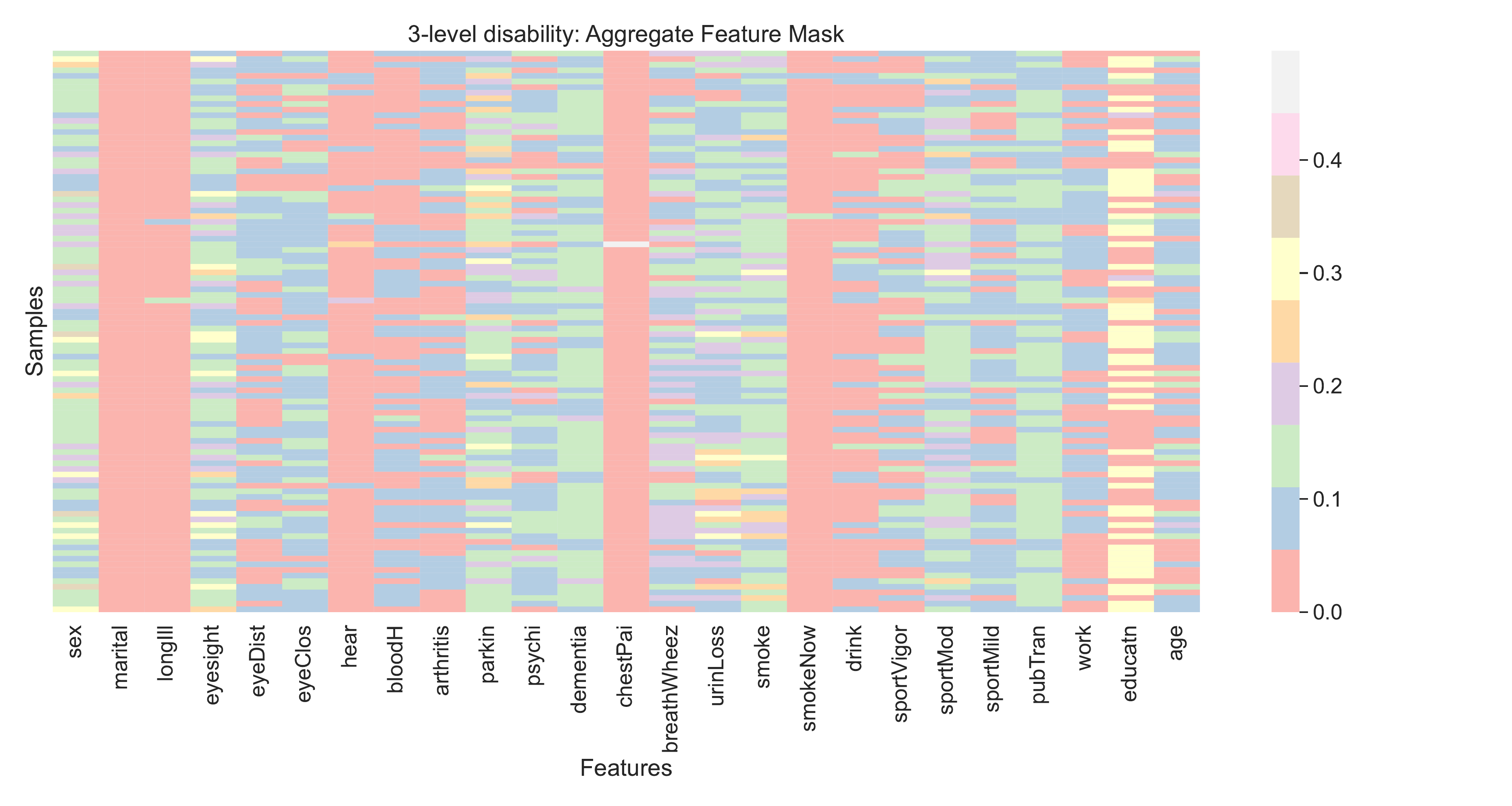}
\caption{TabNet: 3-level disability, features selection at layer 7 and aggregate layers}
\label{fig:TN3level}
\end{figure}

\begin{figure}[H]
\centering
\includegraphics[scale = 0.30]{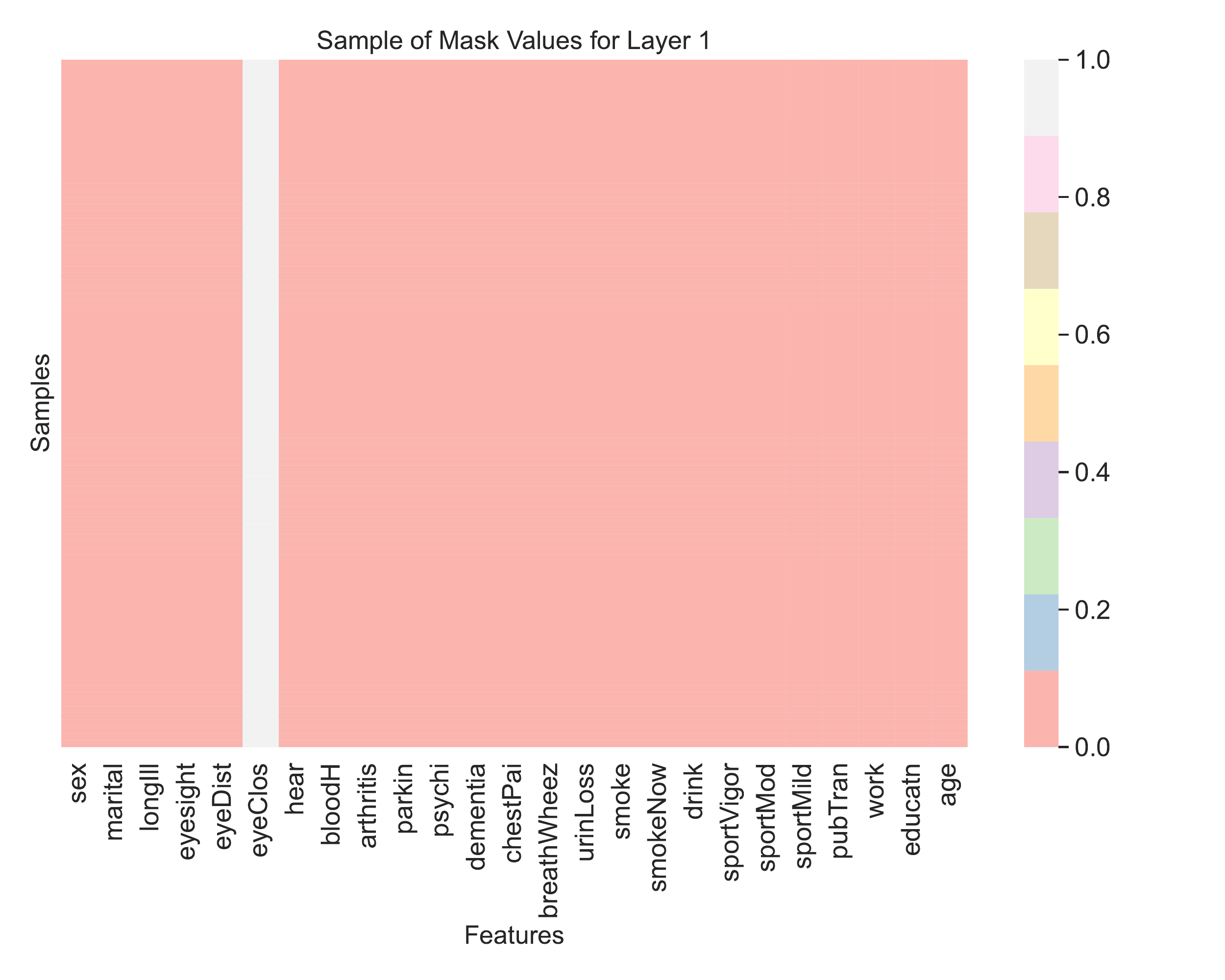}
\includegraphics[scale = 0.30]{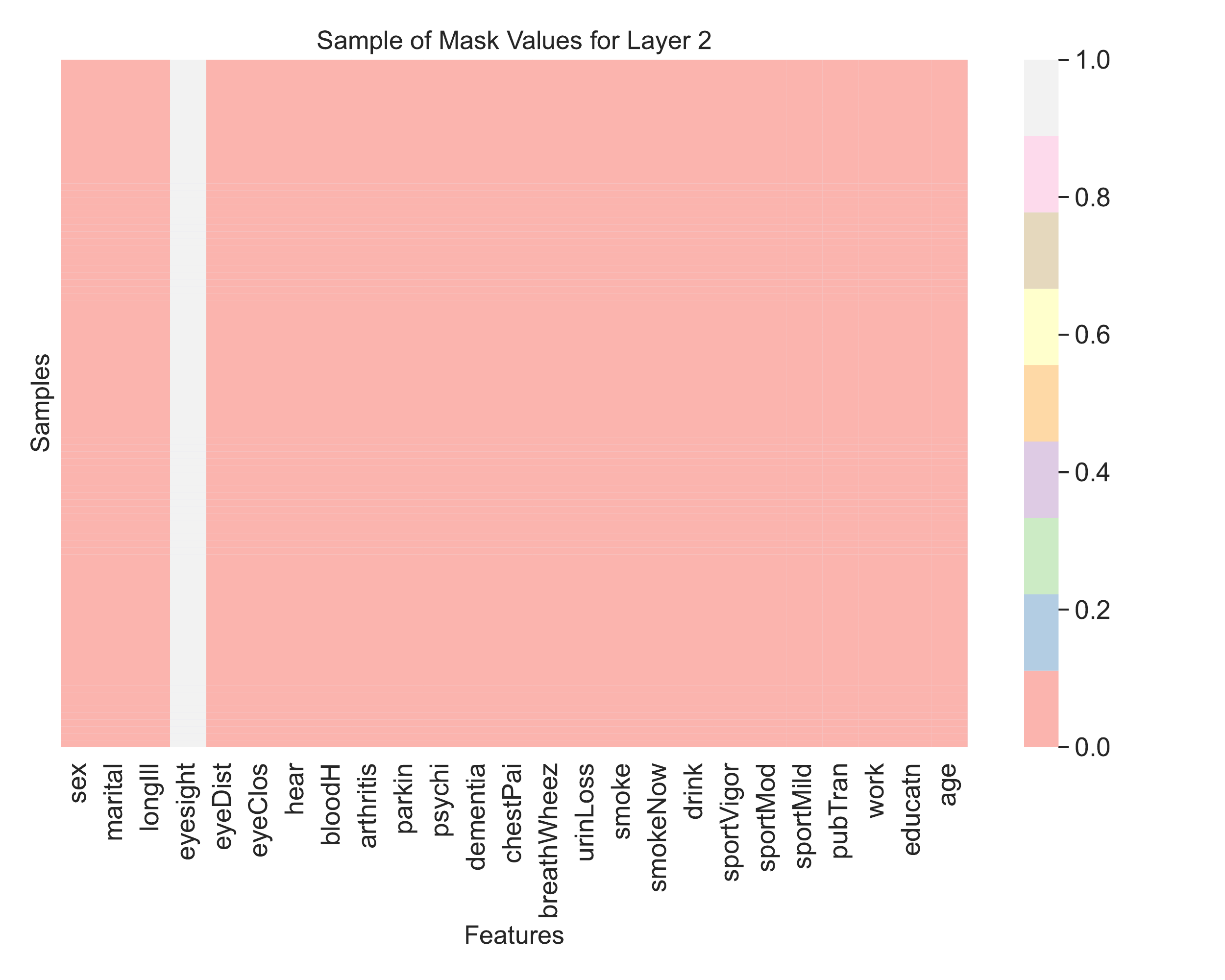}
\includegraphics[scale = 0.30]{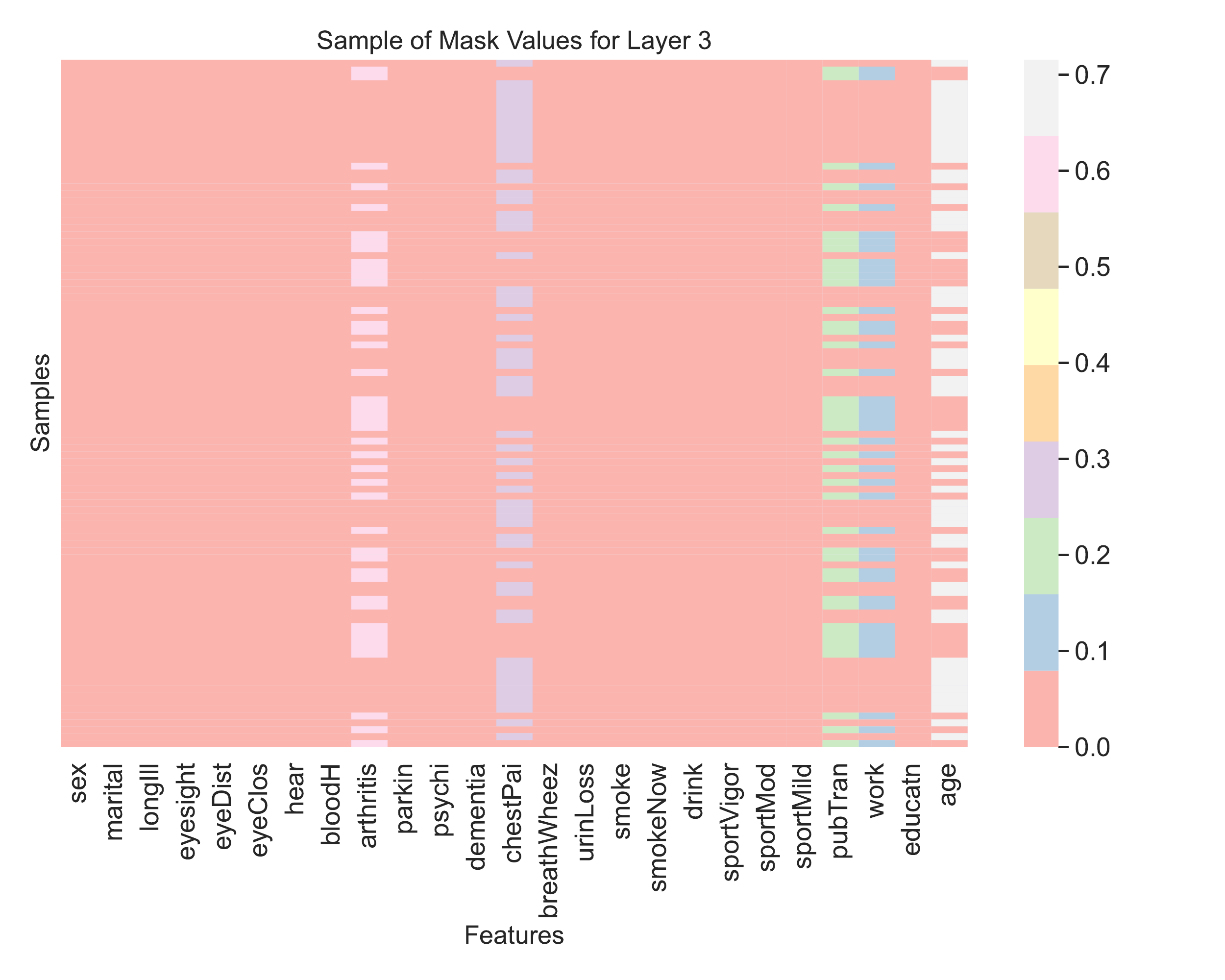}
\includegraphics[scale = 0.30]{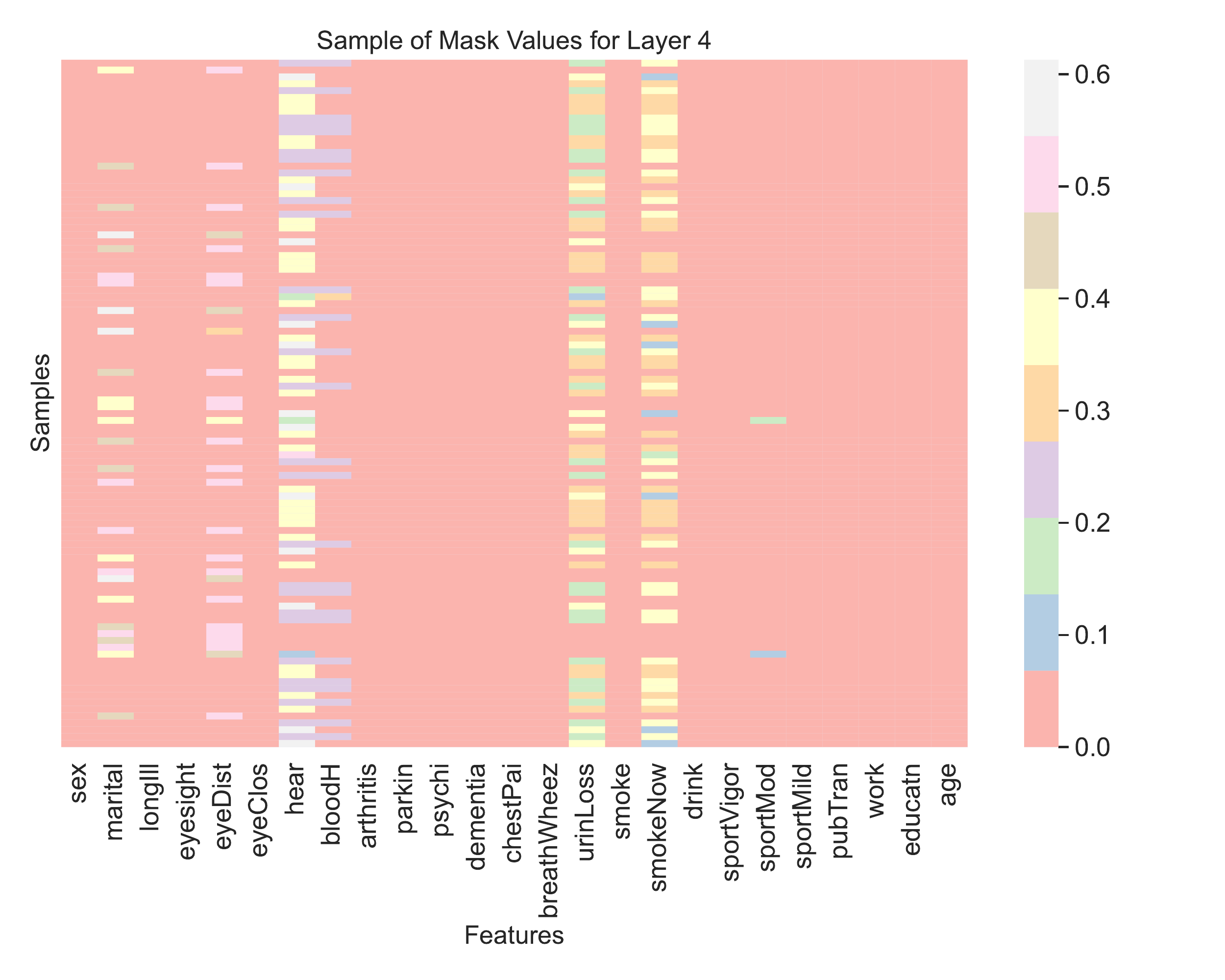}
\includegraphics[scale = 0.30]{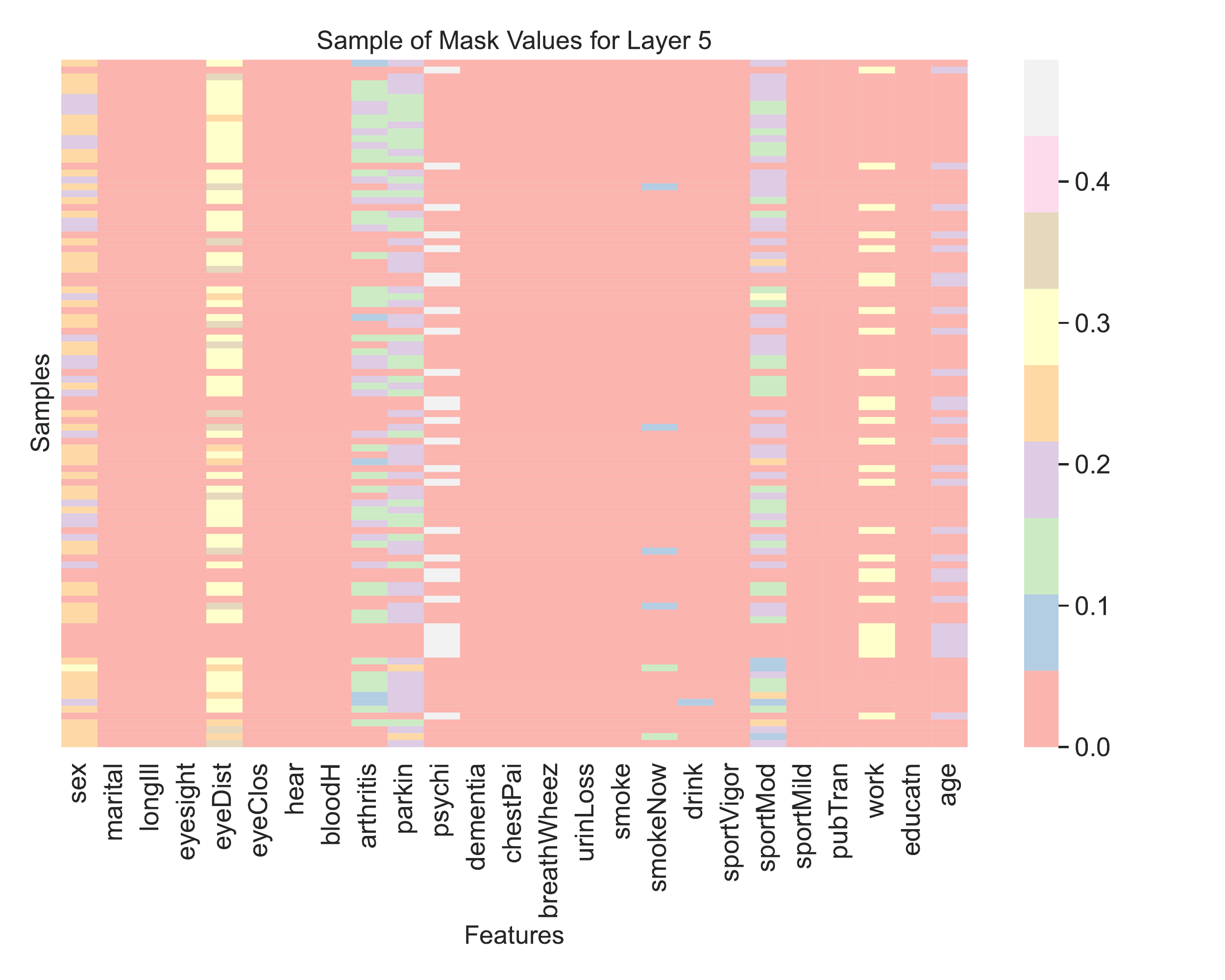}
\includegraphics[scale = 0.30]{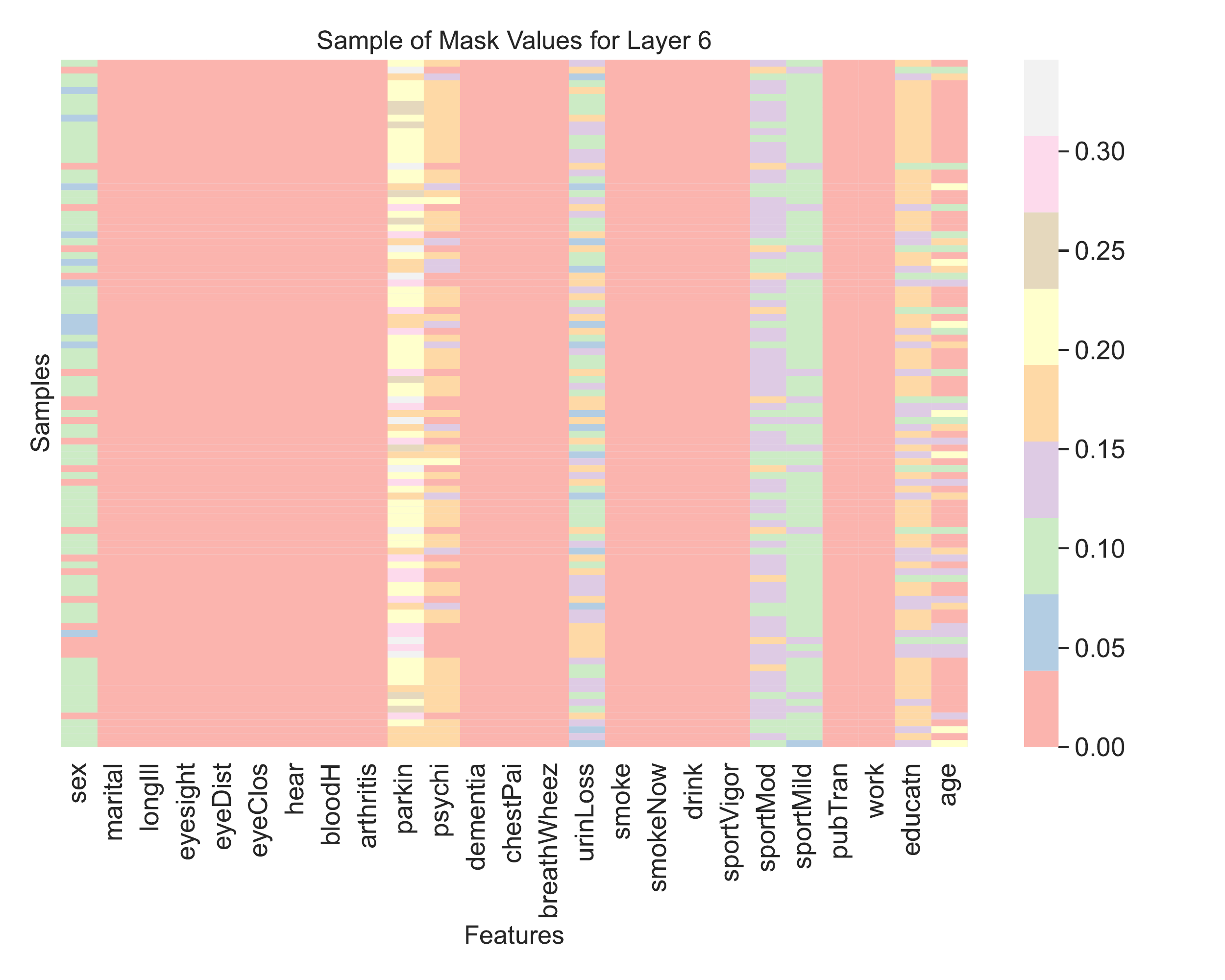}
\caption{TabNet: 4-level disability. features selection at layers 1, 2, 3, 4, 5, 6}
\label{fig:TN4}
\end{figure}

\begin{figure}[H]
\centering
\includegraphics[scale = 0.30]{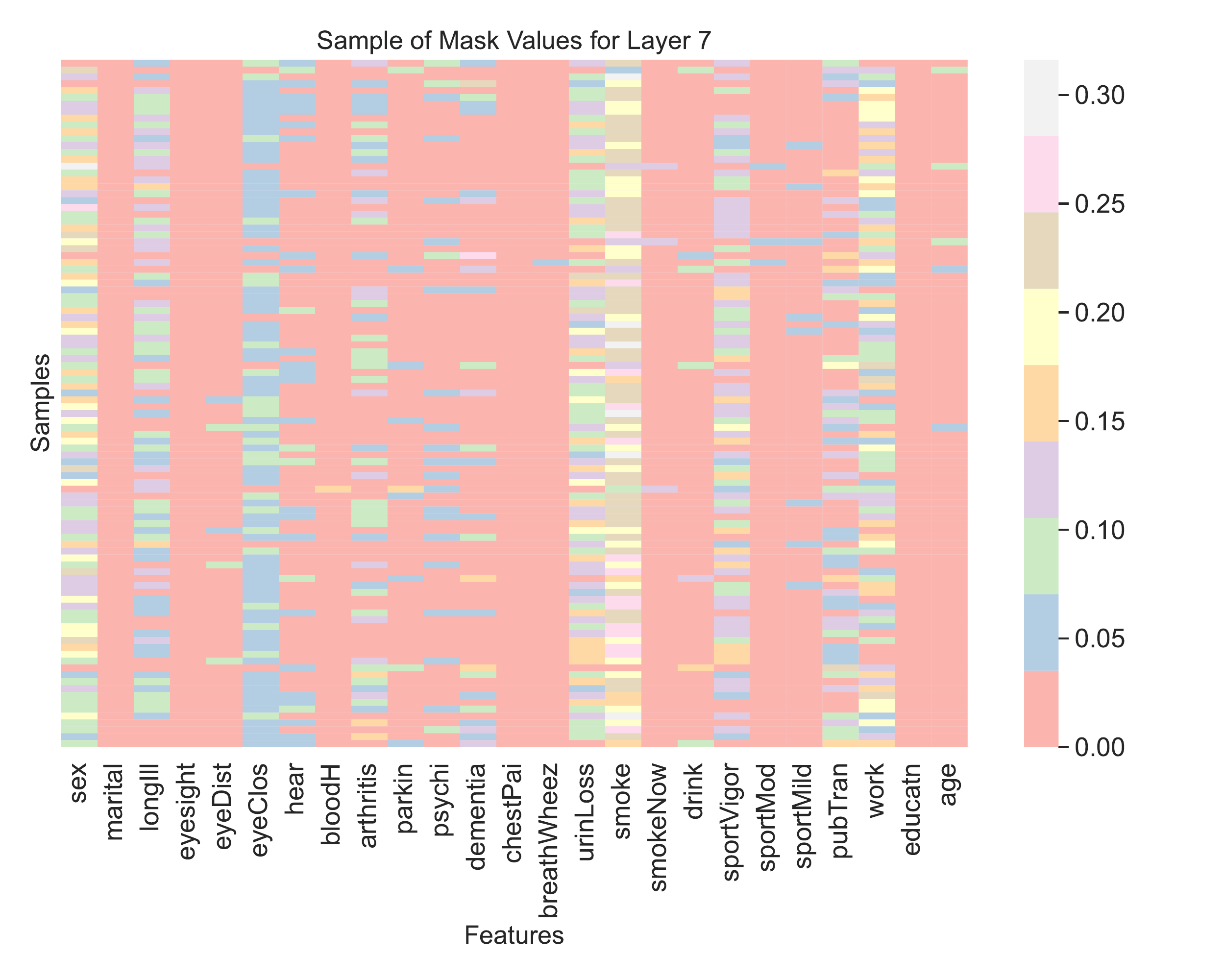}
\includegraphics[scale = 0.30]{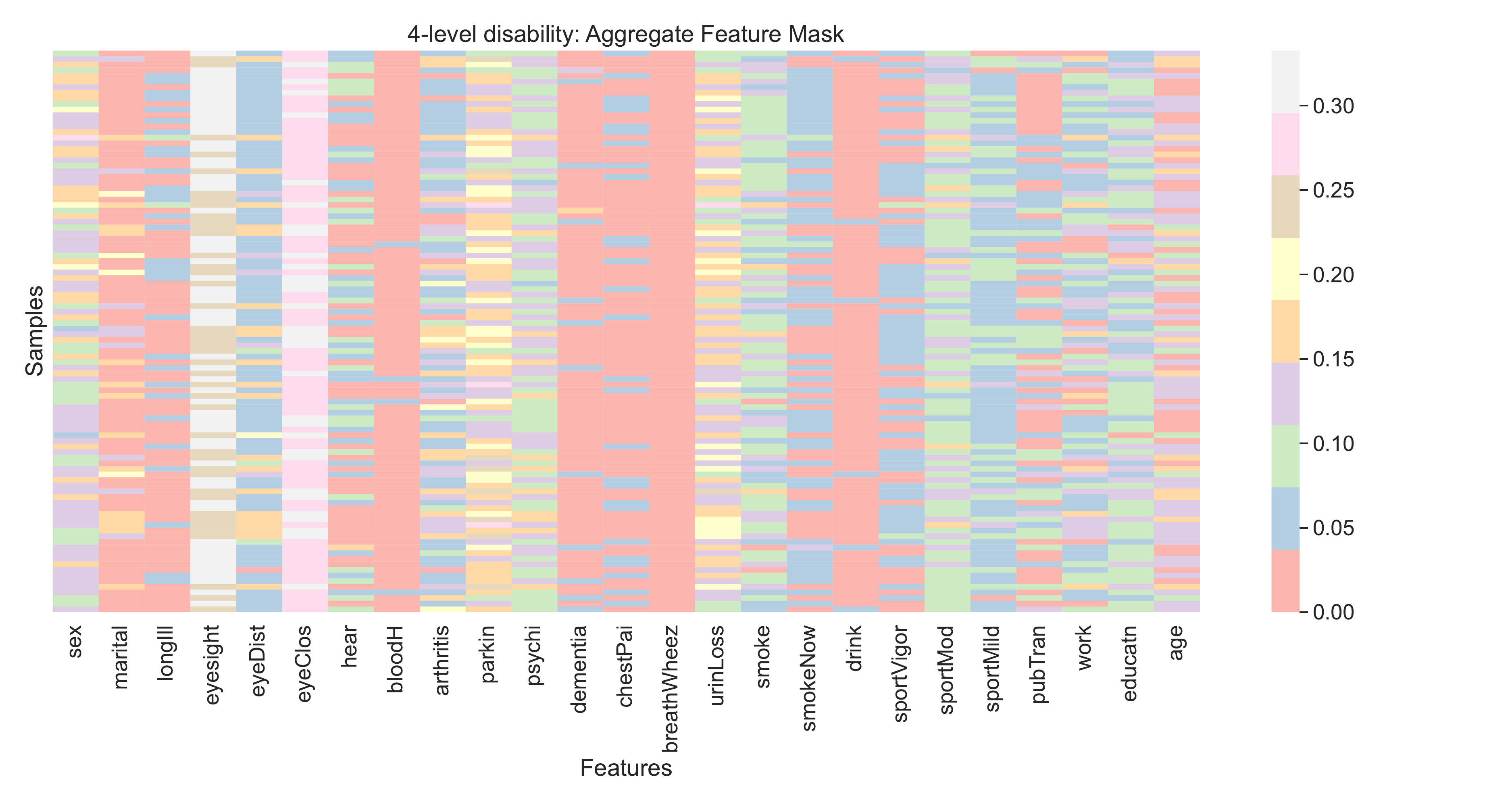}
\caption{TabNet: 4-level disability, features selection at layer 7 and aggregate layers}
\label{fig:TN4level}
\end{figure}

 In general, the models' performance is better in the case of binary disability, with TabNet outperforming TabTransformer and Wide \& Deep models. As expected, micro scores are always higher than macro scores since to calculate macro scores, all classes are equally treated. All models obtain a high level of AUC for all levels of disability, which indicates that they all excel at predicting the levels of disability. We were not able to observe any patterns in models' performance for different disability classes, except that models detect mild and moderate disability more accurately than severe and no disability. In the case of 4-level disability, we can see more instability in the ROC curve for no disability class. But, overall, we can observe consistent patterns in ROC curves among models for specific disability levels. Apart from predictability capacity, another significant property of a model in some industries is its interpretability capacity. As explained before, TabNet has feature selection capacity both at the individual and at the aggregate level. In other words, as we can see in Figures \ref{fig:TN2}-\ref{fig:TN4level} it can select significant features at different decision steps. In these figures, the horizontal axis represents the features, and the vertical axis represents our sample, which includes the first 100 individuals in the test set. The colours show the significance of each feature with the lighter colour being the most relevant features. Some features are represented by pastel red colour. According to the colour bar, these features are not relevant. In Figure \ref{fig:TN2}, we can observe that in the first decision step for binary disability, urinary incontinence is significant in predicting disability for all participants in this sample. Coll-Planas et al. (2008) found that there is a relationship between urinary incontinence and the disablement process. Greer et al. (2015) also studied this relationship among US women above 40. In the second decision step, we can see that high blood pressure is selected as a significant factor for almost all participants in this sample and hearing problems as another important factor among some people. He and MacGregor (2007) found that blood pressure is the most significant cause of death and disability by causing strokes and coronary heart 

\noindent disease. See also Hajjar et al. (2007), who found that people with high blood pressure are at greater risk of disability. In the third decision step, Parkinson's and psychiatric diseases are selected only for some people in this sample, whereas smoking, regardless of whether quitting or not, and vigorous activities are selected for the majority of people. However, looking at the colour bar, we can see that smoking is more significant than vigorous activities. Our findings agree with the report by ASH (2020) that the relationship between ever smoked and disability is stronger than between recent smoking and disability. Using logistic regression, they find that people who have ever smoked are $2.35\%$ more likely to be disabled than those who have never smoked. At decision levels 4, education, at level 5, use of public transport, at level 6, age, work, mild activities, and eyesight, and at level 7, moderate activities and eyesight are identified as important features. Figure \ref{fig:TN2level} shows the aggregate feature importance. Solid pastel red areas represent insignificant features. Smoking (still smoking) is not a significant factor for any participants in this sample. Factors, such as marital status, dementia, and chest pain are selected only for some participants. In this figure, we can see that urine incontinence and ever smoking are represented by light colour. Paterson and Warburton (2010) studied the relationship between different levels of physical activities and disability. They found that greater physical activity is associated with higher functional status, and moderate level of physical activity reduces the risk of functional limitations and disability.

\noindent Chakravarty et al. (2012) also found that lifestyle factors such as smoking and routine exercise can reduce disability and improve survival. Kc and Lentzner (2010) studied the relationship between education and disability in 70 countries and concluded that formal education is associated with lower levels of disability and mortality. Laforge et al. (1992) looked at the relationship between hearing and vision impairments and functional decline, and 1-year mortality. The selection of public transport as an important feature can be justified by the fact that it can be considered as a mild activity that affects disability. In Figure \ref{fig:lifestyle} and Table \ref{tab:lifstyl}, we can also see that most disabled people rarely use public transport. Figures \ref{fig:TN3}, and \ref{fig:TN3level} show the features importance for 3-level disability. As we can see, the results are not consistent with binary disability. In the first decision step, chest pain is selected as the significant factor for all members in our sample, and in the second decision step, the use of public transport, and step 3, education level. Even marital status is selected in step 4. But if we look at the aggregate results in Figure \ref{fig:TN3level}, the results are more consistent. Red pastel area includes marital status, chest pain, and still smoking similar to Figure \ref{fig:TN2level}. In Figure \ref{fig:TN3level}, the aggregate feature mask shows that the most significant factor is education, although not selected for all participants in the sample. Figure \ref{fig:TN4} and \ref{fig:TN4level} illustrate feature importance for 4-level disability. This time, in step 1, close-sightedness and in step 2, eyesight is considered significant factors. Both of them are also represented by bright colour in aggregate results in Figure \ref{fig:TN4level}.       

\section{Conclusion}
In an aging population, forecasting the levels of disability and the factors that affect the severity of disability can help decision makers to better plan for the required medical, health services, and social support. This can help with investment and cost saving. In this study, we try three neural networks on ELSA to predict the levels of disability and the factors that affect disability. We first apply $K$-modes algorithm, to cluster participants into 2, 3, and 4 levels of disability and then use Wide \& Deep, TabTransformer, and TabNet models to predict these levels. Wide \& Deep model is used in recommender systems and combines wide models such as logistic regression and deep model. TabTransformer and TabNet are both Transformer-based models. We compared our models based on Accuracy, Precision, Recall, $F_1$ score, and AUC and found that all three models have high predictability power and their performance in the case of 3-level and 4-level disability is very similar. However, in the case of binary disability, TabNet outperforms the other two models. Another advantage of TabNet over other models is its interpretability due to sequential processing similar to DTs. It can identify significant features at each decision step and can also provide aggregate decisions. However, we found that the results for feature importance are not consistent for different levels of disability. Looking at the aggregate results, we found that factors, such as urinary incontinence, ever smoking, education, and exercise are significant and affect disability.

\newpage

\appendix

\section{Appendix}
\label{app:activities}

\setcounter{table}{0}
\setcounter{figure}{0}
\setcounter{equation}{0}
\renewcommand{\thetable}{A\arabic{table}}
\renewcommand{\thefigure}{A\arabic{figure}}
\renewcommand{\theequation}{A\arabic{equation}}

\begin{table}[H]
\caption{Activities related to mobility, ADLs, and IADLs}
\label{tab:ADL}
\centering
\footnotesize
\scalebox{0.7}{
\begin{tabular}{ll}\toprule
Mobility	& walking 100 yards \\
		& sitting for about two hours\\
		& getting up from a chair after sitting for long periods\\
		& climbing several flights of stairs without resting\\
		& climbing one flight of stairs without resting\\
		& stooping, knelling or crouching\\
		& reaching or extending your arms above shoulder level\\
		& pulling or pushing large objects like a living room chair\\
		& lifting or carrying weights over 10 pounds like a heavy bag\\
		& picking up a 5p coin from a table\\\midrule
ADL		& dressing, including putting on shoes and socks\\
		& walking across a room\\
		& bathing or showering\\
		& eating, such as cutting up your food\\
		& getting in or out of bed\\
		& using the toilet, including getting up or down\\\midrule
IADL		& using a map to figure out how to get around in a strange place\\
		& preparing a hot meal\\
		& shopping for groceries\\
		& making telephone calls\\
		& taking medications\\
		& doing work around the house or garden\\
		& managing money, such as paying bills and keeping track of expenses\\\bottomrule
\end{tabular}}
\end{table}

\begin{table}[h]
\caption{Lifestyle factors and disability levels: 2 levels (disability 1), 3 levels (mild 1, severe 2), 4 levels (mild 1, moderate 2, severe 3), no disability 0, never 0.}
\label{tab:lifstyl}
\centering
\footnotesize
\scalebox{0.6}{
\begin{tabular}{lccccccc}\toprule
					   & Disability $(1)$  &  Mild $(1)$ & Severe $(2)$ & Mild $(1)$ & Moderate $(2)$ & Severe $(3)$ & Total\\\midrule
Smoking (ever)			   & 1,204		       &  4,789	   & 1,071		   & 4,789	       & 1,071		  & 1,126		  & 7,231\\
Nowadays				   & 359		       & 1,335	   & 320		   & 1,335	       & 320			  & 287		  & 1,997\\	\midrule
Drinking (0)		   	   & 429		       & 664		   & 397		   & 664	       & 397			  & 194		  & 1,338\\
Special occasions	           & 468		       & 1,317	   & 405		   & 1,317	       & 405			  & 378		  & 2,193\\
1 or 2 per month	           & 163		       & 800		   & 145		   & 800	       & 145			  & 187		  & 1,156\\
1 or 2  per week	           & 351		       & 2,514	   & 312		   & 2,514	       & 312			  & 467		  & 3,383\\
Daily					   & 272		       & 2,016	   & 237		   & 2,016	       & 237			  & 358		  & 2,660\\
$>$ 1 per day	   		   & 64		       & 344		   & 57		   & 344	       & 57			  & 72		  & 489\\\midrule
Active sports (0)	           & 1,614 		       & 4,184	   & 1,461		   & 4,184	       & 1,461		  & 1,176		  & 7,115\\
1 to 3 per month		   & 52		       & 847		   & 33		   & 847	       & 33			  & 150		  & 1,052\\
1 per week			   & 33		       & 897		   & 21		   & 897	       & 21			  & 138		  & 1,071\\
$>$ once a week		   & 48		       & 1,727	   & 38		   & 1,727	       & 38			  & 192		  & 1,981\\\midrule
Gentle sports (0)      	  	   & 1,089		       & 848		   & 1,024		   & 848	       & 1,024		  & 307		  & 2,315\\
1 to 3 per month		   &145		       & 452		   & 116		   & 452	       & 116			  & 151		  & 754\\
1 per week			   & 178		       & 1,293	   & 151		   & 1,293	       & 151		           & 311		  & 1,811\\
$>$ 1 per week	                    & 335		       & 5,062	   & 262		   & 5,062	       & 262			  & 887		  & 6,339\\\midrule
Mild sports (0)	           	   & 643		       & 568		   & 610		   & 568	       & 610		   	  & 137		  & 1,375\\
1 to 3 per month		   & 84		       & 303		   & 76		   & 303	       & 76		          & 57		  & 448\\
1 per week			   & 261		       & 828		   & 244		   & 828	       & 244			  & 201		  & 1,307\\
$>$ 1 per week	   	           & 759		       & 5,956	   & 623		   & 5,956	       & 623			  & 1,261		  & 8,089\\\midrule
Bus, train,... (0)		           & 912		       & 1,850	   & 848		   & 1,850	       & 848			  & 406		  & 3,216\\
Rarely				   & 301		       & 2,330	   & 256		   & 2330	       & 256			  & 451		  & 3,112\\
Sometimes			   & 226		       & 1,591	   & 188		   & 1,591	       & 188			  & 313		  & 2,160\\
Quite often			   & 156		       & 855		   & 139		   & 855	       & 139			  & 201		  & 1,234\\
A lot					   & 152		       & 1,029	   & 122		   & 1,029	       & 122			  & 285		  & 1,497\\	\bottomrule	 
\end{tabular}}
\end{table}


\newpage

\begin{thebibliography}{999}
\bibitem{} Action on smoking and health (ASH). Smoking, employability, and earnings. September 2020. \url{https://ash.org.uk/uploads/SmokingEmployability.pdf}
\bibitem{} Arik, Sercan Ö., and Tomas Pfister. (2021). TabNet: Attentive interpretable tabular learning. \textit{Proceedings of the AAAI Conference on Artificial Intelligence}, 35(8), 6679--6687.
\bibitem{} Ba, Jimmy Lei, Jamie Ryan Kiros, and Geoffrey E. Hinton. (2016). Layer normalisation. \textit{arXiv preprint arXiv:1607.06450}.
\bibitem{} Blake, Margaret, Sally Bridges, David Hussey and Dhriti Mandalia. (2015). The dynamics of ageing: The 2010 English Longitudinal Study of Ageing (wave 5).\textit{London: NatCen Social Research}.
\bibitem{} Borisov, Vadim, Tobias Leemann, Kathrin Seßler, Johannes Haug, Martin Pawelczyk, and Gjergji Kasneci. (2022). Deep neural networks and tabular data: A survey. \textit{IEEE Transactions on Neural Networks and Learning Systems}.
\bibitem{} Bridges, Sally, David Hussey, and Margaret Blake. (2015). The dynamics of ageing: The 2012 English Longitudinal Study of Ageing (wave 6). \textit{London: NatCen Social Research}.
\bibitem{} Cai, Shaofeng, Kaiping Zheng, Gang Chen, H. V. Jagadish, Beng Chin Ooi, and Meihui Zhang. (2021). ARM-Net: Adaptive relation modelling network for structured data. \textit{Proceedings of the 2021 International Conference on Management of Data}, 207--220.
\bibitem{} Chakravarty, Eliza F., Helen B. Hubert, Eswar Krishnan, Bonnie B. Bruce, Vijaya B. Lingala, and James F. Fries. (2012). Lifestyle risk factors predict disability and death in healty aging adults. \textit{The AmericanJournal of Medicine}, 125(2), 190--197.
\bibitem{} Chen, Brian K., Hawre Jalal, Hideki Hashimoto, Sze-Chuan Suen, Karen Eggleston, Michael Hurley, Lena Schoemaker, and Jay Bhattacharya. (2016). Forecasting trends in disability in a super-aging society: adapting the future elderly model to Japan. \textit{The Journal of Economics of Ageing}, 8, 42--51.
\bibitem{} Cheng, Heng-Tze, Levent Koc, Jeremiah Harmsen, Tal Shaked, Tushar Chandra, Hrishi Aradhye, Glen Anderson, Greg Corrado, Wei Chai, Mustafa Ispir, Rohan Anil, Zakaria Haque, Lichan Hong, Vihan Jain, Xiaobing Liu, Hemal Shah. (2016). Wide and deep learning for recommender systems. \textit{Proceedings of the 1st workshop on deep learning for recommender systems}, 7--10. 
\bibitem{} Chollet, Francois. (2021). \textit{Deep learning with Python}. Simon and Schuster. 
\bibitem{} Coll-Planas, Laura, Michael Denkinger, and Thorsten Nikolaus. (2008). Relationship of urinary incontinence and late-life disability: implications for clinical work and reserach in geriatrics. \textit{Zeitschrift f\"{u}r Gerontologie und Geriatrie}, 41(4).
\bibitem{} Dauphin, Yann N., Angela Fan, Michael Auli, and David Grangier. (2017). Language modeling with gated convolutional networks. \textit{International Conference on Machine Learning}, 933-941. PMLR.
\bibitem{} Denuit, Michel, Jan Dhaene, Marc Goovaerts, and Rob Kaas. (2006). \textit{Actuarial theory for dependent risks: measures, orders and models}. John Wily \& Sons. 
\bibitem{} Fiedler, James. (2021). Simple modifications to improve tabular neural networks. \textit{arXiv preprint arXiv:2108.03214}.
\bibitem{} Fong, Joelle H., Michael Sherris, and James Yap. (2017). Forecasting disability: application of a frailty model. \textit{Scandinavian Actuarial Journal}, 2017(2), 125--147.
\bibitem{} Freedman, Vicki A., Eileen Crimmins, Robert F. Schoeni, Brenda C. Spillman, Hakan Aykan, Ellen Kramarow, Kenneth Land, James Lubitz, Kenneth Manton, Linda G. Martin, Diane Shinberg, and Timothy Waidmann. (2004). Resolving inconsistencies in trends in old-age disability: report from a technical working group. \textit{Demography}, 41(3), 417--441. 
\bibitem{} Freund Yoav and Robert E. Schapire. (1997). A decision-theoretic generalization of on-line learning and an application to boosting. \textit{Journal of Computer and System Sciences}, 55(1): 119-139. 
\bibitem{} Frosst, Nicholas, and Geoffrey Hinton. (2017). Distilling a neural network into a soft decision tree. \textit{arXiv preprint arXiv:1711.09784}.
\bibitem{} Gobbens, Robbert J. J., and Marcel A. L. M. van Assen. (2014). The prediction of ADL and IADL disability using six physical indicators of frailty: A longitudinal study in the Netherlands. \textit{Current gerontology and geriatrics research}, 2014, 1--10.
\bibitem{} Gobbens, Robbert J. J., Marcel A. L. M. van Assen, Katrien G. Luijkx, and Jos MGA Schols. (2012). The predictive validity of the Tilburg Frailty Indicator: disability, health care utilization, and quality of life in a population at risk. \textit{The Gerontologist} 52(5), 619--631.
\bibitem{} Goodfellow, Ian, Yoshua Bengio, and Aaron Courville. (2016). \textit{Deep learning}. MIT Press.
\bibitem{} Greer, Joy A., Rengyi Xu, Kathleen J. Propert, and Lily A. Arya. (2015). Urinary incontinence and disability in community-dwelling women: A cross-sectional study. \textit{Neurourology and Urodynamics}, 34(6), 539--543.
\bibitem{} Hajjar, Ihab, Daniel T. Lackland, L. Adrienne Cupples, and Lewis A. Lipsitz. (2007). Association between concurrent and remote blood pressure and disability in older adults. \textit{Hypertension}, 50(6), 1026--1032.
\bibitem{} Hastie, Trevor, Robert Tibshirani, and Jerome H. Friedman. (2009). \textit{The elements of statistical learning: data mining, inference, and prediction}. 2nd edition, Springer, New York. 
\bibitem{} He, Feng J., and Graham A. MacGregor. (2007). Blood pressure is the most important cause of death and disability in the world. \textit{European Heart Journal Supplements}, 9(suppl\_B). B23--B28.
\bibitem{} He, Zengyou, Shengchun Deng, and Xiaofei Xu. (2006). Approximation algorithms for k-modes clustering. \textit{International Conference on Intelligent Computing, ICIC 2006 Kunming, China, August 16-19, 2006 Proceedings, Part II 2}, 296--302. Springer Berlin Heidelberg.
\bibitem{} He, Kaiming, Xiangyu Zhang, Shaoqing Ren, and Jian Sun. (2016). Deep residual learning for image recognition. \textit{Proceeding of the IEEE conference on computer vision and pattern recognition}, 770--778.
\bibitem{} Huang, Xin, Ashish Khetan, Milan Cvitkovic, and Zohar Karnin. (2020). TabTransformer: Tabular data modelling using contextual embeddings. \textit{arXiv preprint arXiv:2012.06678}.
\bibitem{} Huang, Zhexue. (1998). Extensions to the k-means algorithm for clustering large data sets with categorical values. \textit{Data Mining and Knowledge Discovery}, 2(3), 283--304.
\bibitem{} Huang, Tongwen, Zhiqi Zhang, and Junlin Zhang. (2019). FiBiNET: Combining feature importance and bilinear feature interaction for click-through rate prediction. \textit{Proceedings of the 13th ACM Conference on Recommender Systems}.
\bibitem{} Ioffe, Sergey, and Christian Szegedy. (2015). Batch normalisation: accelerating deep network training by reducing internal covariate shift. \textit{Proceedings of the 32nd International Conference on Machine Learning}, 448-456. PMLR.
\bibitem{} Kadra, Arlind, Marius Lindauer, Frank Hutter, and Josif Grabocka. (2021). Well.tuned simple nets excel on tabular datasets. \textit{35th Conference on Neural Information Processing Systems}, NeurIPS. 
\bibitem{} Kc, Samir, and Harold Lentzner. (2010). The effect of education on adult mortality and disability: A global perspective. \textit{Vienna Yearbook of Population Research}, 201--235.
\bibitem{} Ke, Guolin, Jia Zhang, Zhenhui Xu, Jiang Bian, and Tien-Yan Liu. (2018). TabNN: A universal neural network solution for tabular data. 
\bibitem{} Laforge, Robert G., William D. Spector, and Josef Sternberg. (1992). The relationship of vision and hearing impairment to one-year mortality and functional decline. \textit{Journal of Ageing and Health}, 4(1), 126--148.
\bibitem{} Lian, Jianxun, Xiaohuan Zhou, Fuzheng Zhang, Zhongxia Chen, Xing Xie, and Guangzhong Sun. (2018). DeepDM: Combining explicit and implicit features interactions for recommender systems. \textit{Proceedings of the 24th ACM SIGKDD International Conference on Knowledge Discovery and Data Mining}.
\bibitem{} Majumdar, Somshubra. (2019). tf-TabNet. \url{https://github.com/titu1994/tf-TabNet}.
\bibitem{} Martins, Ansdre, and Ramon Astudillo. (2016). From softmax to sparsemax: A sparse model of attention and multi-label classification. \textit{International Conference on Machine Learning}, 1614--1623. PMLR.
\bibitem{} Manton, Kenneth G., and XiLiang Gu. (2005). Disability declines and trends in medicare expenditures. \textit{Ageing Horizons}, 2(1), 25--34.
\bibitem{} Murphy, Kevin P. (2012). \textit{Machine learning: A probabilistic perspective}. MIT press.
\bibitem{} Nair, Vinod, and Geoffrey E. Hinton. (2010). Rectified linear units improve restricted boltzmann machines. \textit{In Proceedings of the 27th International Conference on Machine Learning (ICML)}, 10, 807--814.
\bibitem{} Palmer, Michael, and David Harley. (2012). Models and measurement in disability: an international review. \textit{Health Policy and Planning}, 27(5), 357--364. 
\bibitem{} Pascanu, Razvan, Tomas Mikolov, and Yoshua Bengio. (2013). On the difficulty of training recurrent neural networks. \textit{International Conference on Machine Learning}, 1310--1318.
\bibitem{} Paterson, Donald H., and Darren E.R. Warburton. (2010). Physical activity and functional limitations in older adults: a systematic review related to Canada's physical activity guidelines. \textit{International Journal of Behavioral Nutrition and Physical Activity}, 7(1), 1--22.
\bibitem{} Pongiglione, Bendetta, Bianca L. De Stavola, Hannah Kuper, and George B. Ploubidis. (2016). Disability and all-cause mortality in the older population: evidence from the English Longitudinal Study of Ageing. \textit{European Journal of Epidemiology}, 31, 735--746.
\bibitem{} Pongiglione, Benedetta, George B. Ploubidis, and Bianca L. De Stavola. (2017). Levels of disability in the older population of England: Comparing binary and ordinal classifications. \textit{Disability and Health Journal}, 10(4), 509--517. 
\bibitem{} Rumelhart, David E., Geoffrey E. Hinton, and Ronald J. Williams. (1986). Learning representations by back-propagating errors. \textit{nature}, 323 6088): 533--536.
\bibitem{} Shavitt, Ira, and Eran Segal. (2018). Regularization learning networks: deep learning for tabular datasets. \textit{Advances in Neural Information Processing Systems}, 31. 
\bibitem{} Somepalli, Gowthami, Micah Goldblum, Avi Schwarzschild, C. Bayan Bruss, and Tom Goldstein. (2021). SAINT: Improved neural networks for tabular data via row attention and contrastive pre-training. \textit{arXiv preprint arXiv:2106.01342}.
\bibitem{} Srivastava, Nitish, Geoffrey Hinton, Alex Krizhevsky, Ilya Sutskever, and Ruslan Salakhutdinov. (2014). Dropout: A simple way to prevent neural networks from overfitting. \textit{Journal of Machine Learning Research}, 15, 1929--1958.
\bibitem{} Steptoe, Andrew, Elizabeth Breeze, James Banks, and James Nazroo. (2013). Cohort profile: The English Longitudinal Study of Ageing. \textit{International Journal of Epidemiology}, 42(6): 1640--1648. 
\bibitem{} Tibshirani, Robert. (1996). Regression shrinkage and selection via the Lasso. \textit{Journal of the Royal Statistical Society. Series B (Methodological)}, 58(1), 267--288.
\bibitem{} Tieleman, Tijmen, and Geoffrey Hinton. (2012). Lecture 6.5 - RMSProp: Divide the gradient by a running average of its recent magnitude. \textit{COURSERA: Neural Networks for Machine Learning}, 4(2): 26--31. 
\bibitem{} Vaswani, Ashish, Noam Shazeer, Niki Parmar, Jacob Uszkoreit, Llion Jones, Aidan N. Gomez, Lukasz Kaiser, and Illia Polosukhin. (2017). Attention is all you need. \textit{Advances in Neural Information Processing Systems}, NIPS. 
\bibitem{} Vermeulen, Joan, Jacques C. L. Neyens, Erik van Rossum, Marieke D. Spreeuwenberg, and Luc P. de Witte. (2011). Predicting ADL disability in community-dwelling elderly people using physical frailty indicators: a systematic review. \textit{BMC Geriatrics}, 11(1), 1--11.
\bibitem{} Yao, Yuan, Lorenzo Rosasco, and Andrea Caponnetto. (2007). On early stopping in gradient descent learning. \textit{Constructive Approximation}, 26(2), 289--315.
\bibitem{} Ye, Andre, and Zian Wang. (2023). Modern deep learning for tabular data Novel approaches to common modelling problems. Apress.
\end{thebibliography}
\end{document}